\documentclass[a4paper,11pt]{article}
 \pdfoutput=1
 \usepackage{jheppub} 
 \usepackage{graphicx}
 \usepackage{amssymb}
\usepackage{dcolumn}
\usepackage{bm}
\usepackage{color}
\usepackage{multirow}
\usepackage{color}
 
\newcommand{\met}{{/\!\!\! E_T}} 
\newcommand{\mpt}{{\;/\!\!\!\! \vec{P}_T}} 
\allowdisplaybreaks

\definecolor{region1}{RGB}{0,0,0}
\definecolor{region2}{RGB}{0,0,0}
\definecolor{region3}{RGB}{0,0,0}
\definecolor{region4}{RGB}{0,0,0}
\definecolor{region5}{RGB}{0,0,0}
\definecolor{region6}{RGB}{0,0,0}
\definecolor{region7}{RGB}{0,0,0}
\definecolor{region8}{RGB}{0,0,0}
\definecolor{region9}{RGB}{0,0,0}

 \newcommand{\lsim}{{\;\raise0.3ex\hbox{$<$\kern-0.75em\raise-1.1ex\hbox{$\sim$}}\;}}
\newcommand{\gsim}{{\;\raise0.3ex\hbox{$>$\kern-0.75em\raise-1.1ex\hbox{$\sim$}}\;}}
\newcommand{\beq}{\begin{equation}}
\newcommand{\eeq}{\end{equation}}
\newcommand{\bea}{\begin{eqnarray}}
\newcommand{\eea}{\end{eqnarray}}

\def\baa{\begin{array}}
\def\eaa{\end{array}}

\mathchardef\minus="002D

\def\met{E_T\hspace{-0.45cm}/\hspace{0.25cm}}

\title{
\boldmath Improving the sensitivity of stop searches with on-shell constrained invariant mass variables}
 
\author[a,b]{Won Sang Cho,}  
\author[a]{James S.~Gainer,}
\author[a]{Doojin Kim,\footnote{Corresponding author.}}
\author[a]{Konstantin T.~Matchev,} 
\author[c]{Filip Moortgat,}
\author[c]{Luc Pape,}
\author[d,e,f]{Myeonghun Park} 

\affiliation[a]{Physics Department, University of Florida, Gainesville, FL 32611, USA}
\affiliation[b]{Center for Theoretical Physics of the Universe, Institute for Basic Science (IBS), Daejeon 305-811, Republic of Korea}
\affiliation[c]{CERN, Geneva CH-1211, Switzerland}
\affiliation[d]{Kavli IPMU (WPI), The University of Tokyo, Kashiwa, Chiba 277-8583, Japan} 
\affiliation[e]{Asia Pacific Center for Theoretical Physics, 77 Cheongam-Ro, Nam-Gu, Pohang 790-784, Korea}
\affiliation[f]{Department of Physics, Postech, Pohang 790-784, Korea}

\abstract{
The search for light stops is of paramount importance, both in general
as a promising path to the discovery of beyond the standard model
physics and more specifically as a way of evaluating the success of
the naturalness paradigm.  While the LHC experiments have
ruled out much of the relevant parameter space, there are ``stop gaps'',
i.e., values of sparticle masses for which existing LHC analyses have
relatively little sensitivity to light stops.  We point out that
techniques involving on-shell constrained $M_2$ variables can do much
to enhance sensitivity in this region and hence help close the stop
gaps.   We demonstrate the use of these variables for several benchmark
points and describe the effect of realistic complications, such as
detector effects and combinatorial backgrounds, in order to provide a
useful toolkit for light stop searches in particular, and new physics
searches at the LHC in general.
 }

\preprint{
\begin{flushright} 
CTPU-14-11
\\
APCTP Pre2014 - 014
\\
IPMU14-0333
\end{flushright} 
}

\date{November 3, 2014}

\begin{document} 
\maketitle
\flushbottom

\section{Introduction}
\label{sec:introduction}

Supersymmetry (SUSY)~\cite{Martin:1997ns} is an important framework
for beyond the Standard Model physics, as, among other features, it
provides an explanation of the relative lightness of the recently
discovered Higgs boson~\cite{Aad:2012tfa, Chatrchyan:2012ufa} and,
potentially, an explanation of dark matter.  The search for SUSY at
the CERN Large Hadron Collider (LHC), is therefore of considerable
importance.  The generic SUSY discovery channel is missing transverse
energy (MET) accompanied by hard jets, which results from the
production of gluinos and/or first generation squarks via the strong
interaction, followed by the subsequent decay of these sparticles to
final states that include an undetected lightest SUSY particle (LSP),
which is a dark matter candidate.
However, the LHC has yet to find evidence of such a
signal~\cite{Aad:2013wta, Aad:2014pda, Aad:2014wea, Aad:2014bva,
Chatrchyan:2013wxa, Chatrchyan:2013fea, Chatrchyan:2014lfa,
Craig:2013cxa}, which strongly motivates looking for SUSY elsewhere.  
Another class of
strong production processes, with somewhat lower cross sections, 
involves the production of third generation squarks: the stop and the sbottom.

\subsection{Motivation for stop searches}
\label{subsec:motivation_for_stop_searches}

Stop production has long been recognized as a viable SUSY discovery
channel~\cite{Altarelli:1984ve,Bigi:1985aq,Hikasa:1987db,Baer:1991cb,Aebischer:2014lfa,Baer:1994xr,Lopez:1994zw,Chou:1999zb,Demina:1999ty}
 and has been looked for at LEP~\cite{ADLO:2004aa, Heister:2002hp,
   Abdallah:2003xe, Achard:2003ge, Abbiendi:2002mp} and the
Tevatron~\cite{Jaffre:2012gx, Aaltonen:2010uf, Mackin:2010zza}, as
well as at the LHC~\cite{Aad:2014bva, Chatrchyan:2013fea, Aad:2013ija, Aad:2014qaa,
Aad:2014mha, Aad:2014kva, Aad:2014kra, Aad:2014lra, Aad:2014nra,
Chatrchyan:2012paa, Chatrchyan:2013lya, Chatrchyan:2013xsw,
Chatrchyan:2013xna, Chatrchyan:2013iqa, Chatrchyan:2013mya,
Khachatryan:2014doa}.  Searches for the stop are especially
well-motivated theoretically because in many models, the stop is
expected to be the lightest squark for three principal reasons:

\begin{enumerate}
\item The beta function for a squark mass contains a {\em positive}
  term proportional to the corresponding Yukawa coupling. The effect of 
  such a positive term is to suppress the mass when evolved from a high energy scale.  
  Since the top Yukawa coupling is the largest Yukawa coupling, one generically
  expects the stop soft mass parameter to emerge as the lightest of the
  squark soft masses after RGE evolution from some high energy
  scale~\cite{Feng:1998iq}.
  (At large $\tan\beta$, similar arguments will apply to the sbottom
  mass as well.)
\item The left-right off-diagonal mixing in the squark mass matrix
  reduces the smaller mass eigenvalue via level repulsion.  The
  smaller of the two eigenvalues is therefore reduced relative to the
  corresponding diagonal element.  The left-right mixing is an
  $SU(2)$-breaking effect, proportional to the Higgs vacuum
  expectation value and hence to the Yukawa coupling. Thus the stop
  is, again, the squark that would be most affected by a mass-lowering effect. 
\item  The radiative corrections to the tree-level relation among
  $m_Z$ and the Higgs soft mass parameters, which sets the electroweak
  scale, are dominated by stop loops.  Hence a large stop mass would
  destabilize the hierarchy (this has become known as the little
  hierarchy problem).  
  The desire to avoid excessive fine-tuning of the electroweak scale
  has spurred interest in ``natural SUSY'' models
  in which the top squark is among the lightest particles in the
  spectrum~\cite{Dimopoulos:1995mi, Pomarol:1995xc, Cohen:1996vb,
    Feng:1998iq, Perelstein:2007nx}.
\end{enumerate}

Thus in this paper we will consider strategies to discover stops.
Although we have in mind searching for the stop in an arbitrary SUSY
model, everything we will say applies equally to other BSM models with
``top partners".\footnote{Our discussion will also apply to situations where 
there are no new particles and instead the top quark itself 
undergoes a rare decay involving more than one invisible
particle~\cite{Kong:2014jwa}.} 
We assume that all other colored sparticles besides the ``stop'' are
either sufficiently heavy to be ignored or nonexistent.  
We are especially interested in what can be done to 
extend sensitivity into regions of parameter space where existing LHC
searches have not had sufficient sensitivity to discover or rule out
the stop. Our approach is complementary to several recent analyses which
have targeted similarly difficult parameter space regions for stop 
discovery~\cite{Han:2008gy, Carena:2008mj, Perelstein:2008zt,
  Plehn:2010st, Bornhauser:2010mw, Bi:2011ha, He:2011tp, Drees:2012dd,
  Bai:2012gs,Plehn:2012pr, Alves:2012ft, Han:2012fw, Kaplan:2012gd,
  Berger:2012an, Ghosh:2012ud, Chen:2012uw, Kilic:2012kw,
  Graesser:2012qy, Krizka:2012ah, Delgado:2012eu, Dutta:2013sta,
  Buckley:2013lpa, Chakraborty:2013moa, Low:2013aza, Bai:2013ema,Ghosh:2013qga,
  Belanger:2013oka, Dutta:2013gga, Buckley:2014fqa, Ortiz:2014iza,
  Czakon:2014fka, Ismail:2014cma, Eifert:2014kea}.

\subsection{Stop decays}
\label{subsec:stop_decays}

Having identified and motivated the production mode that we will
consider, we now must decide on the manner in which the particles will
decay.
Unlike other squarks, for which only the gauge couplings are non-negligible
(thereby reducing the number of potentially relevant decay modes), stops have many 
viable decay modes --- there exist two-body decays of stops
to gluinos ($\tilde t \to \tilde g + t$), neutralinos ($\tilde t \to \tilde \chi^0_i + t$), 
charginos ($\tilde t \to \tilde \chi^+_i + b$), gravitinos ($\tilde t
\to \tilde G + t$)~\cite{Chou:1999zb}, 
or even other stops ($\tilde t_2 \to \tilde t_1 +Z$)~\cite{Perelstein:2007nx}
and sbottoms ($\tilde t \to \tilde b + W^+$)~\cite{Datta:2011ef}. 
When the two-body decays are suppressed, there are several 
three-body decays which may dominate, e.g.~$\tilde t \to b W^+ \tilde \chi^0_i$,
$\tilde t \to b W^+ \tilde G$, $\tilde t \to b \ell^+ \tilde \nu$,
$\tilde t \to b \nu \tilde \ell^+$, etc.~\cite{Hikasa:1987db,
  Baer:1991cb, Chou:1999zb, Demina:1999ty}. Finally, 
there can also be loop-induced two-body decays, e.g.~$\tilde t \to c
\tilde\chi^0_i$ \cite{Hikasa:1987db, Baer:1991cb,Aebischer:2014lfa}.

\begin{figure}[t]
\centering
\includegraphics[scale=0.9]{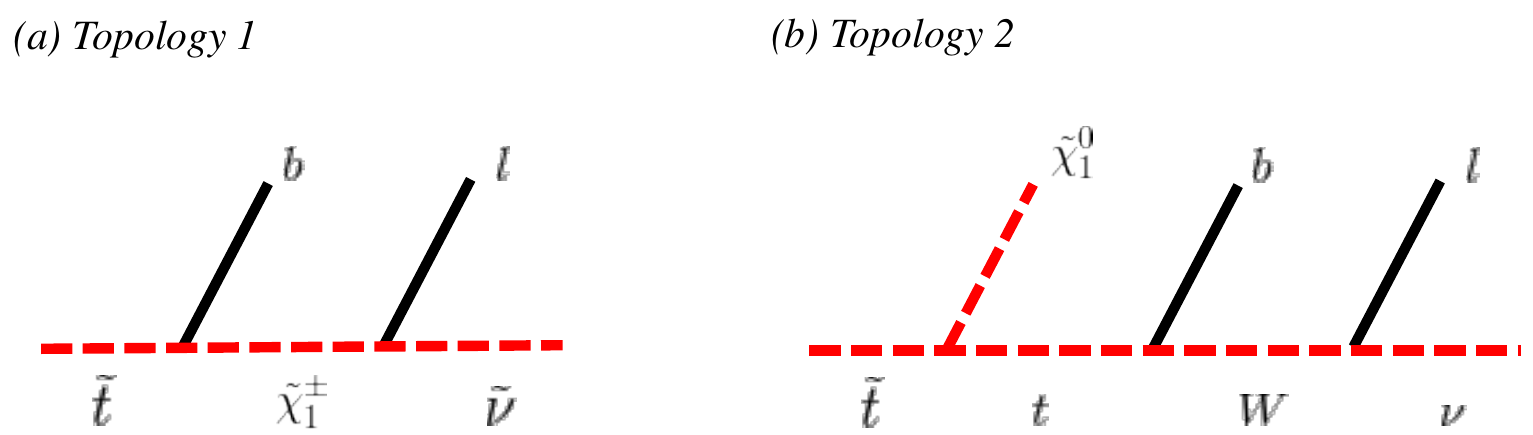}
\caption{\label{fig:process} The two signal decay topologies under
  consideration in this paper. 
In diagram (a), a stop (antistop) decays to a bottom (antibottom)
quark and a positively (negatively) charged on-shell chargino; the
chargino decays into an antilepton
(lepton) and sneutrino (antisneutrino). 
In diagram (b), the stop (antistop) decays to a top (antitop) quark
and a neutralino.  The top (antitop) in turns decays to a bottom
(antibottom) quark and a positively (negatively) on-shell charged $W^\pm$
boson, which decays, in turn, to an antilepton (lepton) and neutrino
(antineutrino).  We refer to the topology in diagram (a) as ``Topology
1'' and the topology in diagram (b) as ``Topology 2''.}
\end{figure}

In this paper we shall focus on the most challenging scenario, in
which the stop produces\footnote{As usual, we assume that the 
stop decay proceeds through a ``decay chain'' of sequential 
decays of on-shell intermediate particles.} the same visible 
particles as a top quark decaying leptonically.  In particular, we
shall consider the two signal decay topologies shown in 
Fig.~\ref{fig:process}, which commonly occur in realistic 
models\footnote{In principle, in addition to the two examples
from Fig.~\ref{fig:process}, there are many other decay topologies
which can mimic a top decay --- for example, there can be additional 
invisible particles emitted in this process~\cite{Agashe:2010gt,
  Agashe:2010tu, Giudice:2011ib, Cho:2012er}, 
or the neutralino in Fig.~\ref{fig:process}(b) can be emitted in
between the bottom quark and the lepton.}.
In the process of Fig.~\ref{fig:process}(a), which we refer to as
``Topology 1", the stop decay chain is identical to the ``leptonic''
decay of a top quark; the only differences are that here the role of
the $W^\pm$ is played by the chargino $\tilde \chi^\pm_1$, and the
role of the neutrino is played by the sneutrino $\tilde \nu$.
(The sneutrino may further decay invisibly to another DM candidate;
since the sneutrino is on-shell, this does not affect our analysis.)
In the process of Fig.~\ref{fig:process}(b), which we shall refer to
as ``Topology 2", the stop decays to a top quark and a neutralino
first; the top then decays leptonically.  The resulting {\em visible}
final state is the same, the difference now is that there are {\em
  two} invisible particles -- a neutralino $\tilde \chi^0_1$ and a
neutrino $\nu$.  When studying Topology 2, we shall assume that the
mass splitting between the stop and the neutralino is large enough
that the top quark produced in this decay is on-shell, as this makes
it \textit{more} difficult to distinguish the signal from top backgrounds.

\begin{figure}[t]
\centering
\includegraphics[scale=0.9]{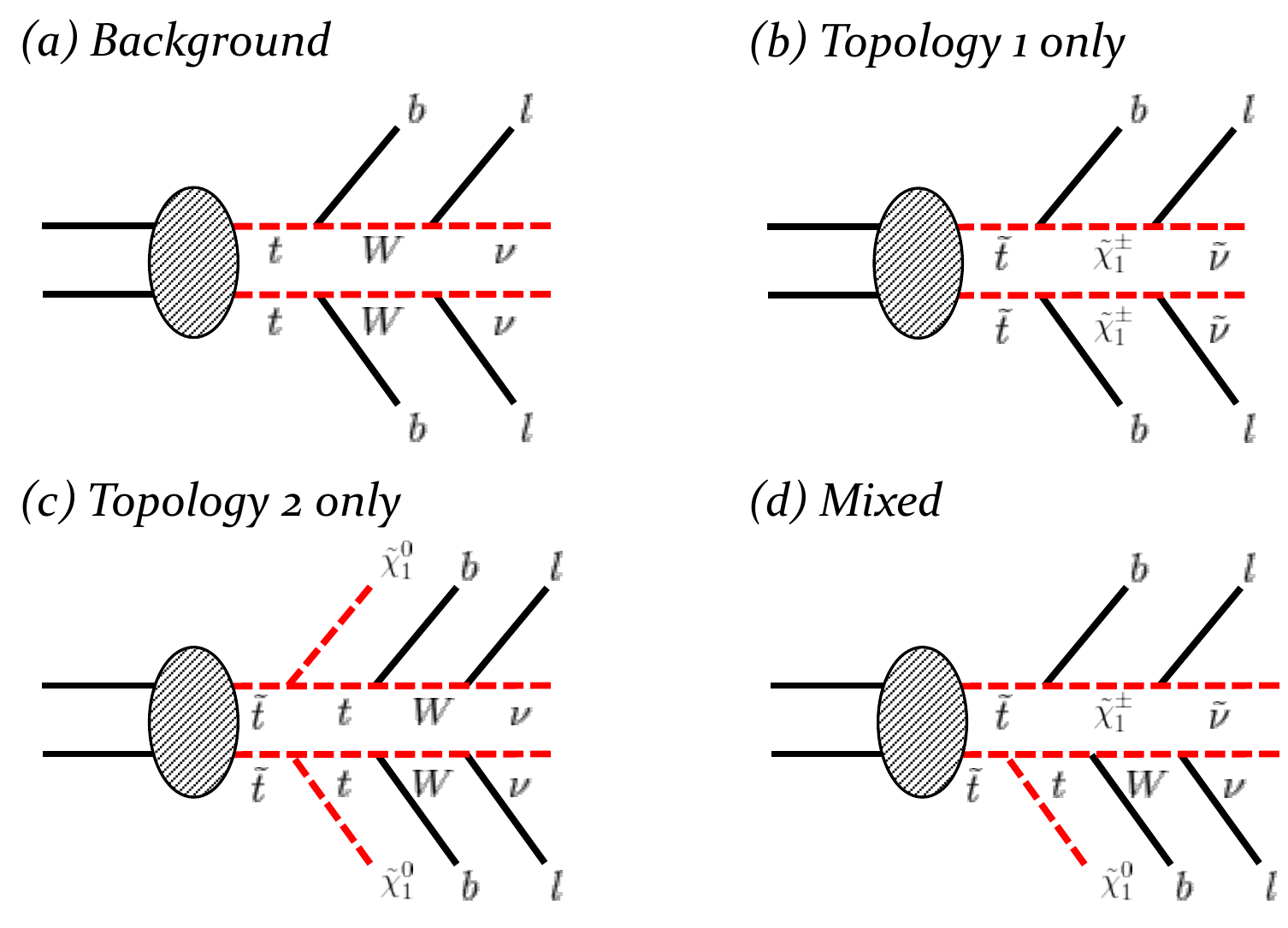}
\caption{\label{fig:DecayTopologies} The four event topologies
  considered in this paper.  Diagram (a) gives the background event topology,
  where the top quarks decay leptonically, i.e.,~as in Fig.~\ref{fig:process}(a),
  but with the charginos replaced by $W^\pm$ bosons and
  the sneutrinos replaced by neutrinos. Diagram (b) is the (signal) event topology
  when both stops decay according to decay Topology 1, diagram (c) is the
  event topology when both stops decay following Topology 2,
  and diagram (d) is the mixed event topology when one stop decays according to
  Topology 1 and the other according to Topology 2.}
\end{figure}

Specifically, we will consider how to discriminate between stop
production, where both stops decay according to one of the topologies
in Fig.~\ref{fig:process}, and the irreducible background from
$t\bar{t}$ dilepton events. Fig.~\ref{fig:DecayTopologies}(a)
illustrates the background event topology, while the corresponding
three possible signal event topologies are depicted in
Figs.~\ref{fig:DecayTopologies}(b-d).  In all these processes, the observed
final state consists of two $b$-jets, two opposite sign (OS) leptons,
and MET, which makes it quite challenging to discover stops in this channel.

Note in particular that our analysis will allow for the mixed event
topology of Fig.~\ref{fig:DecayTopologies}(d).  This is because we shall 
{\em not} make any assumptions about the relative branching
fraction between Topologies 1 and 2 in Fig.~\ref{fig:process}.  If the
two branching fractions are comparable, there is a sizable fraction of
events of the type depicted in Fig.~\ref{fig:DecayTopologies}(d);
their number benefits also from the combinatorial factor of 2 relative
to the events in Fig.~\ref{fig:DecayTopologies}(b) or the events in
Fig.~\ref{fig:DecayTopologies}(c).

\begin{figure}[t]
\centering
\includegraphics[scale=1]{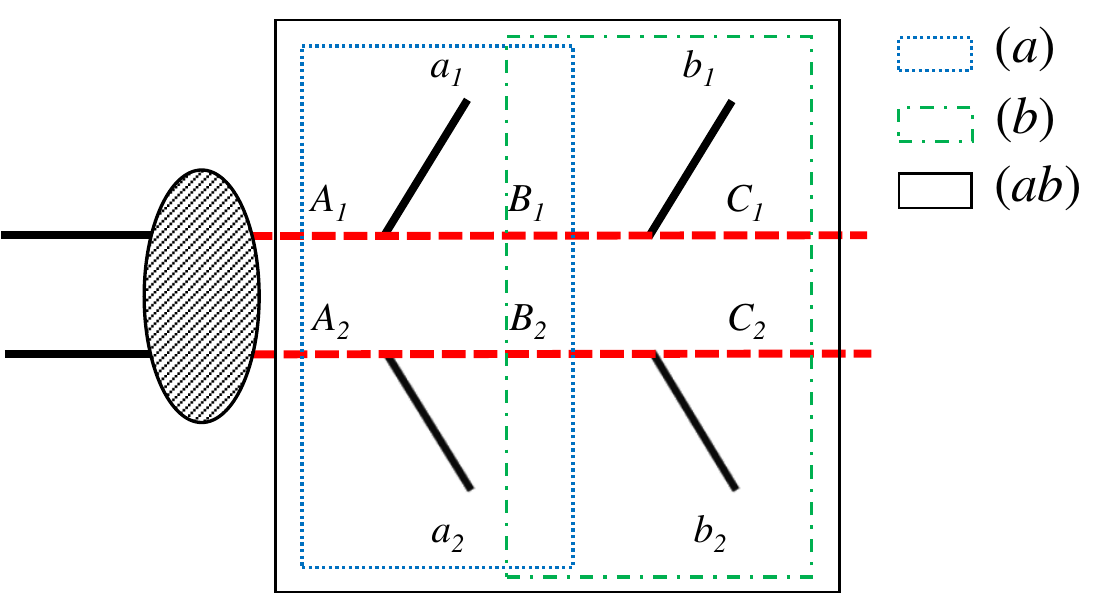}
\caption{\label{fig:DecaySubsystem} The background event
topology of Fig.~\ref{fig:DecayTopologies}(a)
with the corresponding subsystems explicitly delineated. The blue
dotted, green dot-dashed, and black solid lines indicate the
subsystems $(a)$, $(b)$, and $(ab)$, respectively.  See also
Table~\ref{tab:background-subsystems}.}
\end{figure}

\subsection{On-shell constrained $M_2$ variables}
\label{subsec:on_shell_constrained_M2_variables}

In this paper we investigate the benefit of the recently proposed
on-shell constrained $M_2$
variables~\cite{Barr:2011xt,Mahbubani:2012kx,Cho:2014naa} 
in discriminating between the signal events of
Fig.~\ref{fig:DecayTopologies}(b-d) and the main background shown
in Fig.~\ref{fig:DecayTopologies}(a).  The $M_2$ variables are the natural
$3+1$-dimensional generalizations~\cite{Barr:2011xt} of the Cambridge 
$M_{T2}$ variable~\cite{Lester:1999tx,Barr:2003rg}, which is already known
as a useful tool for background
suppression~\cite{Barr:2009wu,Allanach:2011ej,Murayama:2011hj}.
Both $M_2$ and $M_{T2}$ were designed
for events in which particles are pair produced and decay
semi-invisibly.  (I.e., some of the decay products are invisible.)
The variables can then be computed for different subsystems in the event, 
or for the original event as a whole~\cite{Burns:2008va}. 
For example, in the case of the background $t\bar{t}$ events 
from Fig.~\ref{fig:DecayTopologies}(a), there are three possibilities, which
are shown in Fig.~\ref{fig:DecaySubsystem}.  
We shall follow the notation of~\cite{Cho:2014naa} and label these
three possibilities as $(ab)$, $(a)$, and $(b)$.  

In the spirit of many other kinematic variables such as
$M_{T2}$~\cite{Lester:1999tx},
$M_{2C}$~\cite{Ross:2007rm}, and $M_{CT2}$~\cite{Cho:2009ve},
the $M_2$ variables are obtained by minimizing some parent invariant mass
with respect to the momenta of the invisible daughter particles.  (The
exact definition and
basic properties of the constrained $M_2$ variables are reviewed in 
Sec.~\ref{sec:review_of_on_shell_constrained_M2_variables} below.) In
the process of minimization, one may additionally impose certain
kinematic constraints which follow
from the hypothesized event topology. The main advantage of the $M_2$ class
of variables is that, being $3+1$-dimensional, they allow one to incorporate 
all known kinematic
constraints~\cite{Mahbubani:2012kx,Cho:2014naa}. For example, $M_{T2}$
and $M_{CT2}$ are transverse variables, and the only constraint which
can be used in their calculation
is the MET constraint. On the other hand, in calculating an $M_2$ variable,
one is free to impose additional mass shell conditions following either from a
previous measurement (as in the case of $M_{2C}$) or from a
theoretical hypothesis
about the specific nature of the events, e.g. that the two decay chains in 
Fig.~\ref{fig:DecaySubsystem} are the same and thus $M_{A_1}=M_{A_2}$,
$M_{B_1}=M_{B_2}$ and $M_{C_1}=M_{C_2}$. 

The presence of additional on-shell constraints
will generally increase the calculated value of $M_2$ --- the more constraints 
there are, the larger the value of $M_2$. This simple observation will be at the
center of our discussion below, and we shall use it in several different ways:
\begin{itemize}
\item The imposition of the additional constraints raises the value of $M_2$, 
distorting the shape of the $M_2$ distribution by increasing the 
populations of the higher $M_2$ bins. We can use this effect to increase 
the signal efficiency in those cases where the background $M_2$ distribution is
bounded by an upper kinematic endpoint, while the signal $M_2$ distribution extends 
beyond this kinematic endpoint. This effect will be discussed and illustrated 
in Sec.~\ref{sec:M2_endpoint_study_for_topology_1} for signal events 
with Topology 1 only, and in Sec.~\ref{sec:M2_endpoint_study_for_topology_2} 
for signal events with Topology 2 only. In either case, the two top squarks decay identically, 
giving rise to the ``symmetric"  event topologies of Fig.~\ref{fig:DecayTopologies}(b) 
and Fig.~\ref{fig:DecayTopologies}(c), respectively.
\item The kinematic variables $M_{T2}$ and $M_2$ were originally designed 
for symmetric events, and intended to be used for such events only.
But what if the actual event is ``asymmetric" and the two parent particles 
decay in a different manner, e.g.,~as in Fig.~\ref{fig:DecayTopologies}(d)?
There are two possible approaches. First, one could suitably modify the
definition of $M_{T2}$ in order to adapt it to an asymmetric
case~\cite{Barr:2009jv,Konar:2009qr}.
This would still work, provided that {\em both} the signal and background have the same 
asymmetric event topology. However, what if the background events are symmetric (as in Fig.~\ref{fig:DecayTopologies}(a)),
while the signal events are asymmetric (as in Fig.~\ref{fig:DecayTopologies}(d)) 
or vice versa? This case is the subject of
Sec.~\ref{sec:mixed-events}, in which we shall advocate the use of the
conventional ``symmetric" 
on-shell constrained $M_2$ variables for this asymmetric case as well. 
As an illustration, we shall consider a particularly difficult scenario, when
the kinematic endpoints of the symmetric events with Topology 1
(Fig.~\ref{fig:DecayTopologies}(b))
or Topology 2 (Fig.~\ref{fig:DecayTopologies}(c)) are too low and are
both ``buried" inside the background distribution. Nevertheless, when
we compute the $M_2$ variables for the mixed event topology of
Fig.~\ref{fig:DecayTopologies}(d), we shall find that the signal
distribution does extend beyond the background endpoint.  The reason
for this apparent ``endpoint violation" is simply the fact that when
calculating $M_2$, we are applying the ``wrong" constraints on the
signal events, but the ``correct" constraints on the background events.
\item The benefit of the on-shell constrained $M_2$ variables is not
  limited only to events in which the background $M_2$ endpoint is
  violated. We can achieve additional separation of signal from
  background by studying the size of the shift caused by the
  application of the on-shell constraints. In
  Sec.~\ref{sec:detector_resolution_etc} we shall find that this shift
  is generally larger for signal events with the mixed event topology
  of Fig.~\ref{fig:DecayTopologies}(d) when compared to the
  corresponding shift for background events.  This observation is
  valid even for signal events which do not violate the background
  $M_2$ endpoint.  The study presented in
  Sec.~\ref{sec:detector_resolution_etc} also includes the effects
  of detector resolution and combinatorial backgrounds.
\end{itemize}

\subsection{Pr\'{e}cis}
\label{subsec:precis}

In this paper we put forth four ideas for stop discovery.
\begin{enumerate}
\item We propose to use the on-shell constrained $M_2$ variables from
  {\em each} of the three subsystems ($(a)$, $(b)$, and
  $(ab)$)~\cite{Burns:2008va,Bai:2013ema}.  Previous efforts described
  in the literature have relied mostly on the dileptonic
  $M_{T2}$~\cite{Plehn:2012pr,Kilic:2012kw,Chakraborty:2013moa,Cho:2008cu}.
\item We advocate the use of the on-shell constrained
  $(3+1)$-dimensional $M_2$ variables in place of their transverse
  cousin $M_{T2}$, due to their ability to ``push'' more signal events
  beyond the background endpoints, thus increasing the signal
  efficiency\footnote{The variable $M_{T2}^W$ introduced in
    \cite{Bai:2012gs} is a concrete realization of an on-shell
    constrained $M_2$ variable in the case when one of the leptons in
    Fig.~\ref{fig:DecayTopologies} is lost.}.
\item We also propose to use the {\em difference} between the on-shell
  constrained $M_2$ variable and its analogue $M_{T2}$ as an
  additional discriminator against the background.
\item We show that the on-shell constrained $M_2$ variable is able to
  specifically target the mixed signal event topology of
  Fig.~\ref{fig:DecayTopologies}(d) and salvage a certain fraction of
  signal events in difficult scenarios when more conventional cuts
  would fail.
\end{enumerate}

\section{Review of on-shell constrained $M_2$ variables}
\label{sec:review_of_on_shell_constrained_M2_variables}

In this section we provide a brief review of on-shell constrained $M_2$
variables. Readers who are familiar with the terminology and notation 
of Ref.~\cite{Cho:2014naa} may skip directly to
Sec.~\ref{sec:M2_endpoint_study_for_topology_1}.
We begin by reminding the reader that there exists a broad class of kinematic
variables that are useful in the analysis of events with missing energy. 
These variables are defined in two steps \cite{Barr:2011xt}:
\begin{itemize}
  \item One first assumes a particular event topology consistent with the
    final state particles observed in the event.
  \item One then minimizes an invariant mass quantity in the
    hypothesized topology over the
    unknown invisible momenta, subject to certain kinematic constraints.
\end{itemize}
The best known example of such a variable is the usual transverse mass $M_T$,
~\cite{Smith:1983aa,Barger:1983wf}, which applies to the simple case
of one decay chain with
a single invisible particle. Although a transverse quantity, $M_T$ can be 
thought of as the minimum value of the 3+1 dimensional invariant mass,
consistent with the measured missing transverse
momentum~\cite{Barr:2011xt}.
This example is rather trivial in the sense that
imposing the vector constraint of transverse momentum conservation already 
fixes the (transverse) components of the invisible particle momentum,
and there is only one minimization left to do. A much more interesting
case arises
when there are two decay chains, with one invisible particle in each. Then the 
concept of $M_T$ is generalized to the stransverse mass,
$M_{T2}$~\cite{Lester:1999tx},
in which one finds the minimum, with respect to the
momenta of invisible particles, of the maximum transverse mass of a
given particle on either side of a (symmetric) decay topology, again 
subject to the constraint of total transverse momentum conservation.

The on-shell constrained $M_2$ variables described in~\cite{Cho:2014naa}  
are analogous to $M_{T2}$, with two main differences. First, the quantity 
being minimized is the four-dimensional invariant mass
rather than the transverse mass (see also~\cite{Ross:2007rm,
  Barr:2011xt, Mahbubani:2012kx}). 
Second, in addition to transverse momentum conservation, one is free
to apply additional
on-shell constraints which follow from the assumed event topology.
For concreteness, let us use the event topology of
Fig.~\ref{fig:DecaySubsystem}
to illustrate the procedure of defining the different types of $M_2$
variables, denoted as
\beq
M_{2\sqcup\sqcup}(S;\tilde m).
\label{M2notation}
\eeq
Here $S\in \{ (ab), (a), (b) \}$ denotes the subsystem under consideration, while
$\sqcup$ is an index placeholder to be defined shortly.
\begin{table}[t]
\centering
\begin{tabular}{| c | c | c | c |}
\hline
Subsystem & Parents & Daughters & Relatives  \\ 
$S$            &  $P_i$    &  $D_i$       & $R_i$  \\
\hline \hline
$(ab)$  &   $\{A_i\}=\{t,\bar{t}\,\}$   &    $\{C_i\}=\{\nu,\bar{\nu}\, \}$  &  $\{B_i\}=\{W^+, W^-\}$ \\
$(a)$    &   $\{A_i\}=\{t,\bar{t}\,\}$   &    $\{B_i\}=\{W^+,W^-\}$         &   $\{C_i\}=\{\nu,\bar{\nu}\,\}$ \\
$(b)$    &   $\{B_i\}=\{W^+,W^-\}$  &    $\{C_i\}=\{\nu,\bar{\nu}\,\}$ &     $\{A_i\}=\{t,\bar{t}\,\}$ \\
\hline
\end{tabular}
\caption{\label{tab:background-subsystems} The roles played by
  different particles in the background process of $t\bar{t}$ production
   (Fig.~\ref{fig:DecayTopologies}(a)), for each of the three
   subsystems defined in  Fig.~\ref{fig:DecaySubsystem}. }
\end{table}
According to the nomenclature of~\cite{Cho:2014naa}, depending on the 
subsystem being considered,
the intermediate particles $A_i$, $B_i$, and $C_i$ fall into one of the following 
three categories (see Table~\ref{tab:background-subsystems}):
\begin{itemize}
\item {\em Daughters} $D_i$. These are the invisible particles at the end of
the decay chains in the subsystem under consideration. 
Following~\cite{Cho:2014naa}, we shall denote their 3-momenta by
$\vec{q}_1$ and $\vec{q}_2$,
respectively. The value of the $M_2$ variable (\ref{M2notation}) will
be obtained by
minimizing over all possible values of $\vec{q}_1$ and $\vec{q}_2$,
consistent with the applied kinematic constraints.
As usual, we shall take the daughters' masses to be equal:
\begin{equation}
\label{eq:daughters-equal}
M_{D_1} = M_{D_2} \equiv \tilde m
\end{equation}
and denote them with $\tilde m$, which will be an input parameter for
the $M_2$ calculation.
\item {\em Parents} $P_i$. These are the two particles at the top of
  the decay chains in the subsystem,
and their masses will be subject to minimization over the invisible momenta in 
order to obtain the variable (\ref{M2notation}). When performing this
minimization, in addition to the missing transverse momentum constraint
\beq
\vec{q}_{1T}+\vec{q}_{2T} = \mpt, 
\eeq
we can additionally require that the two parent masses 
(when considered as functions of the invisible momenta) are the same:
\begin{equation}
\label{eq:parents-equal}
M^2_{P_1}(\vec{q}_1,\tilde m) = M^2_{P_2}(\vec{q}_2,\tilde m).
\end{equation}
The presence (or absence) of this constraint is indicated by the first
${}_\sqcup$ index in (\ref{M2notation}),
which takes value $C$ if the constraint is applied, and $X$ otherwise:
\bea
M_{2C\sqcup}(S;\tilde m)  &\equiv&
\min_{\substack{\vec{q}_{1},\vec{q}_{2}\\ 
\vec{q}_{1T}+\vec{q}_{2T} = \mpt   \\
M^2_{P_1}(\vec{q}_1,\tilde m) = M^2_{P_2}(\vec{q}_2,\tilde m) } }
\left\{M_{P_1}(\vec{q}_{1},\tilde m) \right\},  
\label{M2Cbla} \\
M_{2X\sqcup}(S;\tilde m)  &\equiv&
\min_{\substack{\vec{q}_{1},\vec{q}_{2}\\ 
\vec{q}_{1T}+\vec{q}_{2T} = \mpt} }
\left\{ \max\left[ M_{P_1}(\vec{q}_{1},\tilde m),
    M_{P_2}(\vec{q}_{2},\tilde m) \right] \right\}.
\label{M2Xbla} 
\eea
\item {\em Relatives} $R_i$. As shown in
  Table~\ref{tab:background-subsystems}, the relatives are the
  remaining particles in the event topology --- they are neither
  parents nor daughters. Depending on the subsystem, relatives can
  appear either inside or outside the subsystem.  Their masses can
  also be written as functions of the respective daughters' momenta,
  so by requiring equal masses\footnote{We note that while a parent
    mass squared is always positive, the mass squared of a {\em
      relative} could be negative: keep in mind that the values for
    the invisible momenta $\vec{q}_1$ and $\vec{q}_2$ found in the
    minimization process are not the true momenta and could be
    unphysical, i.e., there is no guarantee that $M^2_{R_i}>0$.  While
    one has the option of adding the further constraint that the
    squared masses of the relative particles be positive, we will not
    do so in this work.} for the relative particles,
\begin{equation}
\label{eq:relatives-equal}
M_{R_1}^2 (\vec{q}_1,\tilde m)= M_{R_2}^2(\vec{q}_2, \tilde m),
\end{equation} 
we are, in effect, imposing an additional constraint on the
minimization over the invisible momenta $\vec{q}_i$. The applicability
of the constraint (\ref{eq:relatives-equal}) will be indicated by the
second ${}_\sqcup$ index in (\ref{M2notation}): as before, it will be
equal to $C$ if the constraint (\ref{eq:relatives-equal}) is applied
and $X$ otherwise.
Altogether, therefore, we have four possible $M_2$ variables:
\bea
M_{2CC}(S;\tilde m)  &\equiv&
\min_{\substack{\vec{q}_{1},\vec{q}_{2}\\ 
\vec{q}_{1T}+\vec{q}_{2T} = \mpt   \\
M_{P_1}^2(\vec{q}_1,\tilde m) = M_{P_2}^2(\vec{q}_2,\tilde m) \\
M_{R_1}^2(\vec{q}_1,\tilde m) = M_{R_2}^2(\vec{q}_2,\tilde m)} }
\left\{M_{P_1}(\vec{q}_{1},\tilde m) \right\},  
\label{M2CCdef} \\ [2mm]
M_{2CX}(S;\tilde m)  &\equiv&
\min_{\substack{\vec{q}_{1},\vec{q}_{2}\\ 
\vec{q}_{1T}+\vec{q}_{2T} = \mpt \\
M_{P_1}^2(\vec{q}_1,\tilde m) = M_{P_2}^2(\vec{q}_2,\tilde m)} }
\left\{M_{P_1}(\vec{q}_{1},\tilde m) \right\},  
\label{M2CXdef} \\ [2mm]
M_{2XC}(S;\tilde m)  &\equiv&
\min_{\substack{\vec{q}_{1},\vec{q}_{2}\\ 
\vec{q}_{1T}+\vec{q}_{2T} = \mpt \\
M_{R_1}^2(\vec{q}_1,\tilde m) = M_{R_2}^2(\vec{q}_2,\tilde m)} }
\left\{ \max\left[ M_{P_1}(\vec{q}_{1},\tilde m), M_{P_2}(\vec{q}_{2},\tilde m) \right] \right\},
\label{M2XCdef} \\ [2mm]
M_{2XX}(S;\tilde m)  &\equiv&
\min_{\substack{\vec{q}_{1},\vec{q}_{2}\\ 
\vec{q}_{1T}+\vec{q}_{2T} = \mpt } }
\left\{ \max\left[ M_{P_1}(\vec{q}_{1},\tilde m), M_{P_2}(\vec{q}_{2},\tilde m) \right] \right\}.
\label{M2XXdef} 
\eea
\end{itemize}

The definitions (\ref{M2CCdef}-\ref{M2XXdef}) should be contrasted to
the analogous definition of the Cambridge $M_{T2}$ variable
\beq
M_{T2}(S;\tilde m)  \equiv
\min_{\substack{\vec{q}_{1T},\vec{q}_{2T}\\ 
\vec{q}_{1T}+\vec{q}_{2T} = \mpt  } }
\left\{ \max\left[ M_{TP_1}(\vec{q}_{1T},\tilde m),
    M_{TP_2}(\vec{q}_{2T},\tilde m) \right] \right\},
\label{MT2def} 
\eeq
where only the transverse components $\vec{q}_{iT}$ are used, and the 
objective function (the function that is minimized) is the larger of
the two {\em transverse} masses of the parents.

The main goal of this paper is to investigate and contrast the ability of the
variables (\ref{M2CCdef}-\ref{MT2def}) to discriminate between the $t\bar{t}$
background of Fig.~\ref{fig:DecayTopologies}(a) and the three types of
stop signal events of Fig.~\ref{fig:DecayTopologies}(b-d). We shall
consider the respective variables for all three subsystems of
Fig.~\ref{fig:DecaySubsystem}. Since we know the mass spectrum for the
background event topology, we shall choose the test mass $\tilde m$ to
be equal to the correct, SM value for the respective daughter
particle. In particular,
\begin{itemize}
\item In subsystem $(ab)$, the parent particle is the top quark;
the daughter particle is the neutrino, whose mass is taken to vanish
($\tilde m=0$).  The other, ``relative'', particle is the $W^\pm$ boson. 
Then {\em for background events}, all 5 variables (\ref{M2CCdef}-\ref{MT2def}) 
are bounded from above by the top mass:
\beq
M_{T2}(ab;0), 
M_{2XX}(ab;0),
M_{2CX}(ab;0),
M_{2XC}(ab;0),
M_{2CC}(ab;0)
\le m_t,
\label{mtab}
\eeq
while for signal events, this bound can be violated.
\item In subsystem $(a)$, the parent particle is again the top quark.
The daughter particle is now the $W^\pm$ boson, with mass $\tilde m= m_W$.
The relative particles are the two neutrinos, which in this case appear
downstream outside the subsystem. For background events, the 
variables are again bounded by the top mass
\beq
M_{T2}(a;m_W), 
M_{2XX}(a;m_W),
M_{2CX}(a;m_W),
M_{2XC}(a;m_W),
M_{2CC}(a;m_W)
\le m_t.
\label{mta}
\eeq
\item In subsystem $(b)$ the parent particles are the $W^\pm$ bosons, 
the daughter particles are the two neutrinos with mass $\tilde m=0$,
and the relative particles are the top quarks appearing upstream
outside the subsystem. The background events obey
\beq
M_{T2}(b;0), 
M_{2XX}(b;0),
M_{2CX}(b;0),
M_{2XC}(b;0),
M_{2CC}(b;0)
\le m_W.
\label{mwb}
\eeq
\end{itemize}

In principle, {\em each} of the bounds (\ref{mtab}-\ref{mwb}) allows
us to cut $100\%$ of the
$t\bar{t}$ background events by removing events with values of the
respective subsystem $M_2$ or $M_{T2}$ variable below the appropriate
threshold (``high pass cut''). 
Therefore, as far as just the background is concerned,
we have 15 alternative choices\footnote{Each of the five variables
  (\ref{M2CCdef}-\ref{MT2def}) can be applied for each of the three
  subsystems $(ab)$, $(a)$, and $(b)$. The explicit definitions of the
  resulting 15 variables can be found in Appendix~\ref{sec:appendix}.}
for reducing it, and they should perform comparably well.
The differences between the five variables
(\ref{M2CCdef}-\ref{MT2def}) begin to emerge when we consider the
effect of such a high pass cut on signal events.  It has been
shown~\cite{Cho:2014naa} that the variables
(\ref{M2CCdef}-\ref{MT2def}) obey the following hierarchy
\begin{equation}
    \label{eq:inequality}
  M_{T2}(S; \tilde{m}) =  M_{2XX}(S; \tilde{m}) = M_{2CX}(S;
  \tilde{m})  \leq M_{2XC}(S; \tilde{m}) \leq M_{2CC}(S; \tilde{m})
\end{equation} 
for any subsystem $S$ and any value of $\tilde m$.
We should therefore expect the distributions of $M_{2XC}$ and $M_{2CC}$ 
to be more populated at higher values and in particular near their endpoints.
As a result, signal events are more likely to pass if the cut is applied on 
the additionally constrained variables, $M_{2XC}$ and $M_{2CC}$,
as opposed to the less constrained variables $M_{T2}$, $M_{2XX}$, and $M_{2CX}$.
This expectation is borne out by the explicit studies below.  
The property (\ref{eq:inequality}) offers an opportunity to increase 
the sensitivity of the LHC experiments to the presence of stop signals of the 
type described in Fig.~\ref{fig:DecayTopologies}(b-d).

\section{$M_2$ endpoint study for topology 1}
\label{sec:M2_endpoint_study_for_topology_1}

We begin by studying the effectiveness of the on-shell constrained
subsystem $M_2$ variables in the case where both stops in the signal
event decay according to Topology 1. This yields the event topology
pictured in Fig.~\ref{fig:DecayTopologies}(b). The background, as always in
this paper, consists of dileptonic top production and is shown in 
Fig.~\ref{fig:DecayTopologies}(a).
We remind the reader that when applied to {\em background events},
all of the on-shell constrained subsystem $M_2$ variables
(as well as $M_{T2}$) exhibit very well-defined kinematic endpoints given
by Eqs.~(\ref{mtab}-\ref{mwb}). Therefore, the effectiveness of the $M_2$ 
variables in identifying signal events is determined by how many signal events
violate the bounds (\ref{mtab}-\ref{mwb}).

The signal events considered in this section (those of
Fig.~\ref{fig:DecayTopologies}(b)) 
have exactly the same topology as the background. Therefore, when applied to 
{\em signal events}, the $M_2$ variables will have well-defined upper
kinematic endpoints as well. The precise value of those endpoints will
depend on the underlying signal mass spectrum, i.e., on the true
values of the stop mass $m_{\tilde t}$, 
the chargino mass $m_{\tilde\chi^\pm}$, and the sneutrino mass $m_{\tilde \nu}$.
For any given point $\{ m_{\tilde t}, m_{\tilde\chi^\pm}, m_{\tilde \nu}\}$ 
in the mass parameter space, using the formulas given in
Ref.~\cite{Burns:2008va}, one can compute the expected $M_2$ kinematic
endpoints for the signal, in each 
of the three subsystems $(ab)$, $(a)$, and $(b)$.\footnote{We remind 
the reader that in this work, the $M_2$ variables are always computed
with test masses corresponding to the background hypothesis; see
Appendix~\ref{sec:appendix}.} 
Depending on the SUSY mass spectrum, some, all, or none of these
signal endpoints will exceed the corresponding background
endpoints. In the second half of this section, we shall illustrate
each of these three scenarios with specific study points. But first we
shall analyze the relevant mass parameter space and categorize the
different regions, which are defined by the location of the signal endpoints
relative to the background endpoints.  This will be the subject of the
next subsection.

\subsection{Anatomy of the mass parameter space for Topology 1}
\label{subsec:regions-and-study-points}

\begin{figure}[t]
\centering
\includegraphics[scale=0.6]{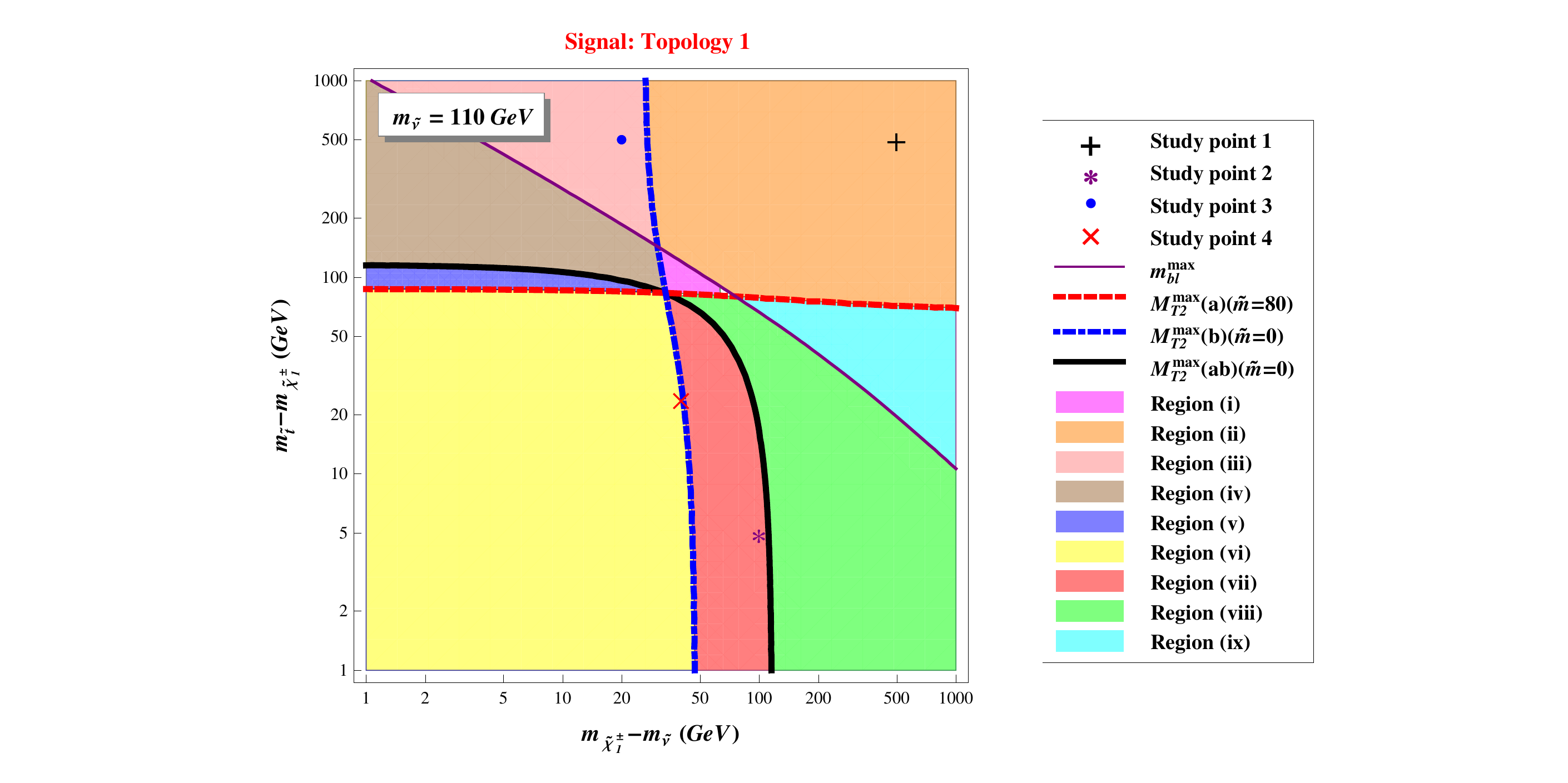}
\caption{\label{fig:map}  The division of stop-chargino-sneutrino mass
  parameter
space into regions, which are defined by which of the background
endpoints are violated by signal events in which both stops decay
according to Topology 1.  The sneutrino mass has been set to $110$ GeV.  
The four study points considered later in this section are also indicated.}
\end{figure}

In order to divide the stop-chargino-sneutrino mass parameter space
into regions, we start with the analytical expressions for the $M_{T2}$ signal 
endpoints~\cite{Cho:2007qv,Cho:2007dh,Burns:2008va}. 
(The endpoints for the corresponding $M_{2\sqcup\sqcup}$ variables are
given by the exact same expressions~\cite{Cho:2014naa}.)
\bea
M_{T2}^{max}(ab;\tilde m=0) & = &  \frac{m^2_{\tilde t}-m^2_{\tilde
    \nu}}{m_{\tilde t}}, 
\label{susymtab}\\
M_{T2}^{max}(a;\tilde m=m_W) & = & \frac{m^2_{\tilde t}-m^2_{\tilde
    \chi^\pm}}{2m_{\tilde t}}
+\sqrt{\left(\frac{m^2_{\tilde t}-m^2_{\tilde \chi^\pm}}{2m_{\tilde
        t}}\right)^2+m_W^2},
\label{susymta} \\
M_{T2}^{max}(b;\tilde m=0) & = & 
 \sqrt{ \frac{(m^2_{\tilde t}-m^2_{\tilde \nu})(m^2_{\tilde
       \chi^\pm}-m^2_{\tilde \nu})}{m^2_{\tilde t}}  }.
\label{susymwb}
\eea
Now combining, e.g., Eqs.~(\ref{mtab}) and (\ref{susymtab}), we find that 
for the variables in the $(ab)$ subsystem, the signal endpoints exceed the
background endpoints if
\beq
\frac{m^2_{\tilde t}-m^2_{\tilde \nu}}{m_{\tilde t}} > m_t.
\label{condition_ab}
\eeq
The corresponding region is delineated by the solid black line in
Fig.~\ref{fig:map},
which shows a slice through the 3-dimensional mass parameter space 
for fixed $m_{\tilde \nu}=110$ GeV. For convenience we choose to represent
the remaining two degrees of freedom as the mass differences 
$m_{\tilde\chi^\pm}-m_{\tilde \nu}$ and $m_{\tilde t}-m_{\tilde\chi^\pm}$.
The region satisfying the condition (\ref{condition_ab}) is above and 
to the right of the solid black line in Fig.~\ref{fig:map}. If the SUSY mass 
spectrum happens to be in this region, the mass splitting between
the stop and the sneutrino is sufficiently large to cause some number of 
signal events to ``leak" beyond the background endpoint (\ref{mtab}).
This means that the subsystem $(ab)$ variables are promising
variables to cut on in order to separate signal from background.

We can apply similar reasoning to the invariant mass variables in subsystems 
$(a)$ and $(b)$. For example, comparing Eqs.~(\ref{mta}) and (\ref{susymta}),
we find that the bound (\ref{mta}) applicable to background events will be violated if
the mass spectrum is such that
\beq
\frac{m^2_{\tilde t}-m^2_{\tilde \chi^\pm}}{m_{\tilde t}} >
\frac{m^2_t-m^2_W}{m_t}.
\label{condition_a}
\eeq
The corresponding region extends above the red dashed line in
Fig.~\ref{fig:map}.
Finally, the condition for violating the background endpoints in
subsystem $(b)$ follows from Eqs.~(\ref{mwb}) and (\ref{susymwb}):
\beq
\frac{(m^2_{\tilde t}-m^2_{\tilde \nu})(m^2_{\tilde
    \chi^\pm}-m^2_{\tilde \nu})}{m^2_{\tilde t}}  > m_W^2.
\label{condition_b}
\eeq
The region where this condition is satisfied is located to the right of the 
blue dot-dashed line in Fig.~\ref{fig:map}.

For completeness, we shall also consider the variable $m_{bl}$, 
the invariant mass of the lepton and $b$ quark from a given branch of
the decay.  
The endpoint of this quantity for signal events is given by
\beq
m_{b\ell}^{max} = 
\sqrt{\frac{(m_{\tilde{t}}^2 -
    m_{\tilde{\chi}^\pm}^2)(m_{\tilde{\chi}^\pm}^2-
    m_{\tilde{\nu}}^2)}{m_{\tilde{\chi}^\pm}^2 }},
\label{susymbl}
\eeq
while background events will obey the bound
\beq
m_{b\ell} \le \sqrt{m_t^2 - M_W^2}.
\label{mblmaxbknd}
\eeq
Therefore the condition for violating the background $m_{b\ell}$ endpoint is
\beq
\frac{(m_{\tilde{t}}^2 -
  m_{\tilde{\chi}^\pm}^2)(m_{\tilde{\chi}^\pm}^2-
  m_{\tilde{\nu}}^2)}{m_{\tilde{\chi}^\pm}^2 }
> m_t^2 - M_W^2.
\label{condition_bl}
\eeq
The corresponding region is found above and to the right of the
diagonal magenta thin solid line in Fig.~\ref{fig:map}.

\begin{table}[t]
\centering
\begin{tabular}{| c | c | c | c | c |}
\hline
~ & $m_{b\ell}$  & $M_{T2}(ab)$ & $M_{T2}(a)$ & $M_{T2}(b)$ \\
\textbf{Region} & endpoint  & endpoint & endpoint & endpoint \\
~ & violation  & violation & violation & violation \\ 
\hline \hline
\textcolor{region1}{i}          &  No   &  Yes & Yes & Yes \\
\textcolor{region2}{ii}         &  Yes  &  Yes & Yes & Yes \\
 \textcolor{region3}{iii}       &  Yes  &  Yes & Yes  & No \\
 \textcolor{region4}{iv}       &  No   &  Yes & Yes  & No \\
 \textcolor{region5}{v}        &  No   &  No & Yes  & No \\
 \textcolor{region6}{vi}       &  No   &  No  & No  & No \\
 \textcolor{region7}{vii}      &  No   &  No  & No & Yes \\
 \textcolor{region8}{viii}     &  No   &  Yes  & No & Yes \\
 \textcolor{region9}{ix}       &  Yes  &  Yes  & No & Yes \\
\hline
\end{tabular}
\caption{\label{tab:regions} The regions of stop-chargino-sneutrino
  mass space, as depicted in Fig.~\ref{fig:map}.  The four columns
  describe whether in the given region, the specified background
  endpoint can be violated by signal events where both stops decay
  according to Topology 1. }
\end{table}

The conditions implied by Eqs.~(\ref{condition_ab}-\ref{condition_b})
and (\ref{condition_bl}) divide the mass parameter space of Fig.~\ref{fig:map}
into nine distinct color-coded regions, which are defined in
Table~\ref{tab:regions}.
  It is easy to verify analytically that the boundaries of the three
  regions defined by conditions
(\ref{condition_ab}-\ref{condition_b}), i.e., the red, blue, and black curves in 
Fig.~\ref{fig:map}, cross at a single point. For any given sneutrino mass, 
$m_{\tilde \nu}$, the values of the stop mass, $m_{\tilde t}$, and the
chargino mass,
$m_{\tilde{\chi}^\pm}$, corresponding to the triple crossing point are
found from the relations
$$
\frac{m^2_{\tilde t}-m^2_{\tilde \nu}}{m_{\tilde t}} = m_t, \qquad
\frac{m_{\tilde{\chi}^\pm}^2- m_{\tilde{\nu}}^2}{m_{\tilde t}} = \frac{m_W^2}{m_t}. 
$$

The map in Fig.~\ref{fig:map} serves a dual purpose.  First, it
singles out the regions which might be easier to discover,
as well as the regions which may pose challenges. Second, within each
region, it identifies the variables which might be useful in the
analysis (see Table \ref{tab:regions}). For example, in region
\textcolor{region2}{ii}, the mass spectrum is sufficiently split, and
all four variables exhibit endpoint violations for signal events.
This in turn suggests that separating signal from background should be
relatively easy, since we can use any of the four types of invariant
mass variables in Table~\ref{tab:regions} to suppress the background
without much signal loss. To illustrate the expected phenomenology of
region \textcolor{region2}{ii}, in Sec.~\ref{subsec:point-1} we
shall analyze in detail a specific study point from this region. Its
mass spectrum is given in Table~\ref{tab:study-points} and its exact
location on the map of Fig.~\ref{fig:map} is marked with a black cross
($+$).

\begin{table}[t]
\centering
\begin{tabular}{| c | c | c | c | c |}
\hline
Study Point & Stop Mass & Chargino Mass & Sneutrino Mass & Region \\
\hline \hline
1  & $1110$ GeV & $610$ GeV & $110$ GeV & \textcolor{region2}{ii} \\
2  & $215$   GeV & $210$ GeV & $110$ GeV & \textcolor{region7}{vii} \\
3  & $630$   GeV & $130$ GeV & $110$ GeV & \textcolor{region3}{iii} \\
4  & $174$   GeV & $150$ GeV & $110$ GeV & \textcolor{region6}{vi} \\
\hline
\end{tabular}
\caption{\label{tab:study-points} The stop, chargino, and sneutrino masses
  for the four study points considered in this section, as well as the
  region of mass parameter space from Fig.~\ref{fig:map} that contains
  each point.}
\end{table}
 
In all but one of the remaining regions of Fig.~\ref{fig:map}, some of
the endpoints are violated while others are not, thus some variables
are expected to perform better than others.  We pick two
representative study points in regions \textcolor{region7}{vii} and
\textcolor{region3}{iii} and study them in
Secs.~\ref{subsec:point-2} and \ref{subsec:point-3}, respectively.
In Fig.~\ref{fig:map}, these two study points are marked with the
magenta asterisk $(\ast)$ and the blue circle ($\bullet$). The
corresponding mass spectra are also listed in
Table~\ref{tab:study-points}. 
 
Finally, region \textcolor{region6}{vi} deserves a special mention,
since it represents a particularly challenging scenario.  Here none of
the four types of invariant mass variables exhibits an endpoint
violation, and the signal events are populating the same kinematic
region as the background events.  Our fourth study point 
in Table~\ref{tab:study-points} 
(denoted in Fig.~\ref{fig:map} with the red ($\times$) symbol)
belongs to this challenging region and is considered in
Sec.~\ref{subsec:point-4}.

\subsection{Study point 1: split spectrum in region ii}
\label{subsec:point-1}

In this section we shall illustrate the properties of region
\textcolor{region2}{ii} in Fig.~\ref{fig:map}
with the study point 1 which is marked with the black ($+$) symbol.
As seen in Table \ref{tab:study-points}, this study point 
has a widely split spectrum; both mass differences 
$m_{\tilde t}-m_{\tilde \chi^\pm}$ and $m_{\tilde \chi^\pm} -
m_{\tilde \nu}$ are $500$ GeV.
\begin{figure}[t]
\centering
\includegraphics[width=7.0cm]{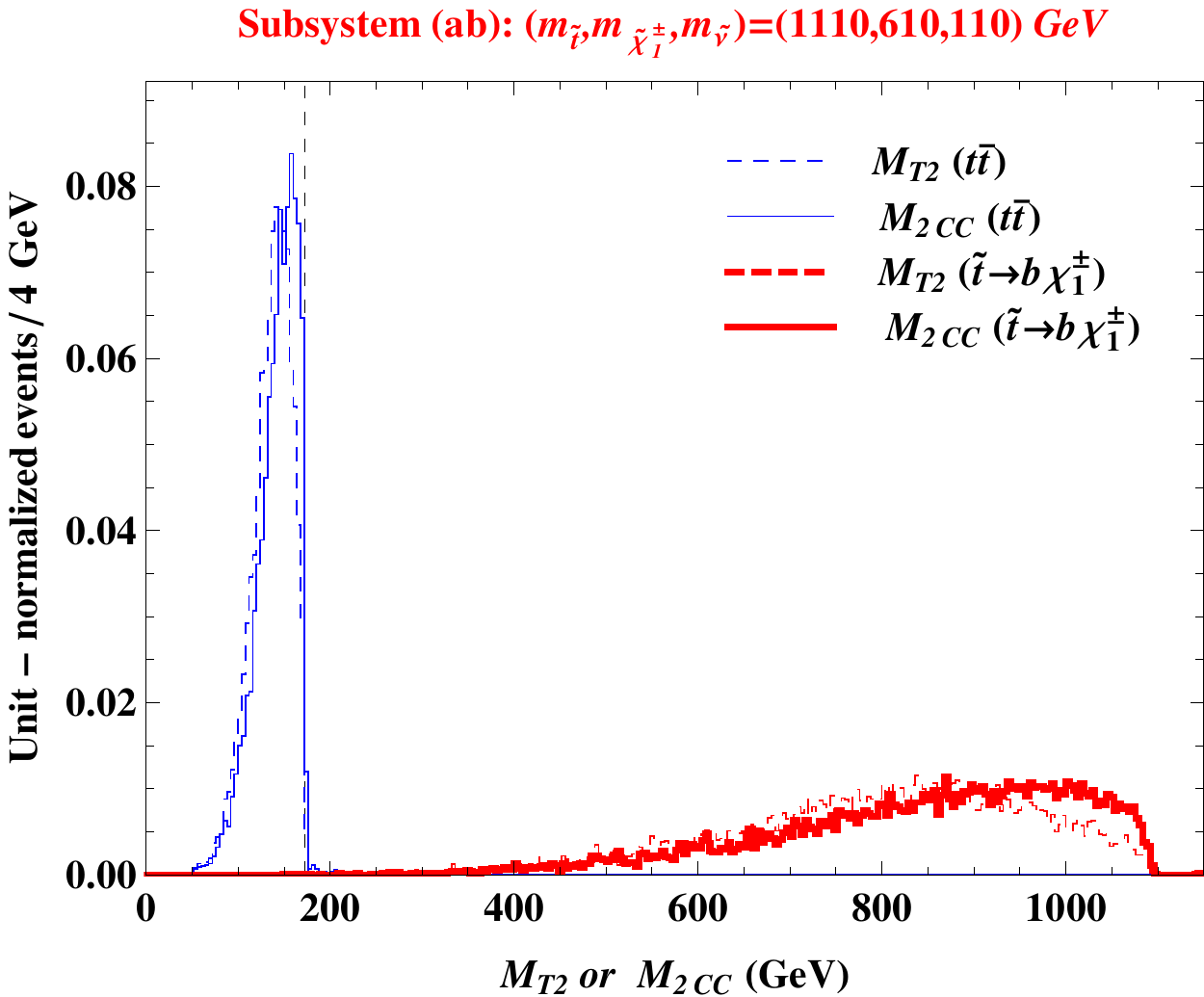}
\includegraphics[width=7.0cm]{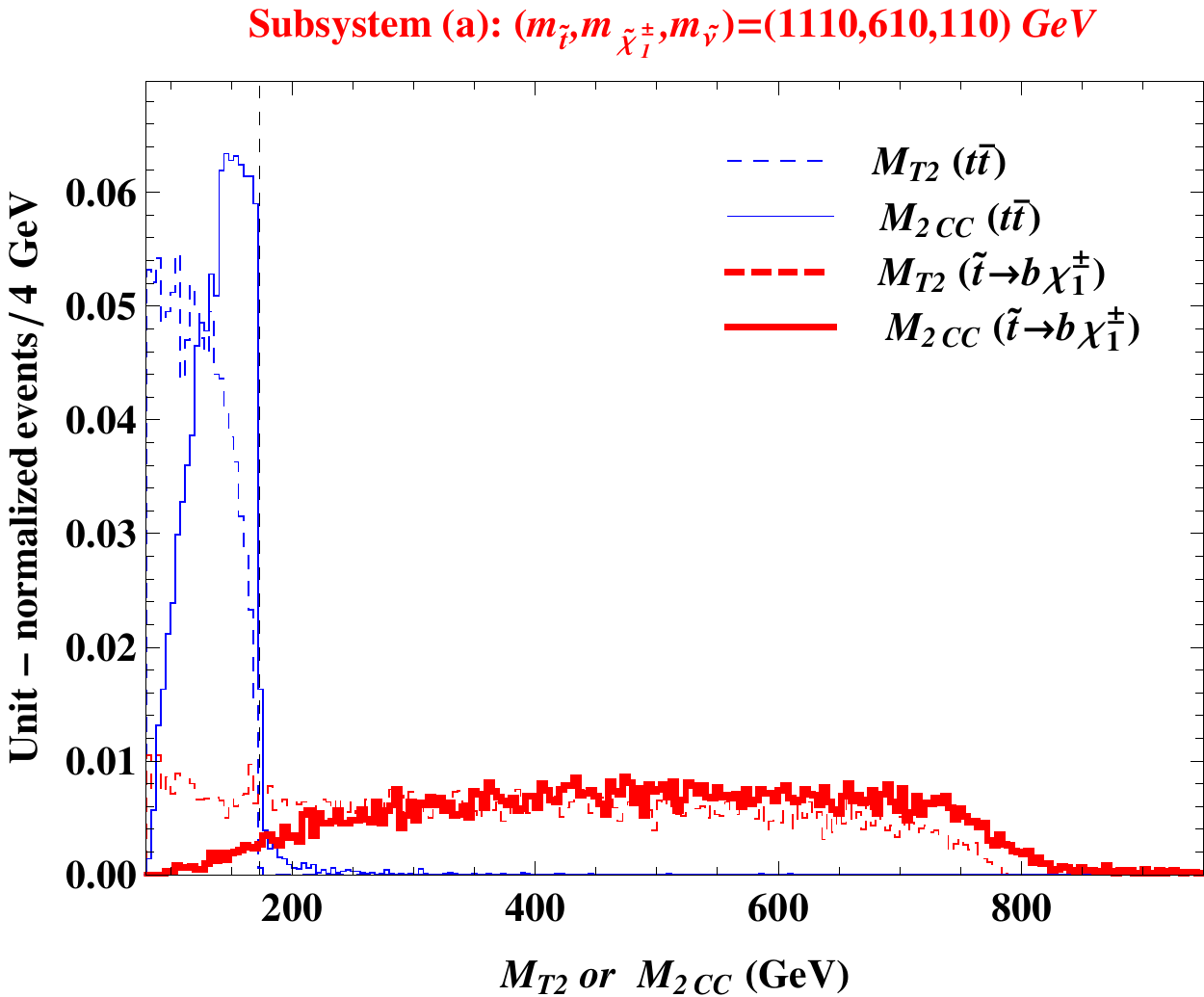}\\
\includegraphics[width=7.0cm]{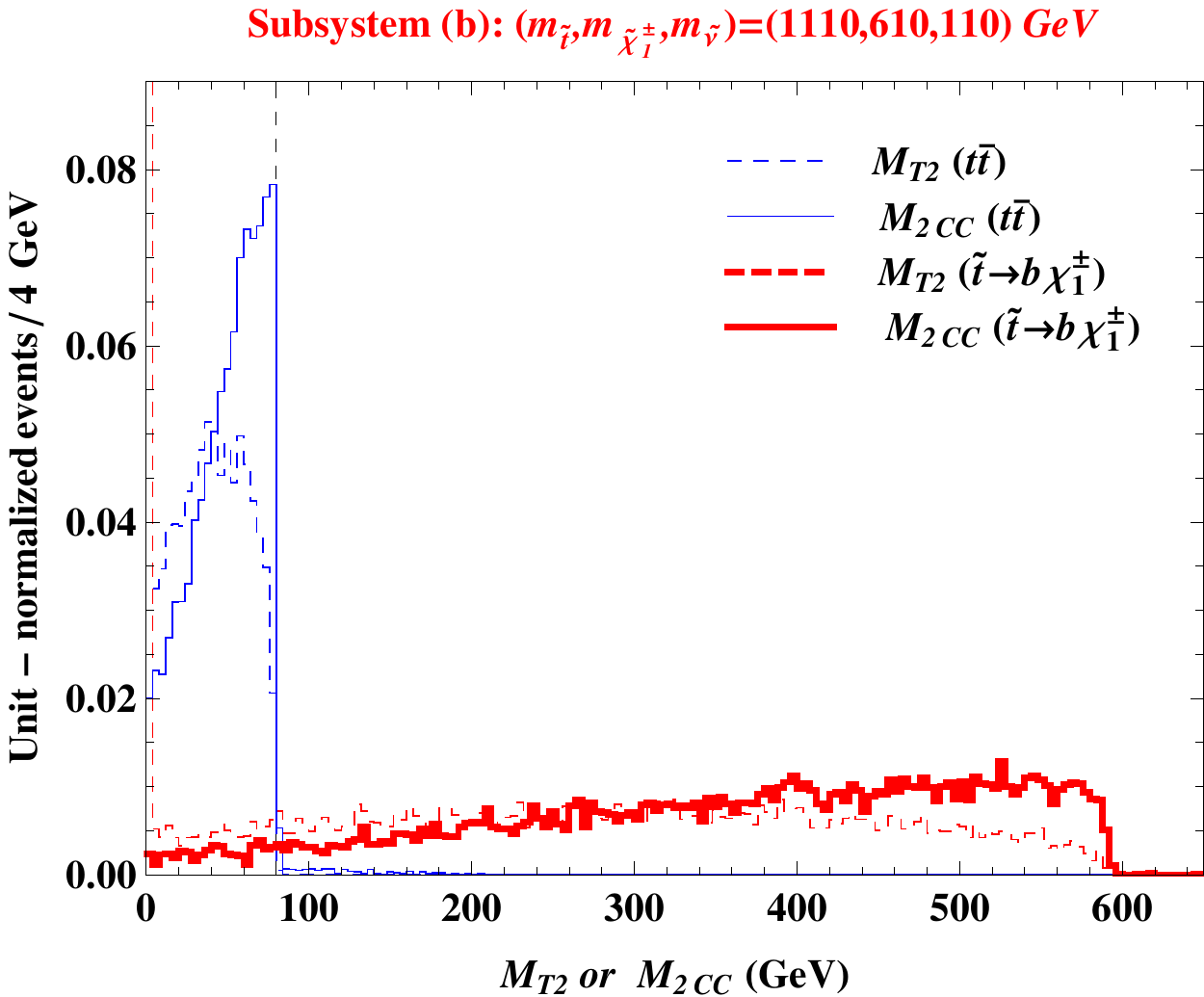}
\includegraphics[width=7.0cm]{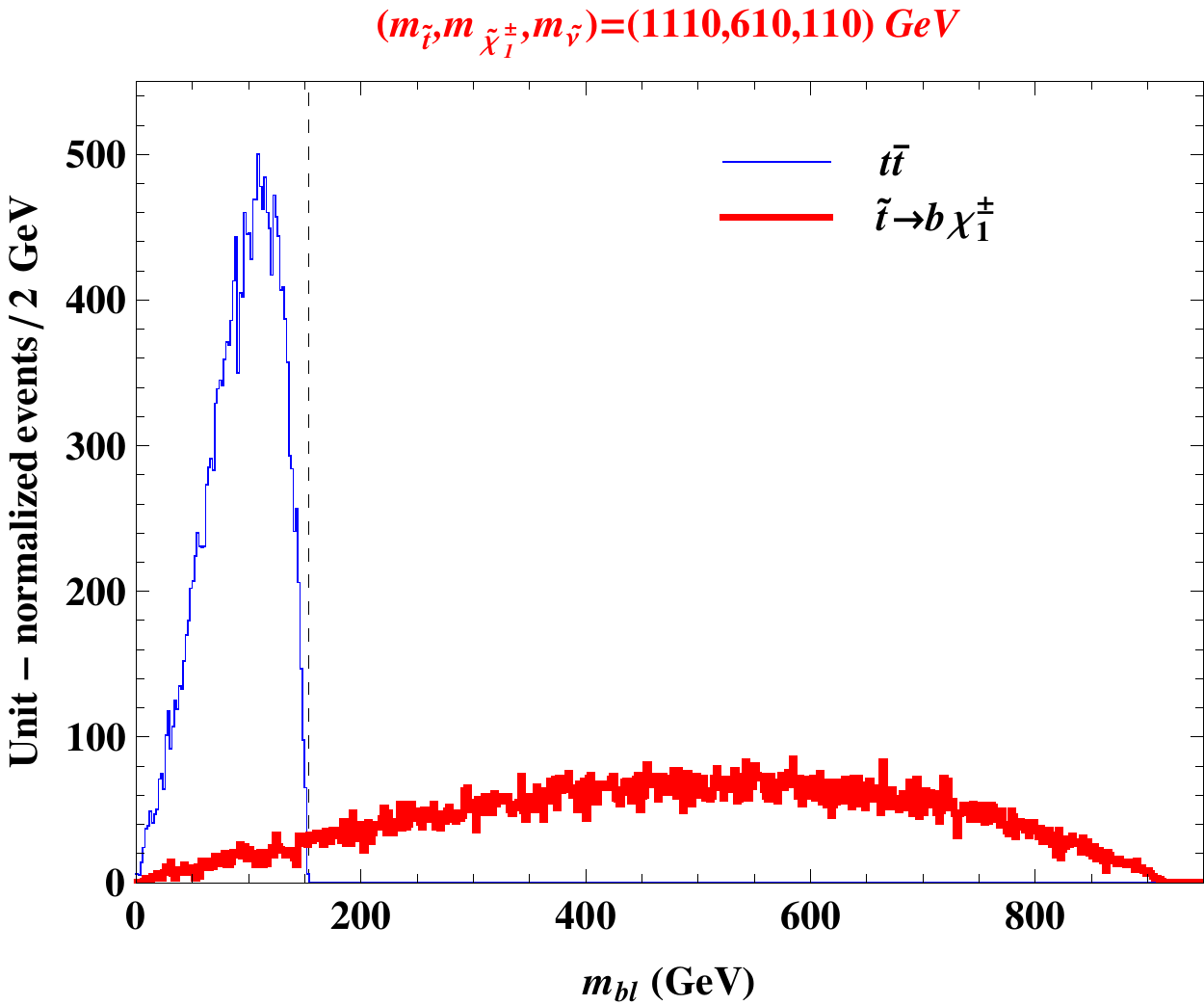}
\caption{\label{fig:MT2cuts3} 
Signal (thick red lines) and background (thin blue lines)
distributions for the subsystem variables $M_{T2}(S;\tilde m)$ (dashed lines)
and $M_{2CC}(S;\tilde m)$ (solid lines). We show results for each of the three 
possible subsystems: $S=(ab)$ (upper left panel), $S=(a)$ (upper right
panel), and
$S=(b)$ (lower left panel). The lower right panel shows the analogous
distributions for the invariant mass $m_{b\ell}$. The signal events are for
study point 1 
($m_{\tilde t}=1110$ GeV,  $m_{\tilde \chi^\pm}=610$ GeV and $m_{\tilde \nu}=110$ GeV)
and have the event topology shown in
Fig.~\ref{fig:DecayTopologies}(b).} 
\end{figure}
One can therefore expect that the signal distributions for our
invariant mass variables will extend well beyond the corresponding
background distributions.  This is confirmed in
Fig.~\ref{fig:MT2cuts3}, where we plot distributions for signal events
(red lines) and background events (blue lines) for $M_{T2}(S;\tilde
m)$ (dashed lines), $M_{2CC}(S;\tilde m)$ (solid lines), and
$m_{b\ell}$.   As always in this section, the signal events are
assumed to have the symmetric event topology of
Fig.~\ref{fig:DecayTopologies}(b), i.e., both stops decay according to
Topology 1.  Fig.~\ref{fig:MT2cuts3} shows results for all three
subsystems: $S=(ab)$ (upper left panel), $S=(a)$ (upper right panel)
and $S=(b)$ (lower left panel).  In each case, the trial mass $\tilde
m$ for the respective daughter particle has been set to the correct
value for SM events, as explained in more detail in
Appendix~\ref{sec:appendix}.  In each panel of
Fig.~\ref{fig:MT2cuts3}, the vertical dashed line marks the location
of the kinematic endpoint for background events, as given by
Eqs.~(\ref{mtab}-\ref{mwb}) and (\ref{mblmaxbknd}). As expected, the
blue (background) distributions always obey these kinematic endpoints.
(The figure has been constructed using parton-level events with no
detector modeling or combinatorial backgrounds --- these effects will
be added later on in Sec.~\ref{sec:detector_resolution_etc}.)

On the other hand, the signal events, shown in red, significantly
violate the endpoints.  In fact, for all four variables considered in
Fig.~\ref{fig:MT2cuts3}, the vast majority of signal events violate
the background endpoints. This is also to be expected, since study
point 1 was chosen specifically in region \textcolor{region2}{ii},
where all four endpoints are expected to be violated (see
Table~\ref{tab:regions}).   This also means that discovery (or
exclusion) for this study point should be relatively straightforward.  

We have noted above that, on an event by event basis,
\begin{equation}
M_{T2}(S; \tilde{m}) \le M_{2CC}(s; \tilde{m}). 
\label{eq:MT2-le-M2CC}
\end{equation}
This is confirmed in Fig.~\ref{fig:MT2cuts3}, where the (solid line)
$M_{2CC}$ distributions can be seen to be somewhat harder than the
respective $M_{T2}$ dashed line distributions.  As a result, cutting
on $M_{2CC}$ instead of $M_{T2}$ will result in a slightly higher signal
efficiency.  (The effect is most easily seen for subsystems $(a)$ and $(b)$.)

\begin{figure}[t]
\centering
\includegraphics[width=4.7cm]{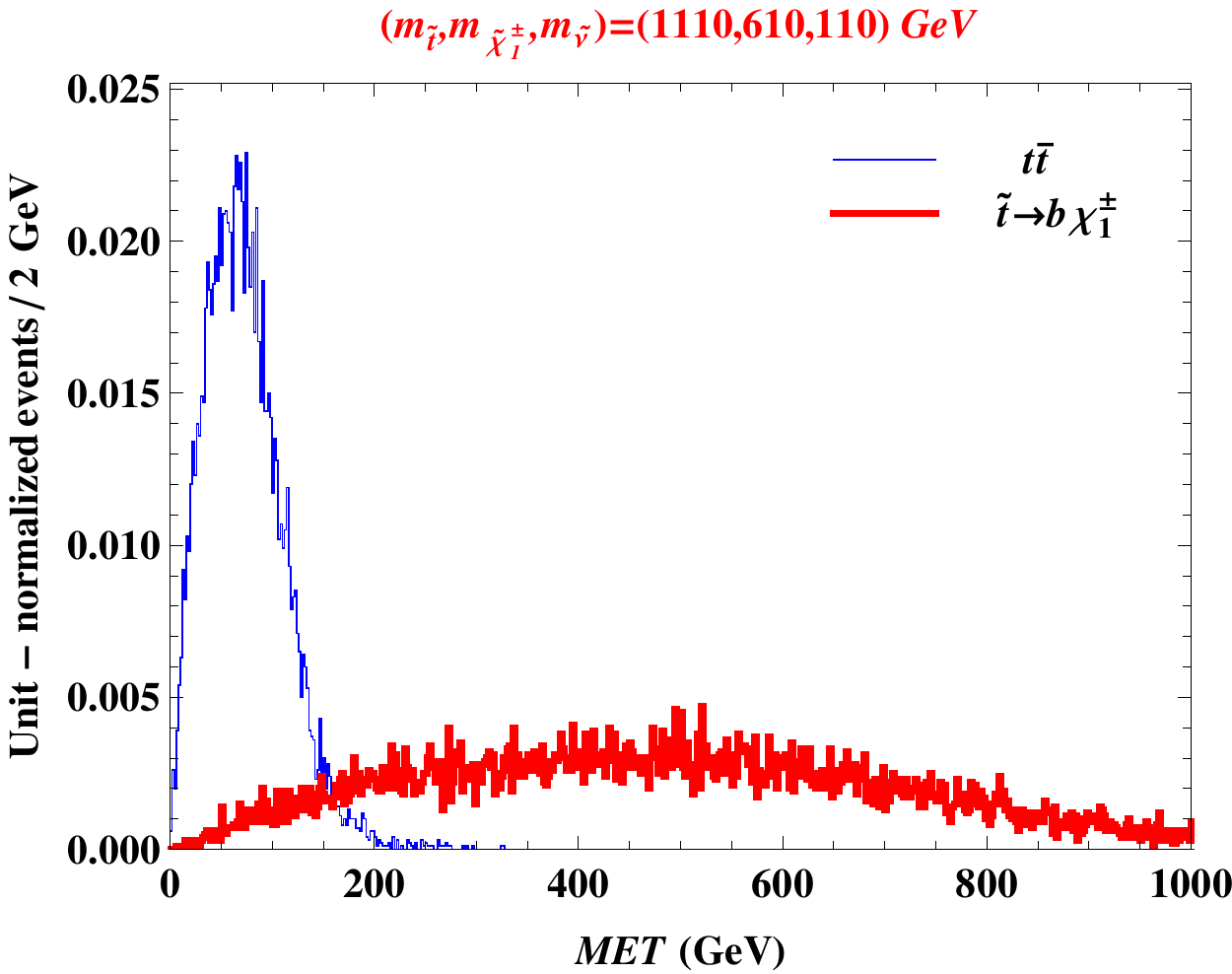}
\includegraphics[width=4.7cm]{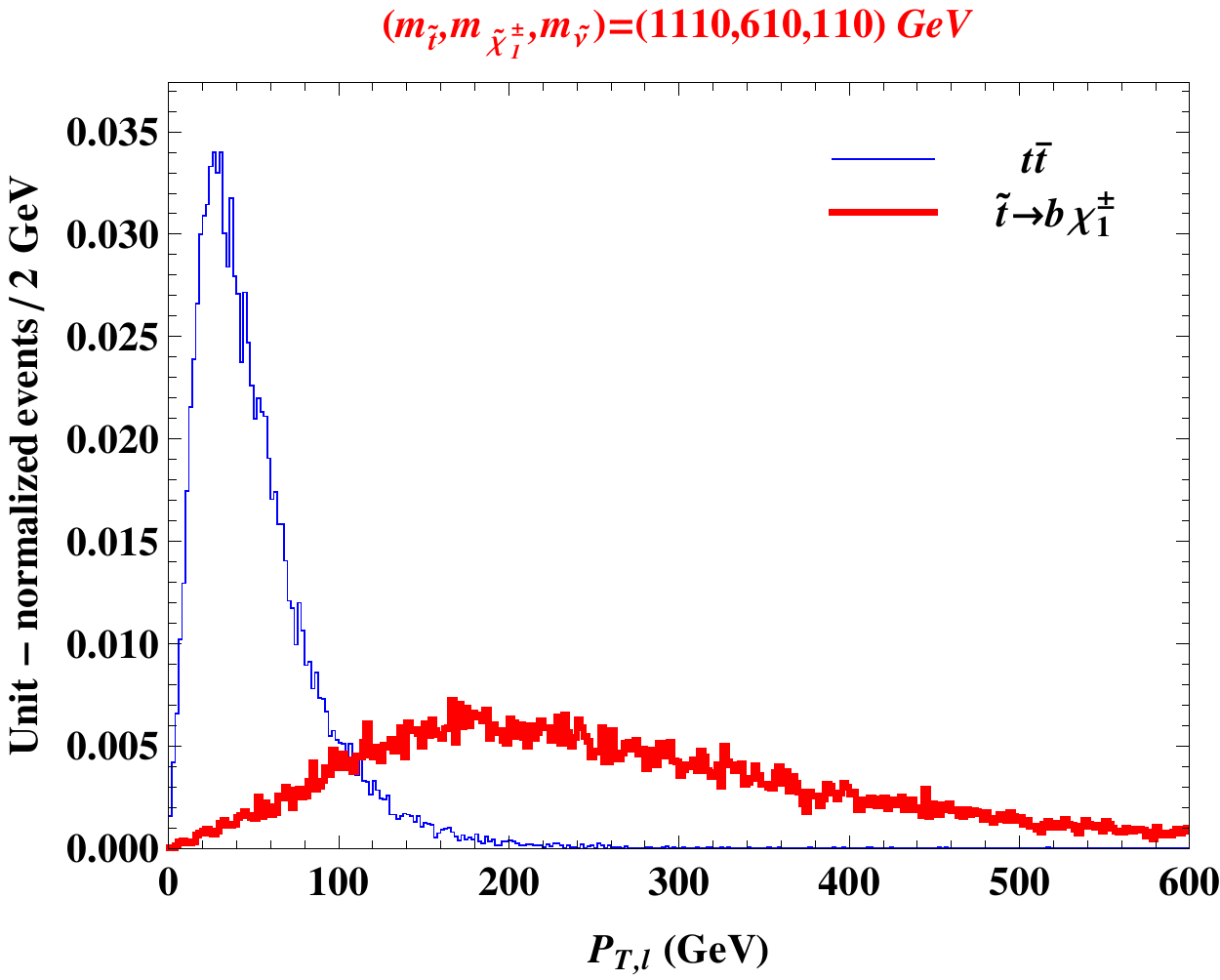}
\includegraphics[width=4.7cm]{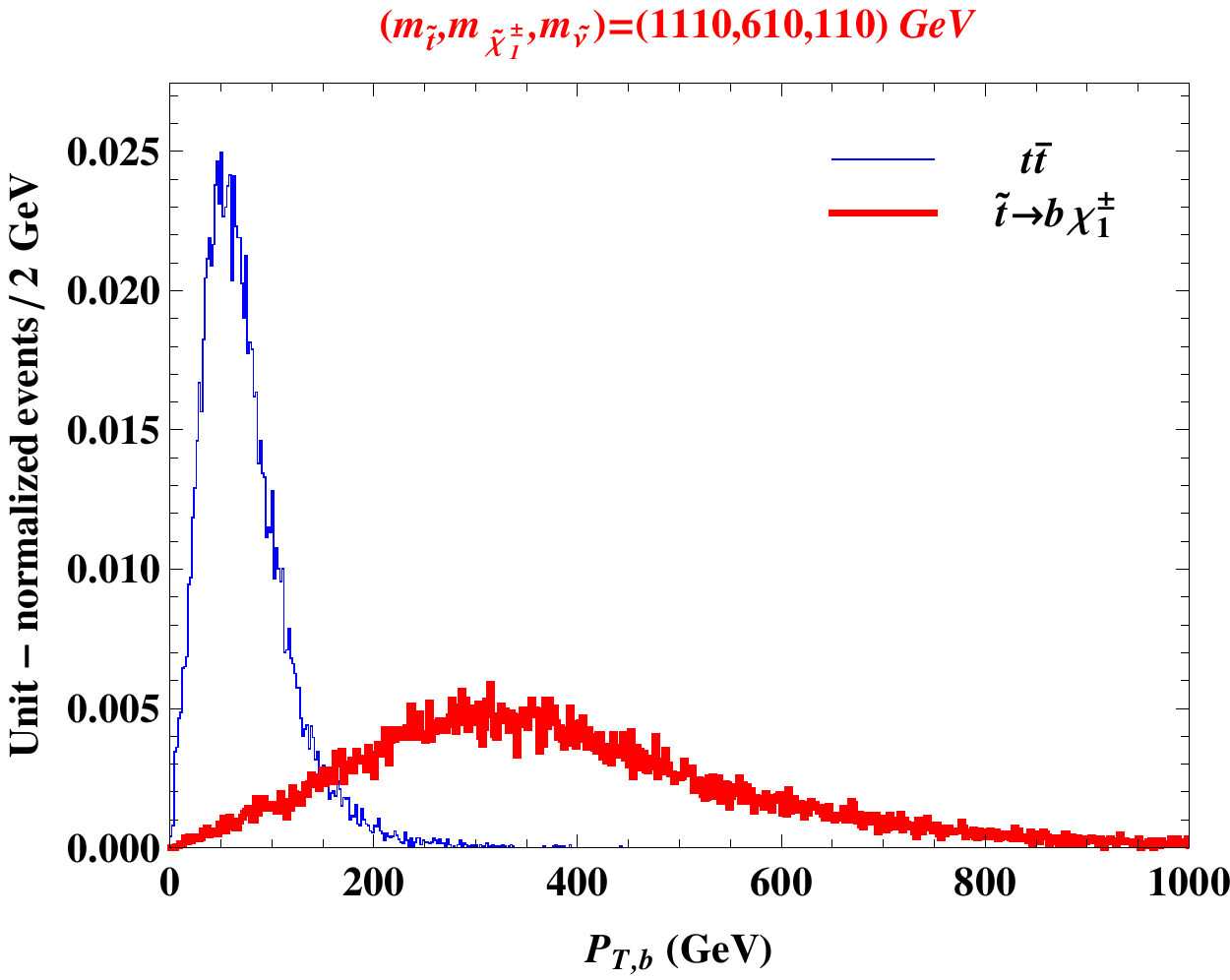}
\caption{\label{fig:cuts3} 
Signal (thick red lines) and background (thin blue lines)
distributions of some standard variables for study point 1
($m_{\tilde t}=1110$ GeV,  $m_{\tilde \chi^\pm}=610$ GeV and $m_{\tilde \nu}=110$ GeV): 
the missing transverse energy $MET$ (left panel), 
the lepton transverse momentum $p_{T,\ell}$ (middle panel), 
and the $b$-quark transverse momentum $p_{T,b}$ (right panel).
} 
\end{figure}

While the variables shown in Fig.~\ref{fig:MT2cuts3} already
allow study point 1 to be discovered or ruled out rather trivially,
the same can be accomplished with more conventional variables like
the missing transverse energy ($MET$), the lepton transverse momentum
$p_{T,\ell}$, or the $b$-quark transverse momentum $p_{T,b}$,
whose distributions for study point 1 are plotted in Fig.~\ref{fig:cuts3}.
In all three cases, even though the background distribution does not 
exhibit a strict endpoint, the signal and background distributions 
are very well separated, so the signal can be easily isolated.
Furthermore, as we can measure four endpoints in Fig.~\ref{fig:MT2cuts3}
and there are only three input mass parameters, full mass reconstruction 
in this case is also possible~\cite{Burns:2008va}.
In conclusion, the analysis of study point 1 demonstrates that region
\textcolor{region2}{ii}
is ``easy" in the sense that the experimenter has a plethora of 
useful tools available for a discovery. It is therefore of interest 
to consider the other, more challenging, regions of Fig.~\ref{fig:map}.

\subsection{Study point 2: soft $b$-jets in region vii}
\label{subsec:point-2}

Our second example is in region \textcolor{region7}{vii}, in which
only the variables in subsystem $(b)$ have endpoint violations. The
mass spectrum for study point 2 is given in
Table~\ref{tab:study-points}.  We notice the relative degeneracy
between the stop and chargino masses, which causes the endpoints of
the signal $M_{T2}(a)$ and $m_{b\ell}$ distributions to be relatively
low.  (See Eqs.~(\ref{susymta}) and (\ref{susymbl}).)  In addition, the
sneutrino mass has been chosen so that the signal endpoint
(\ref{susymtab}) of the $M_{T2}(ab)$ variable is also below the
standard model expectation of (\ref{mtab}).  This leaves the
$M_{T2}(b)$ variable as the only viable alternative in region
\textcolor{region7}{vii}.

\begin{figure}[t]
\centering
\includegraphics[width=7.0cm]{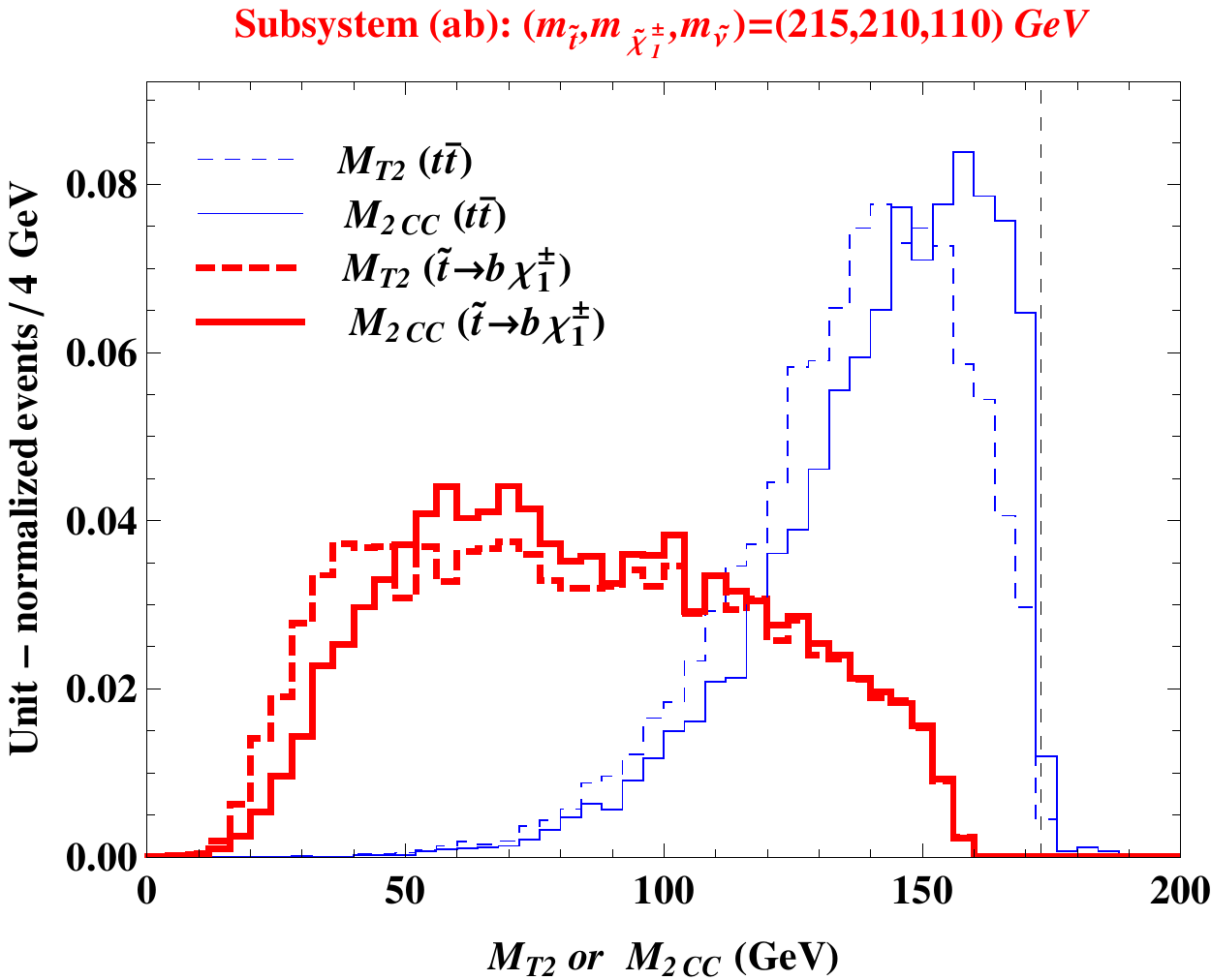}
\includegraphics[width=7.0cm]{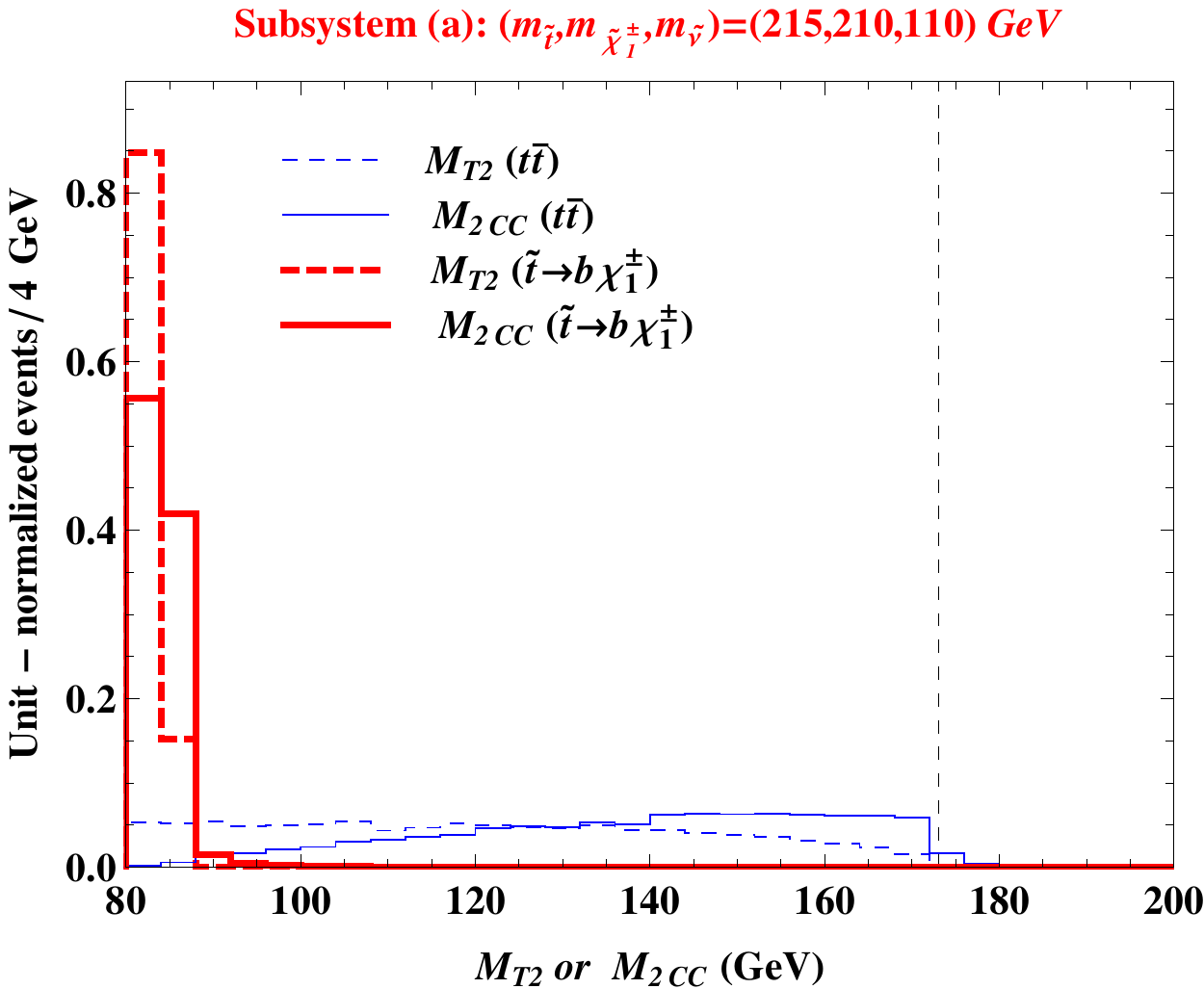}
\includegraphics[width=7.0cm]{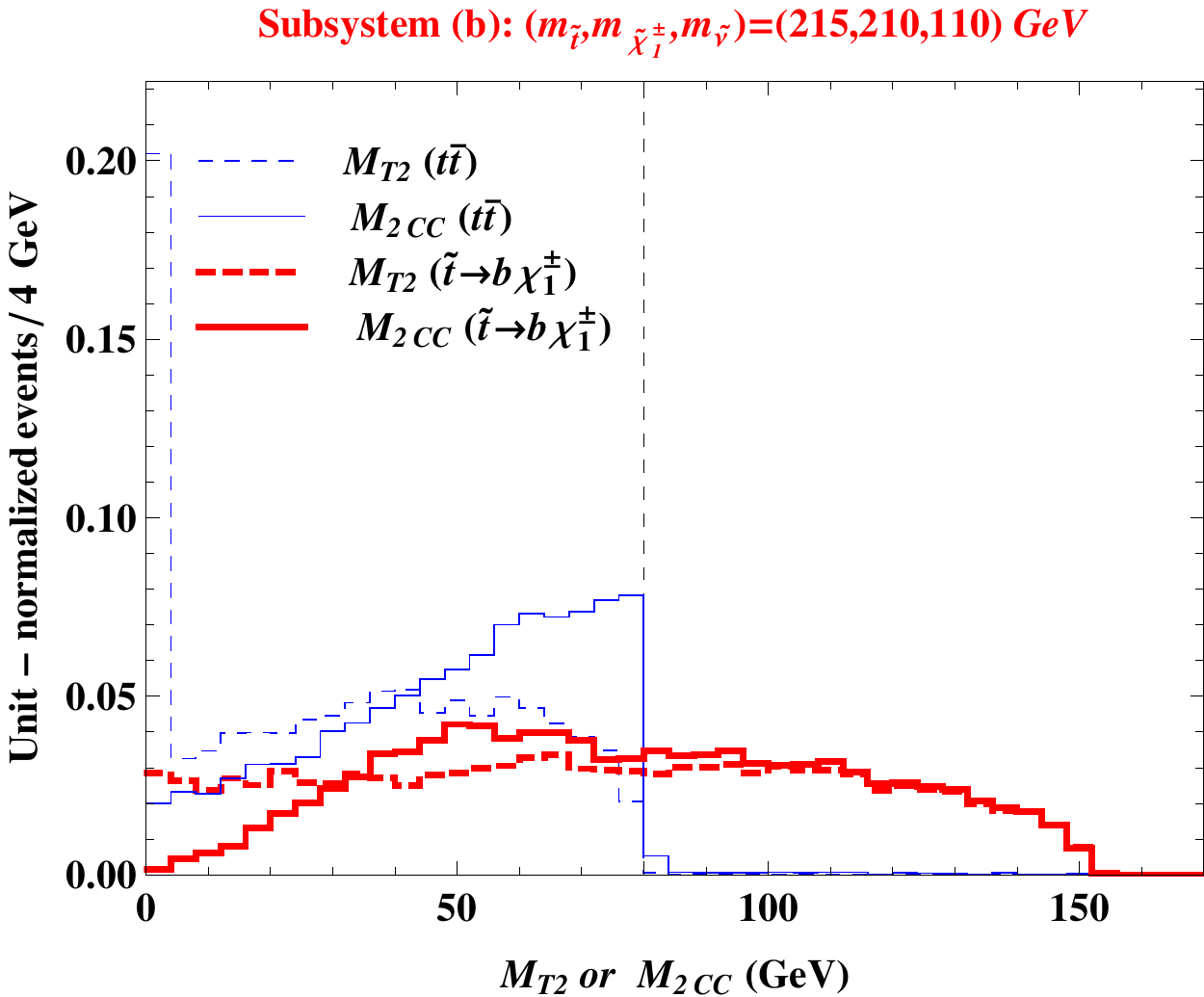}
\includegraphics[width=7.0cm]{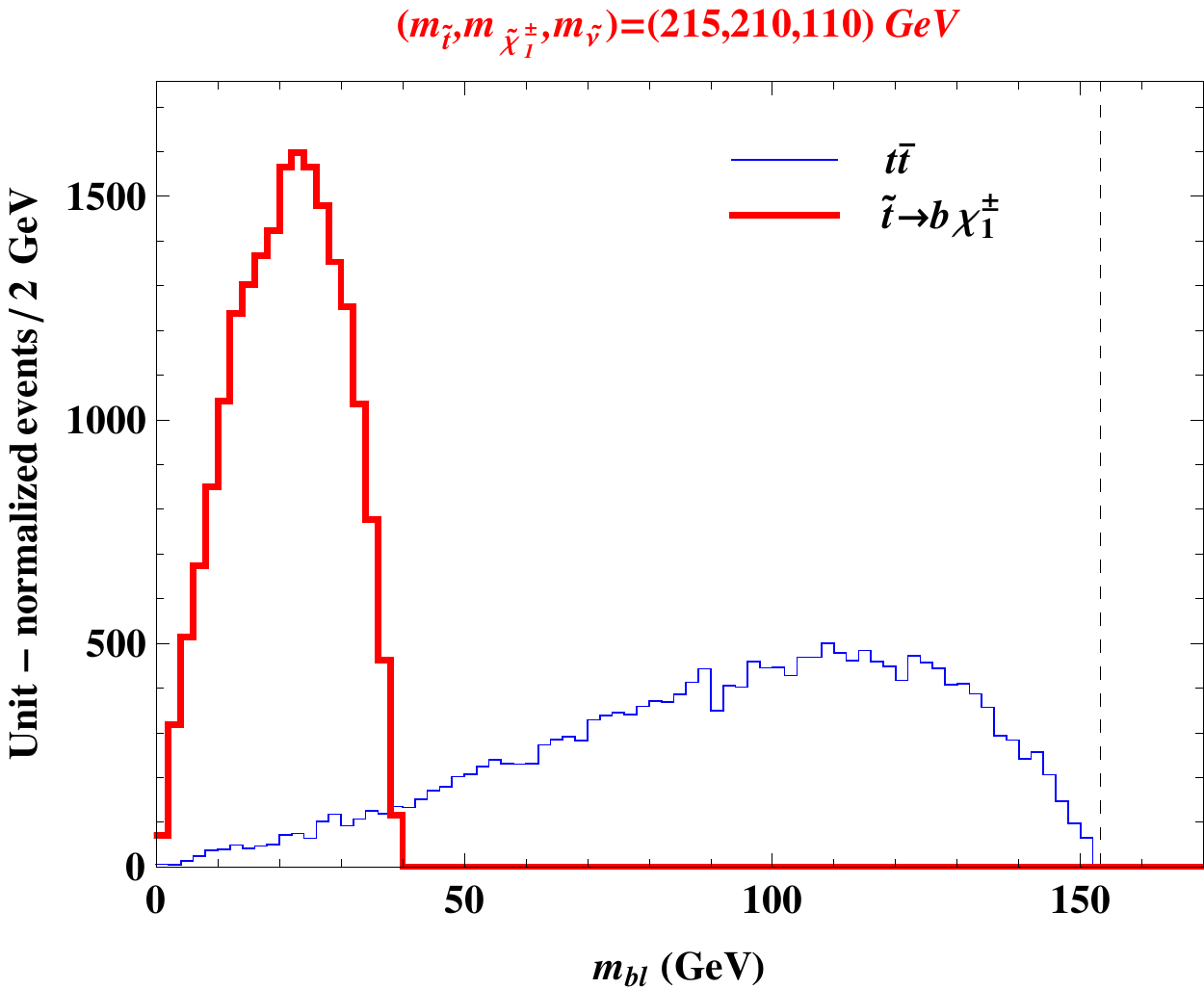}
\caption{\label{fig:MT2cuts1} 
The same as Fig.~\ref{fig:MT2cuts3}, but with signal events for
study point 2 ($m_{\tilde t}=215$ GeV,  $m_{\tilde \chi^\pm}=210$ GeV and $m_{\tilde \nu}=110$ GeV).} 
\end{figure}
\begin{figure}[h]
\centering
\includegraphics[width=4.7cm]{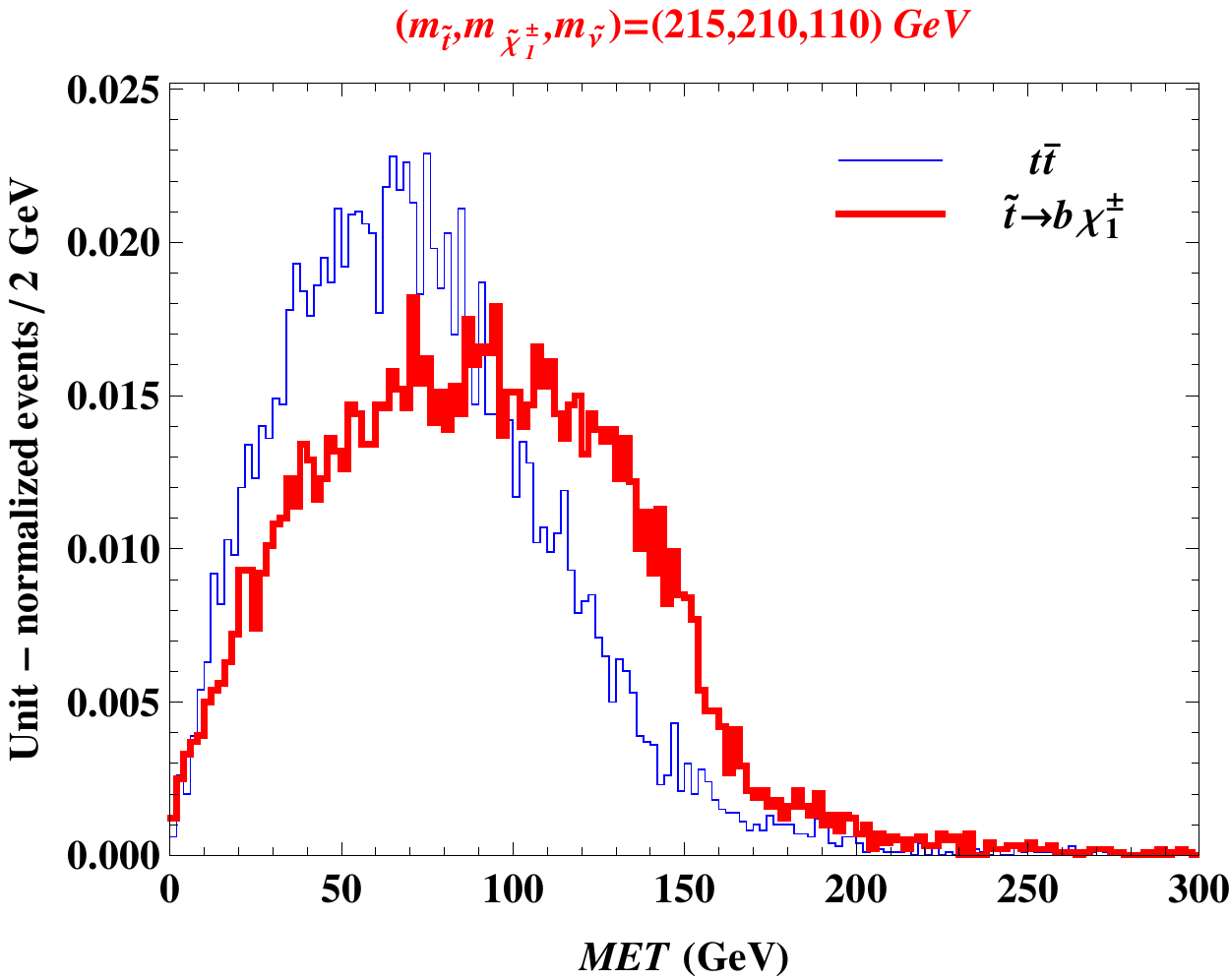}
\includegraphics[width=4.7cm]{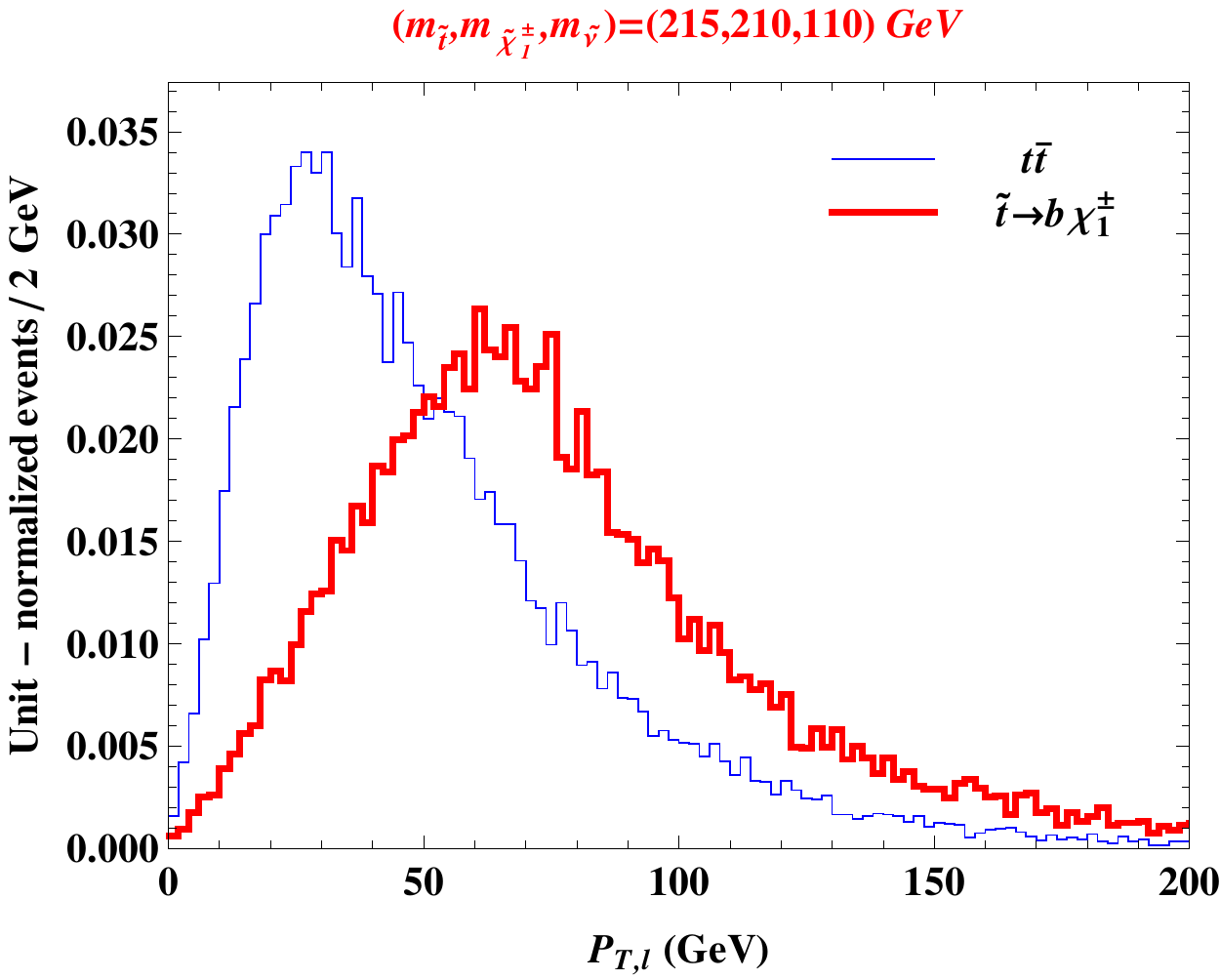}
\includegraphics[width=4.7cm]{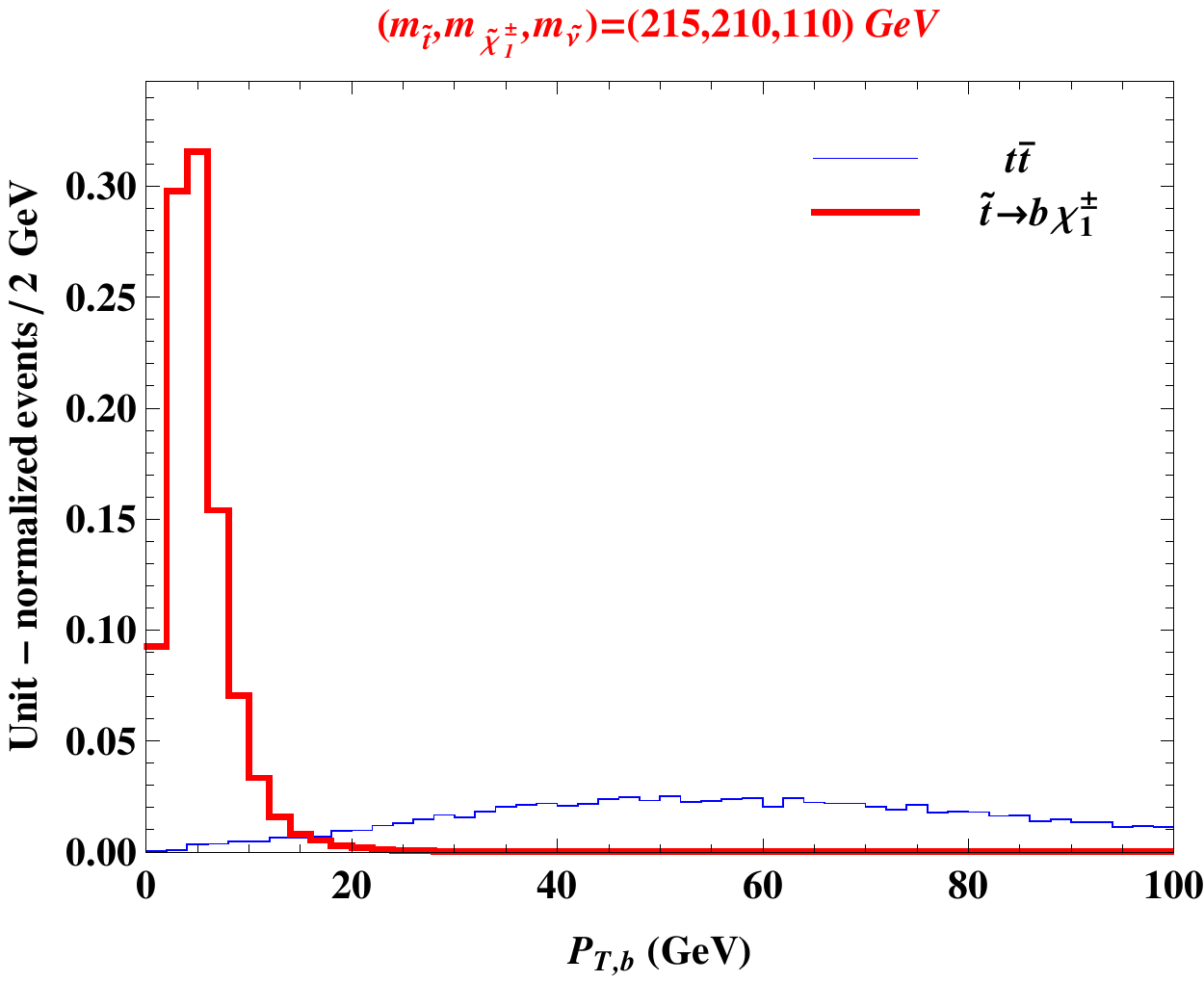}
\caption{\label{fig:cuts1} 
The same as Fig.~\ref{fig:cuts3}, but for signal events for
study point 2 ($m_{\tilde t}=215$ GeV,  $m_{\tilde \chi^\pm}=210$ GeV and $m_{\tilde \nu}=110$ GeV).} 
\end{figure}

These observations are illustrated in Fig.~\ref{fig:MT2cuts1} where we
plot the relevant invariant mass distributions for study point 2 using
the same conventions as in Fig.~\ref{fig:MT2cuts3} above.  Again, we
assume that all signal events have the event topology shown in
Fig.~\ref{fig:DecayTopologies}(b), i.e., both stops decay according to
Topology 1.  Fig.~\ref{fig:MT2cuts1} confirms that $M_{T2}(b)$ is a
good variable to cut on: placing a high pass cut with threshold just
above $m_t$ would eliminate all of the background, while leaving
almost half of the signal.  
To determine the optimal value of the threshold and the effectiveness
of the cut requires realistic detector simulation (see
Sec.~\ref{sec:detector_resolution_etc}), but it is clear that such
a cut will often be useful. This observation is not new --- the
variable $M_{T2}(b)$ has been discussed in the literature
under various names, e.g., $M_{T2}^{(210)}$~\cite{Burns:2008va} and
dileptonic
$M_{T2}$~\cite{Plehn:2012pr,Kilic:2012kw,Chakraborty:2013moa,Cho:2008cu}.  Here
we would like to contrast $M_{T2}(b)$ to the alternative on-shell
constrained variable $M_{2CC}(b)$. The advantage of the latter is the
slightly higher signal efficiency.  On the other hand, the advantage
of the traditional $M_{T2}$ is its simplicity --- in its calculation,
one does not have to identify the $b$-jets, thus, one avoids
combinatorial ambiguities and the additional penalty due to
$b$-tagging.

Note that the signal and background distributions for the other three
variables in Fig.~\ref{fig:MT2cuts1}: $M_{T2}(ab)$, $M_{T2}(a)$, and
$m_{b\ell}$, also appear to be quite different, so one might wonder
whether they could be useful if the cut were inverted (i.e., if one
performs a low pass cut). However, we expect other background
processes besides $t\bar{t}$ to contribute events at low values and
swamp the signal~\cite{Barr:2009wu,Allanach:2011ej,Murayama:2011hj}. 
Such backgrounds may not be as well-understood,
which is why in this study we shall only consider {\em high pass} cuts 
on the invariant mass variables.\footnote{Additionally we remind 
the reader that Fig.~\ref{fig:MT2cuts1} (and analogous) figures 
throughout the work, depict unit normalized
distributions for signal and background; in reality the background
rates will be far higher than the signal rates in any realistic
model.  This also makes it more challenging to utilize the differences
in shape for a given variable \textit{below} the endpoint; endpoint
violation is the preferred feature for discovery.}

Having identified $M_{T2}(b)$ and $M_{2CC}(b)$ as promising variables,
one might wonder how the more conventional variables would perform in
this case.  In Fig.~\ref{fig:cuts1}, we show parton-level signal and
background distributions for study point 2 for the three more
traditional variables considered in Fig.~\ref{fig:cuts3}: $MET$,
$p_{T,\ell}$, and $p_{T,b}$. Since the stop-chargino mass splitting is
rather small, the $b$-jets are quite soft and would often not be
reconstructed.  On the other hand, the $MET$ and $p_{T,\ell}$
distributions show some separation between signal and background,
but the separation is less clear than we observed in the case of $M_{T2}(b)$
(the lower left panel of Fig.~\ref{fig:MT2cuts3}).  Therefore, placing
cuts on $MET$ and $p_{T,\ell}$ would not be as effective as cutting on
$M_{T2}(b)$.

\subsection{Study point 3: soft leptons in region iii}
\label{subsec:point-3}

Our next example illustrates the complementarity of the subsystem
invariant mass variables.  In the previous subsection
(\ref{subsec:point-2}), we considered a signal study point with soft
$b$-jets and relatively hard leptons, as seen in
Fig.~\ref{fig:cuts1}. Now we shall discuss the opposite situation,
when the leptons are relatively soft, while the jets are hard.  For
this purpose, we focus on study point 3 in region
\textcolor{region3}{iii}, where according to Table \ref{tab:regions}
we expect endpoint violations for $M_{T2}(ab)$, $M_{T2}(a)$, and
$m_{b\ell}$.
\begin{figure}[t]
\centering
\includegraphics[width=7.0cm]{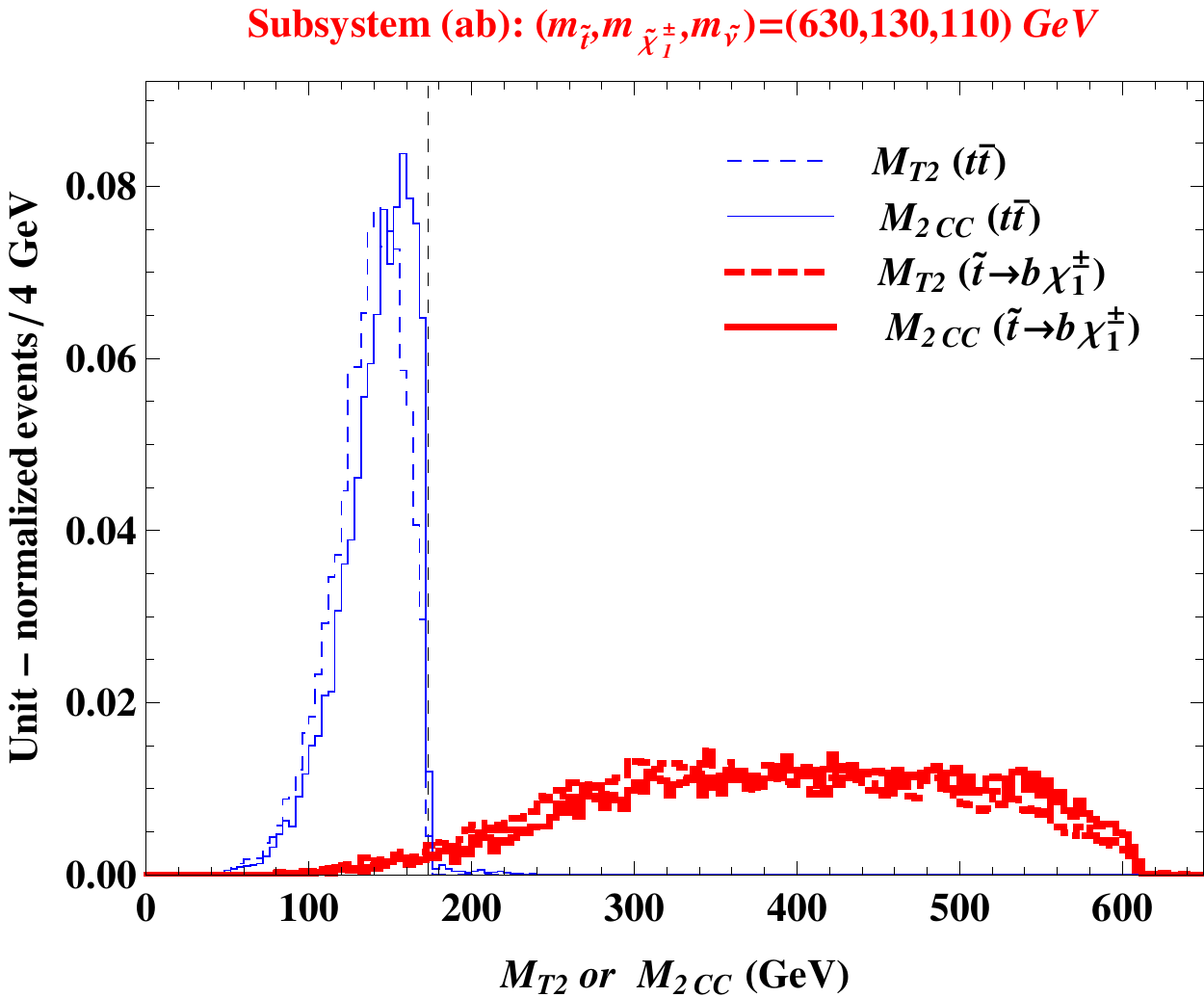}
\includegraphics[width=7.0cm]{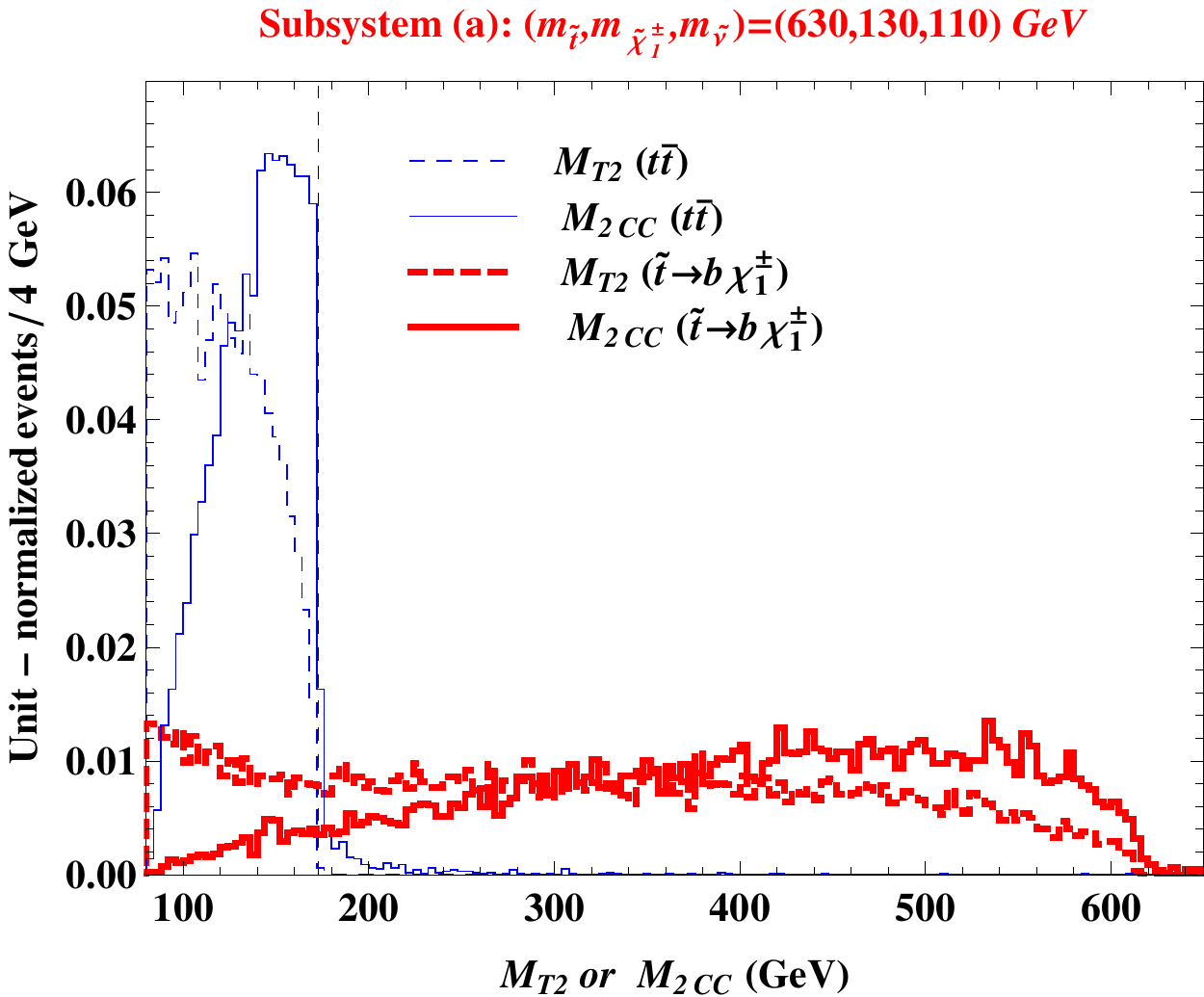}\\
\includegraphics[width=7.0cm]{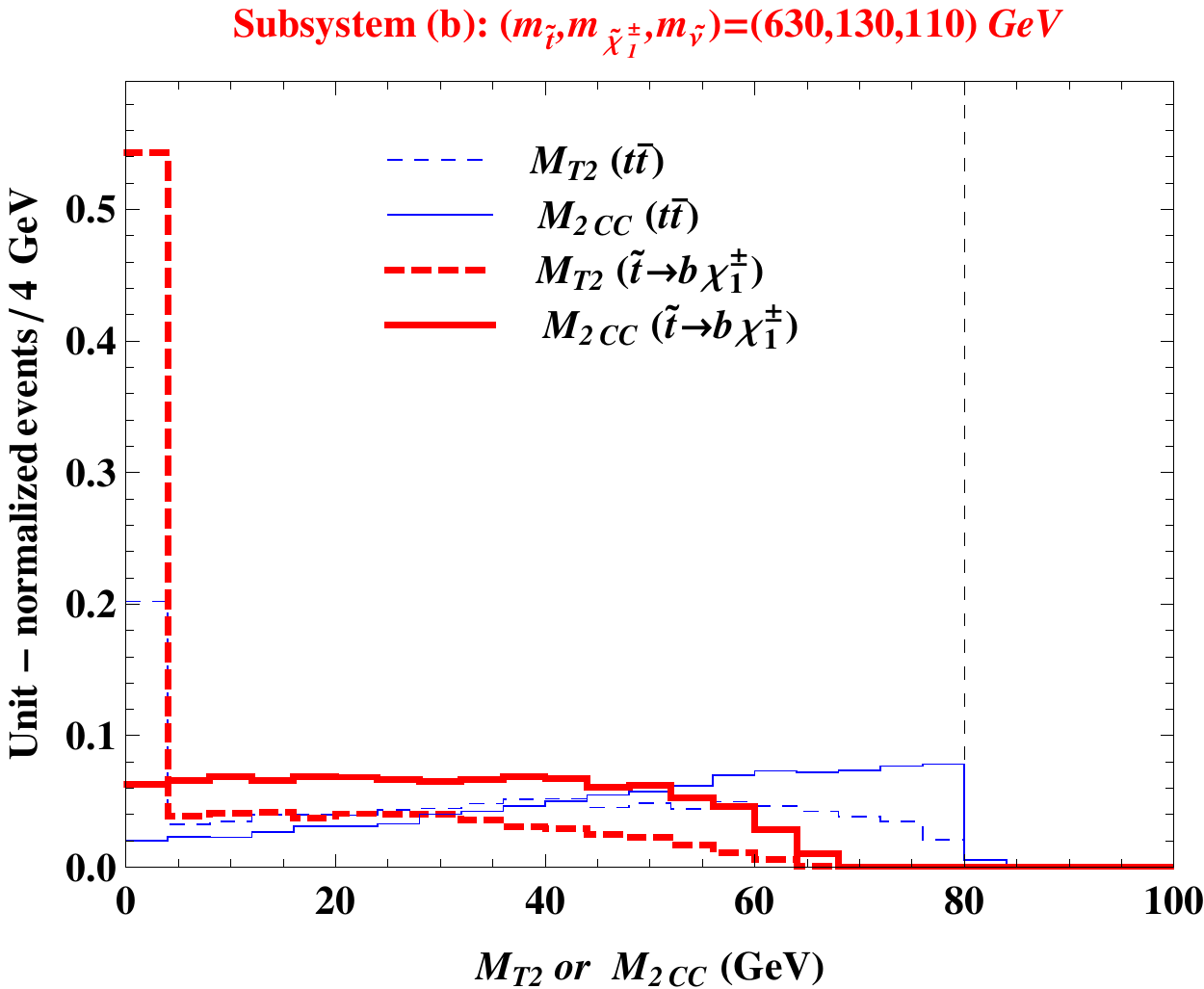}
\includegraphics[width=7.0cm]{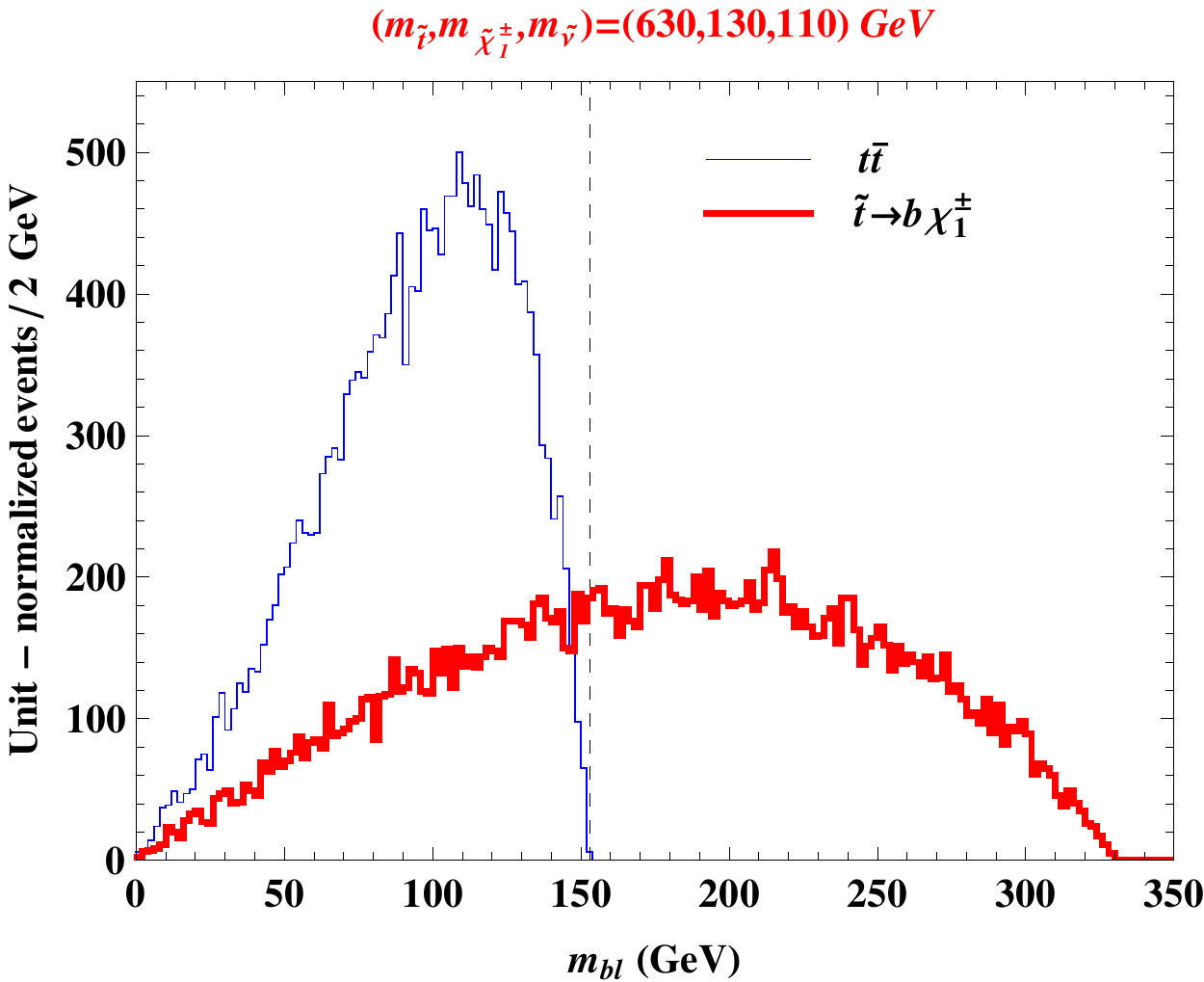}
\caption{\label{fig:MT2cuts2} 
The same as Fig.~\ref{fig:MT2cuts3}, but with signal events for
study point 3 ($m_{\tilde t}=630$ GeV,  $m_{\tilde \chi^\pm}=130$ GeV and $m_{\tilde \nu}=110$ GeV). } 
\end{figure}
\begin{figure}[h]
\centering
\includegraphics[width=4.7cm]{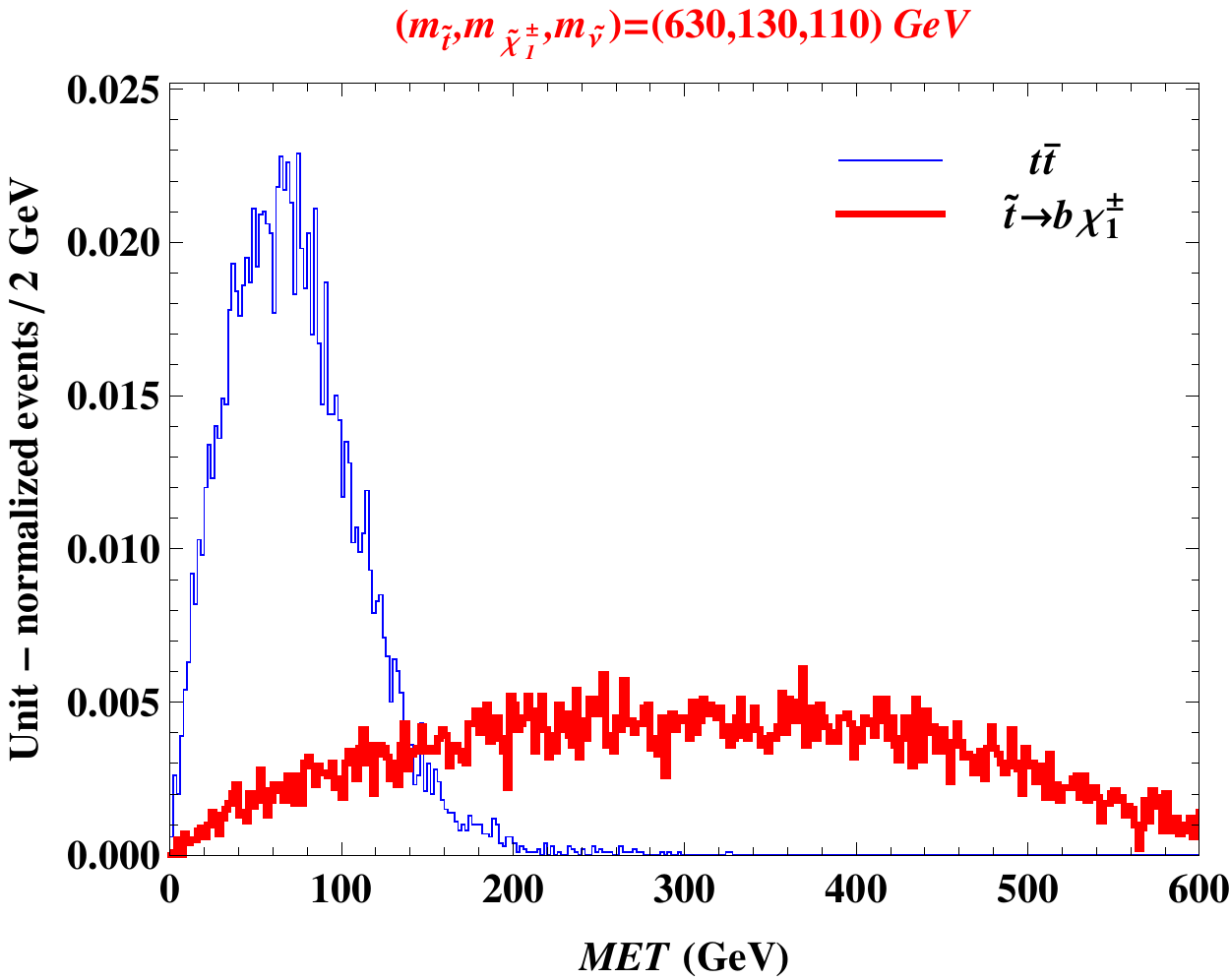}
\includegraphics[width=4.7cm]{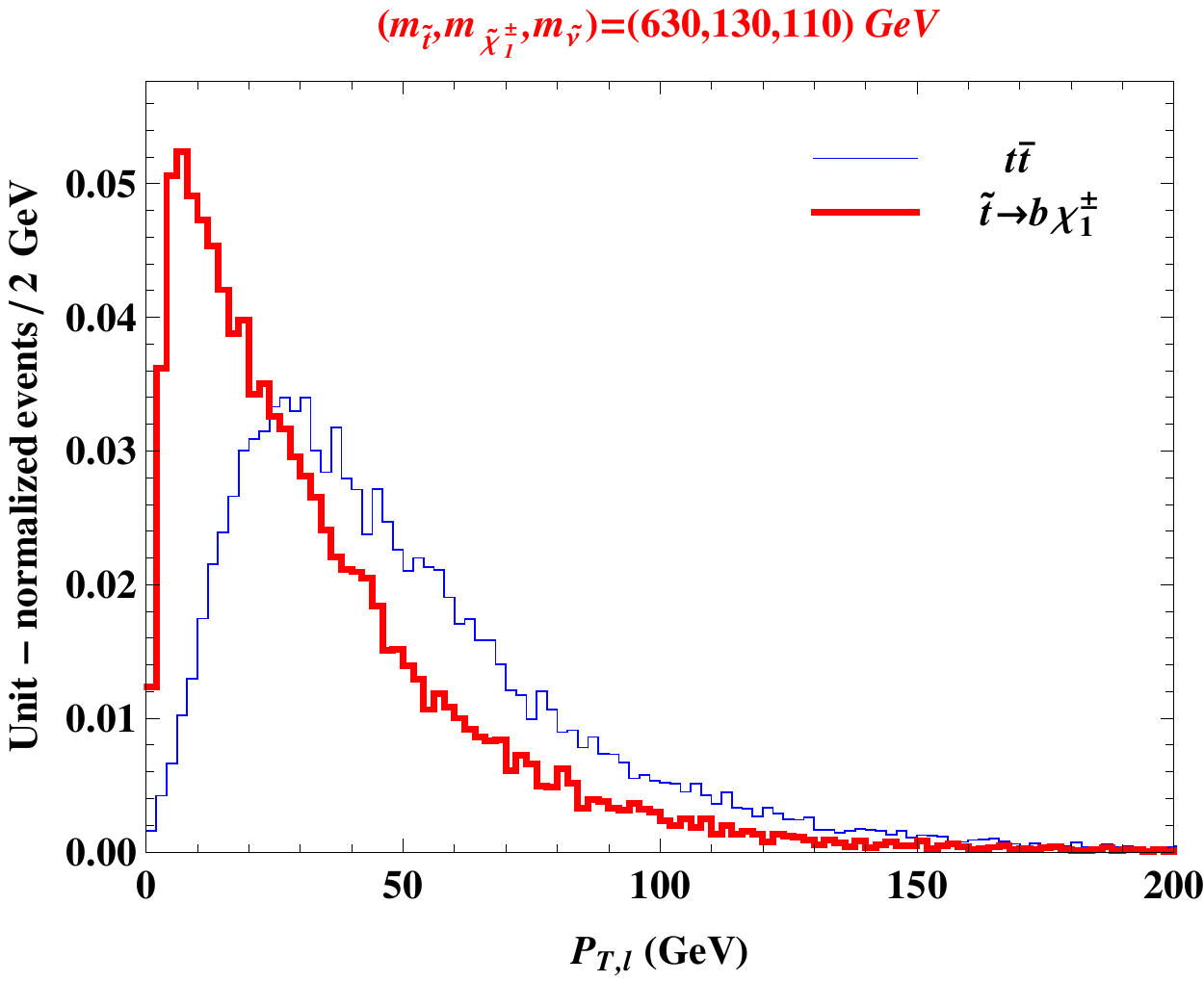}
\includegraphics[width=4.7cm]{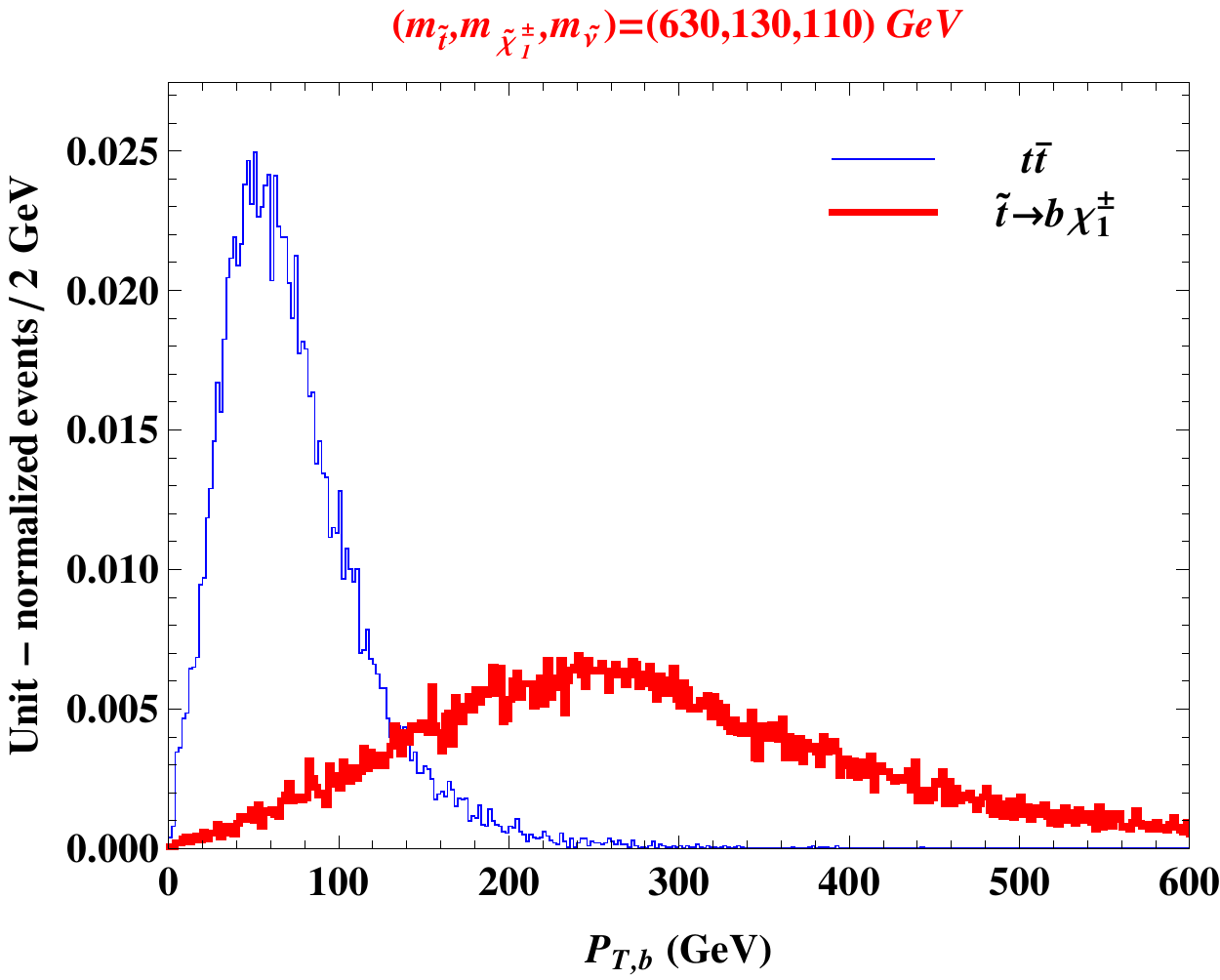}
\caption{\label{fig:cuts2} 
The same as Fig.~\ref{fig:cuts3}, but for signal events for
study point 3 ($m_{\tilde t}=630$ GeV,  $m_{\tilde \chi^\pm}=130$ GeV and $m_{\tilde \nu}=110$ GeV).} 
\end{figure}
This feature is demonstrated in Fig.~\ref{fig:MT2cuts2}, where we
again compare the signal and background distributions for the same
four types of variables as in Figs.~\ref{fig:MT2cuts3} and
\ref{fig:MT2cuts1}.  We see that this time, as expected, the dilepton
$M_{T2}$ variable ($M_{T2}(b)$ in our notation) is suboptimal; due
to the softness of the leptons, the signal $M_{T2}(b)$ and
$M_{2CC}(b)$ distributions lie entirely within the background region.
On the other hand, the other three variables perform very well, unlike
the case in Sec.~\ref{subsec:point-2}. In particular, the subsystem
$(ab)$ variables alone could possibly remove the background with
virtually no loss of signal.  The subsystem $(a)$ variables are also
promising; the use of $M_{2CC}(a)$ seems slightly more effective than
the use of $M_{T2}(a)$. Finally, the usual invariant mass, $m_{b\ell}$,
allows one to separate signal and background, but the signal loss is
more significant for this variable.

The lesson from Figs.~\ref{fig:MT2cuts1} and \ref{fig:MT2cuts2} is
that in order to efficiently probe the full mass parameter space of
Fig.~\ref{fig:map}, one would have to design an analysis which
utilizes the full complement of subsystem invariant mass variables,
since different variables are optimal in different regions. 
Of course, one should not overlook the more conventional variables.
In Fig.~\ref{fig:cuts2} we show the signal and background
distributions of $MET$, $p_{T,\ell}$, and $p_{T,b}$ for study point 3.
As expected, the lepton $p_T$ distribution for the signal is rather
soft, but the signal distributions for both $MET$ and $p_{T,b}$ have
long tails which extend to the right of the bulk of the corresponding
background distribution.  This suggests that $MET$ and $p_{T,b}$ could
also play a useful role in the analysis.

\subsection{Study point 4: a difficult case in region vi}
\label{subsec:point-4}

\begin{figure}[t]
\centering
\includegraphics[width=7.0cm]{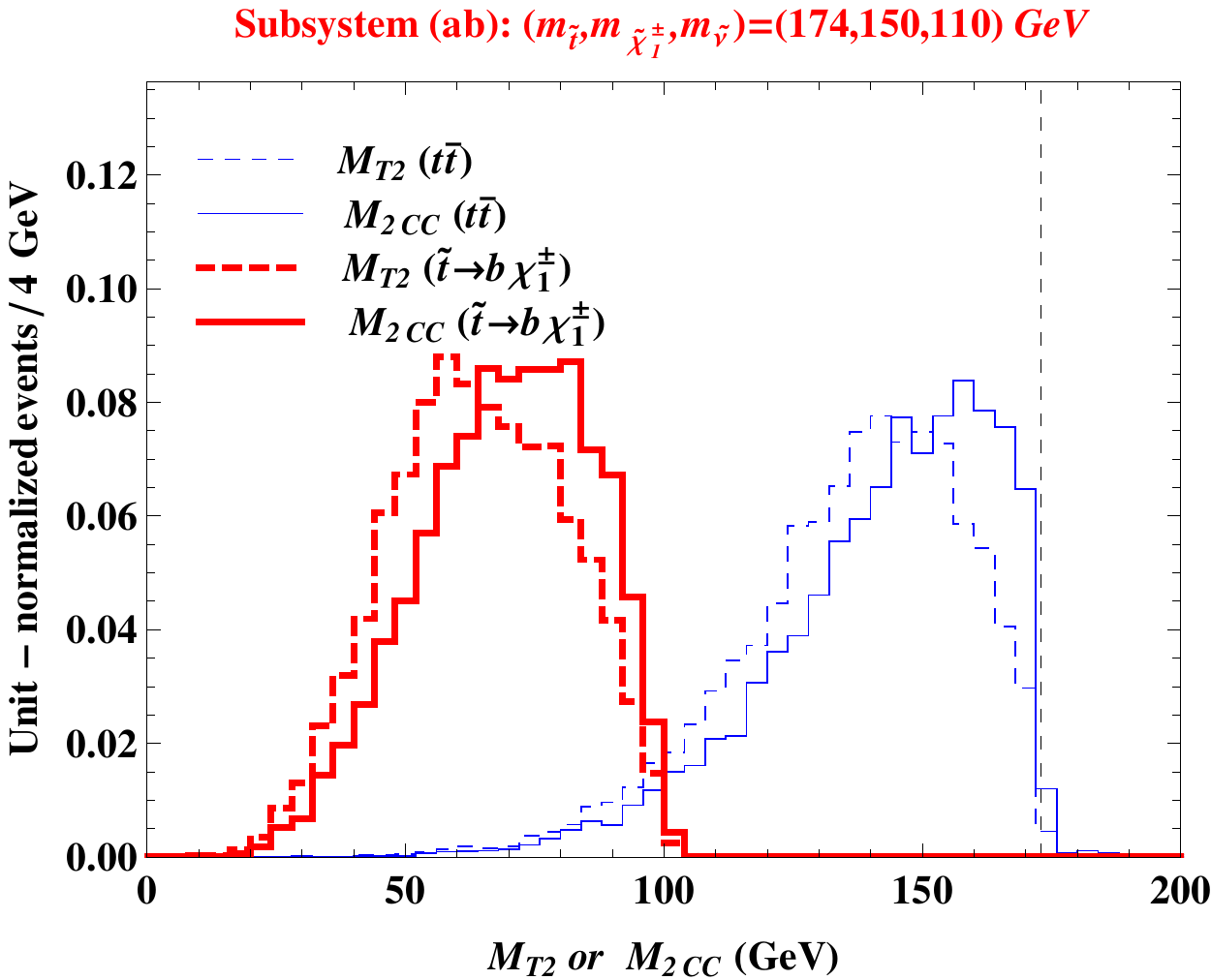}
\includegraphics[width=7.0cm]{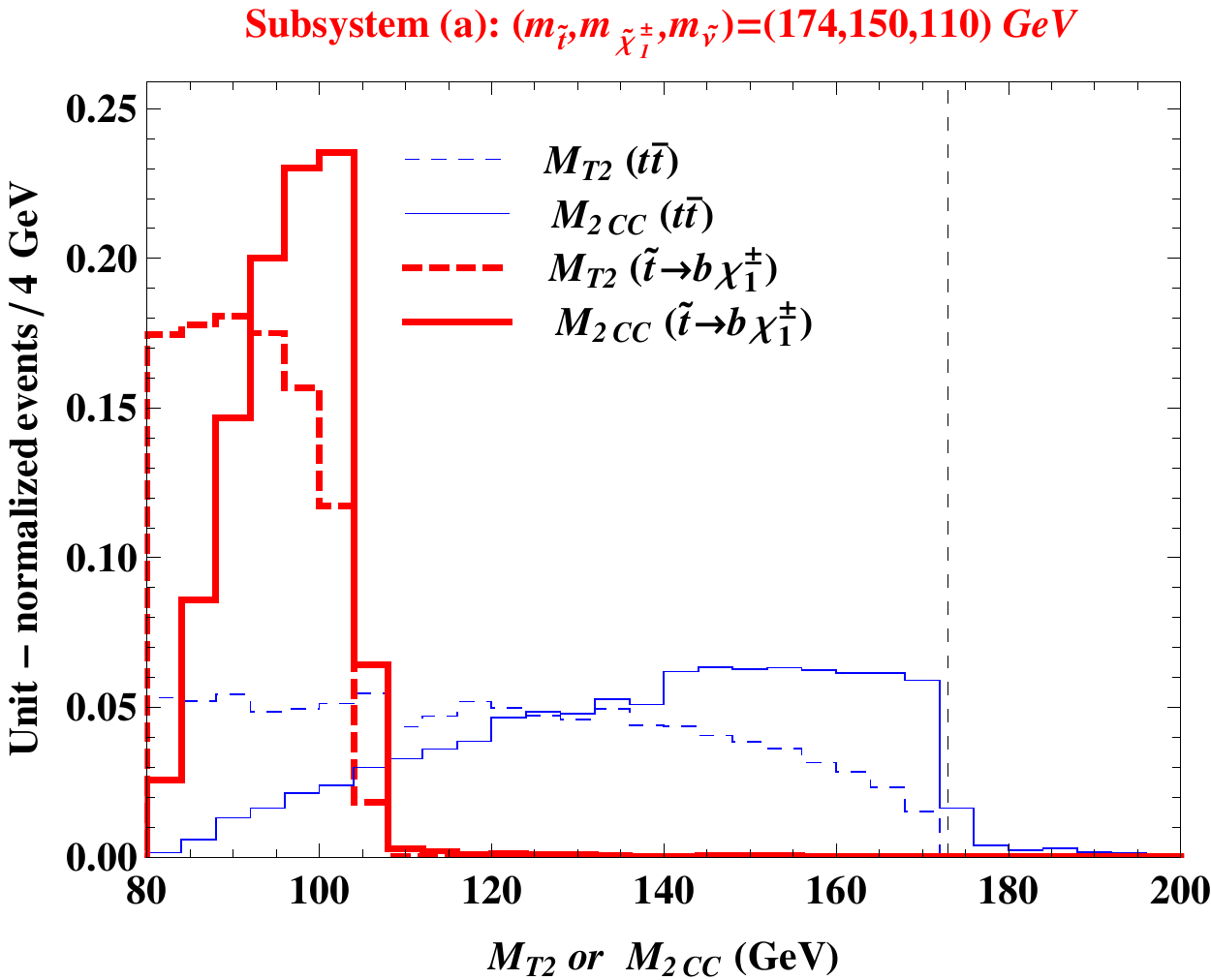}\\
\includegraphics[width=7.0cm]{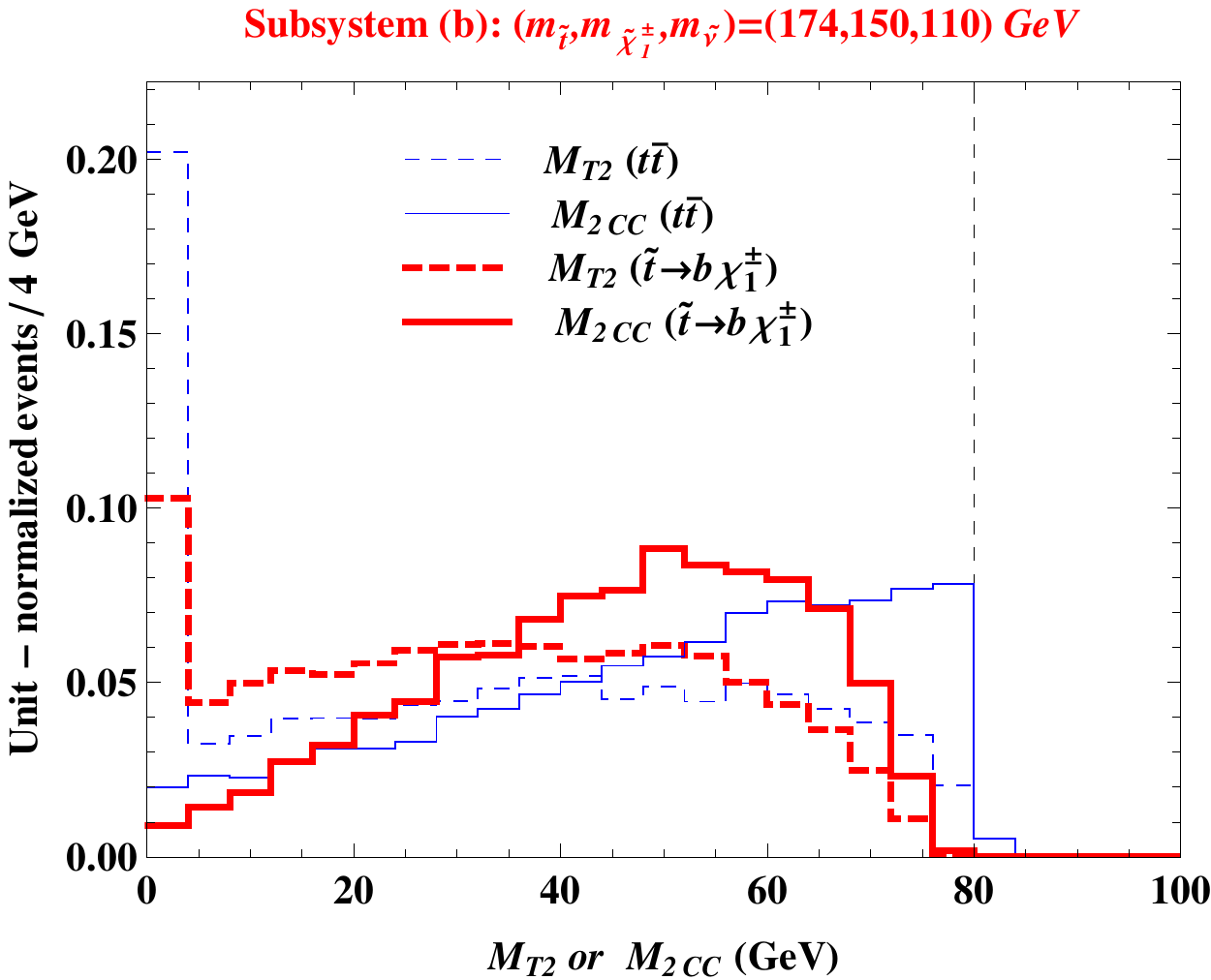}
\includegraphics[width=7.0cm]{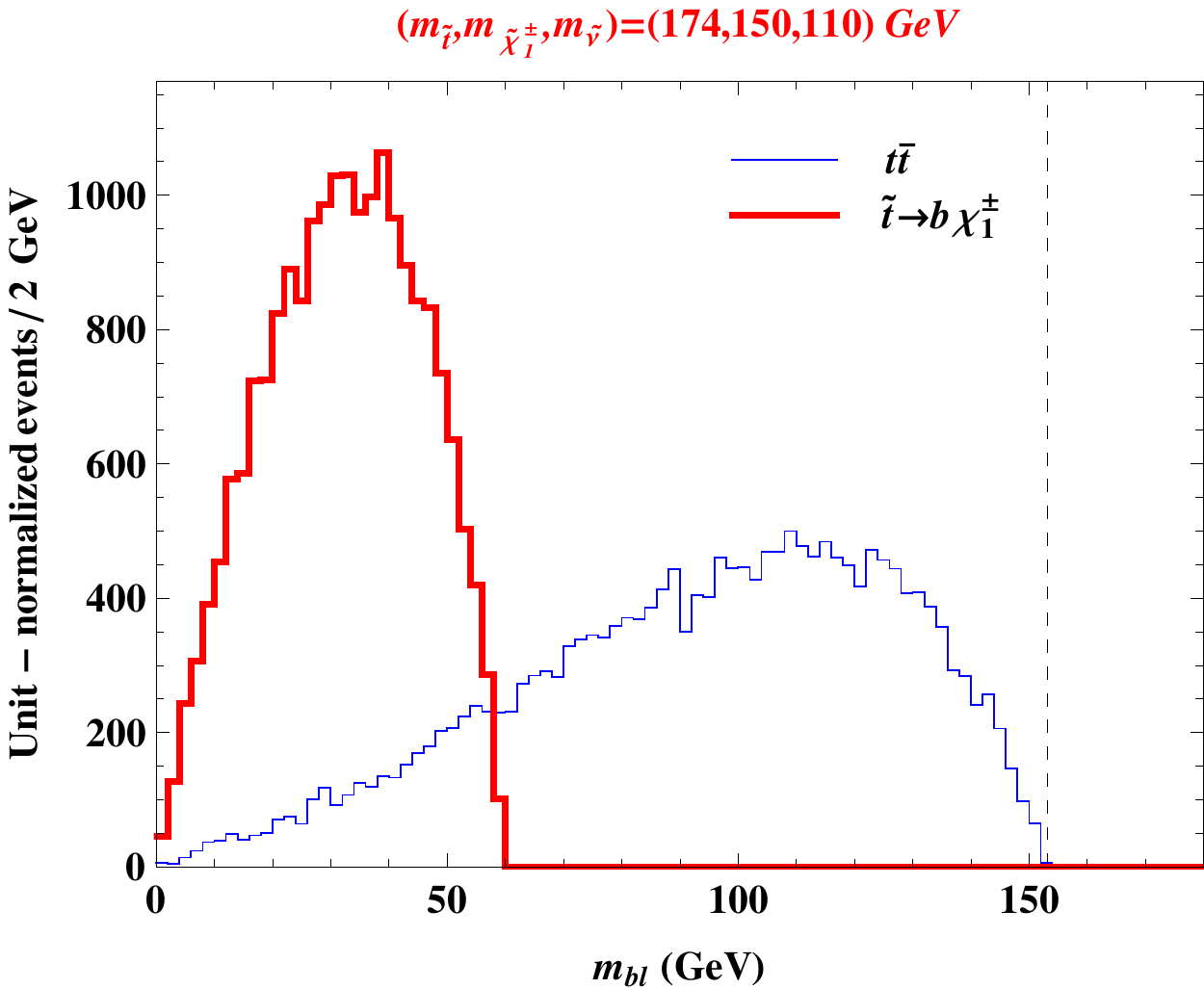}
\caption{\label{fig:MT2cuts4} 
The same as Fig.~\ref{fig:MT2cuts3}, but with signal events for
study point 4 ($m_{\tilde t}=174$ GeV,  $m_{\tilde \chi^\pm}=150$ GeV and $m_{\tilde \nu}=110$ GeV). } 
\end{figure}

\begin{figure}[h]
\centering
\includegraphics[width=4.7cm]{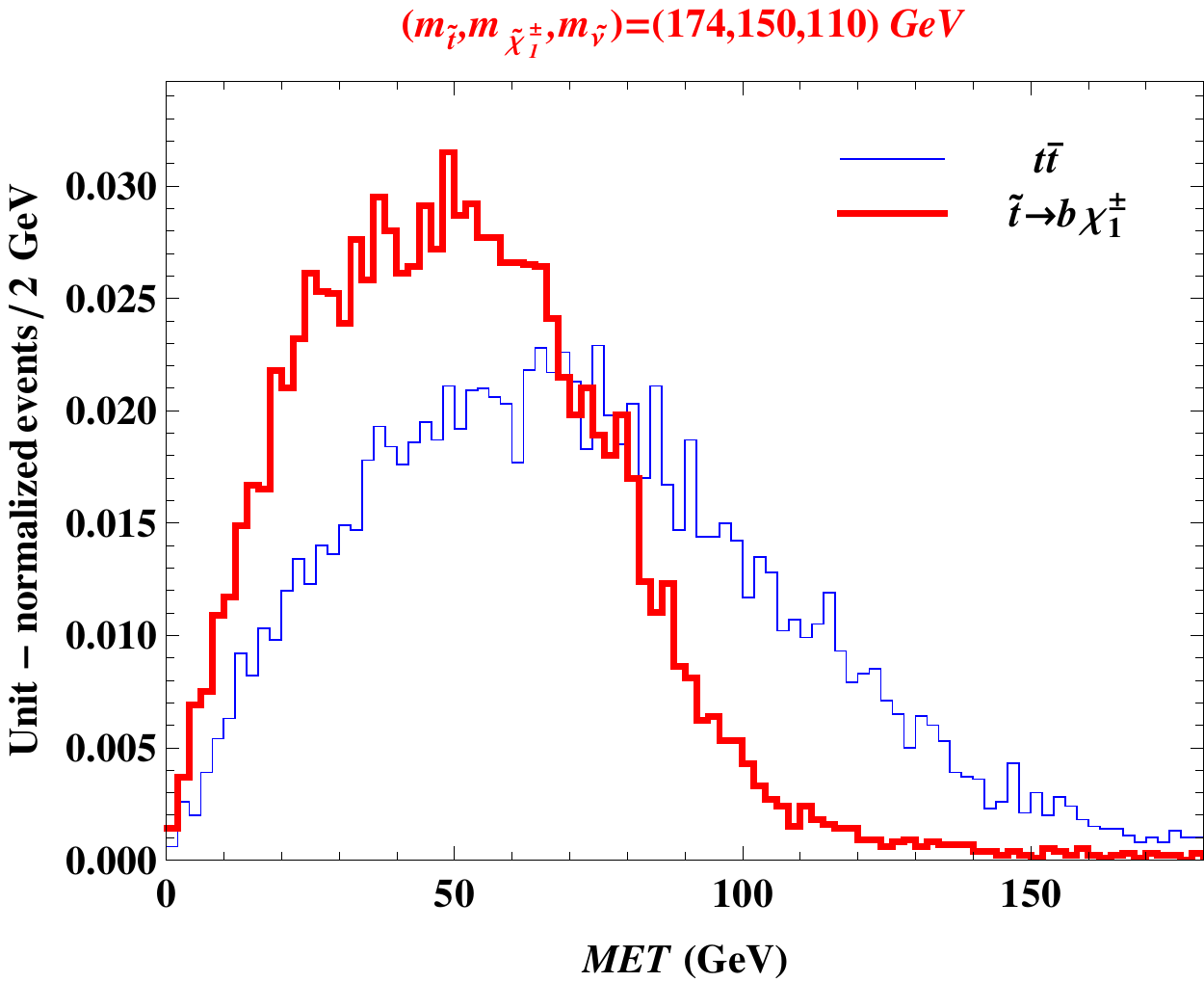}
\includegraphics[width=4.7cm]{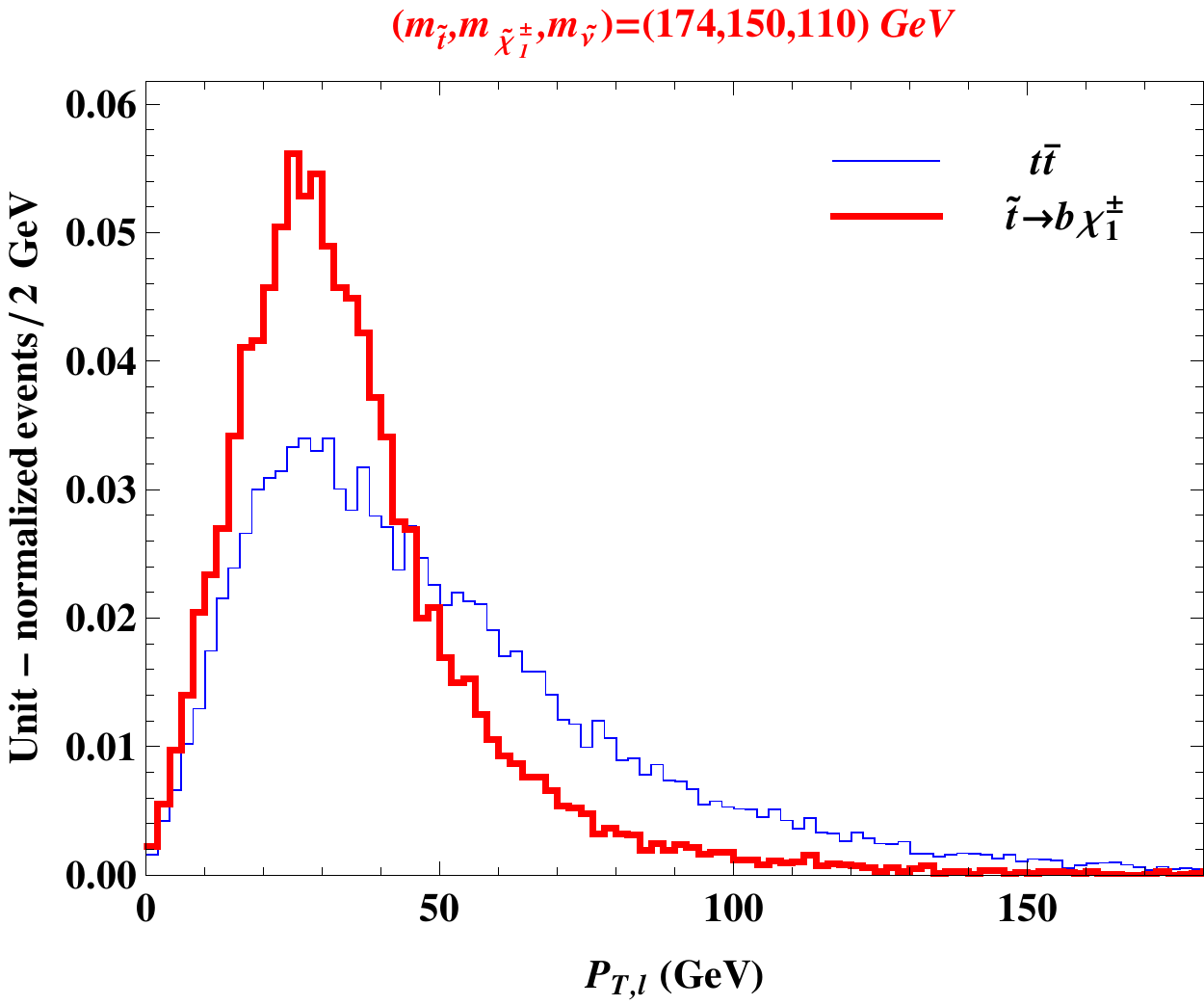}
\includegraphics[width=4.7cm]{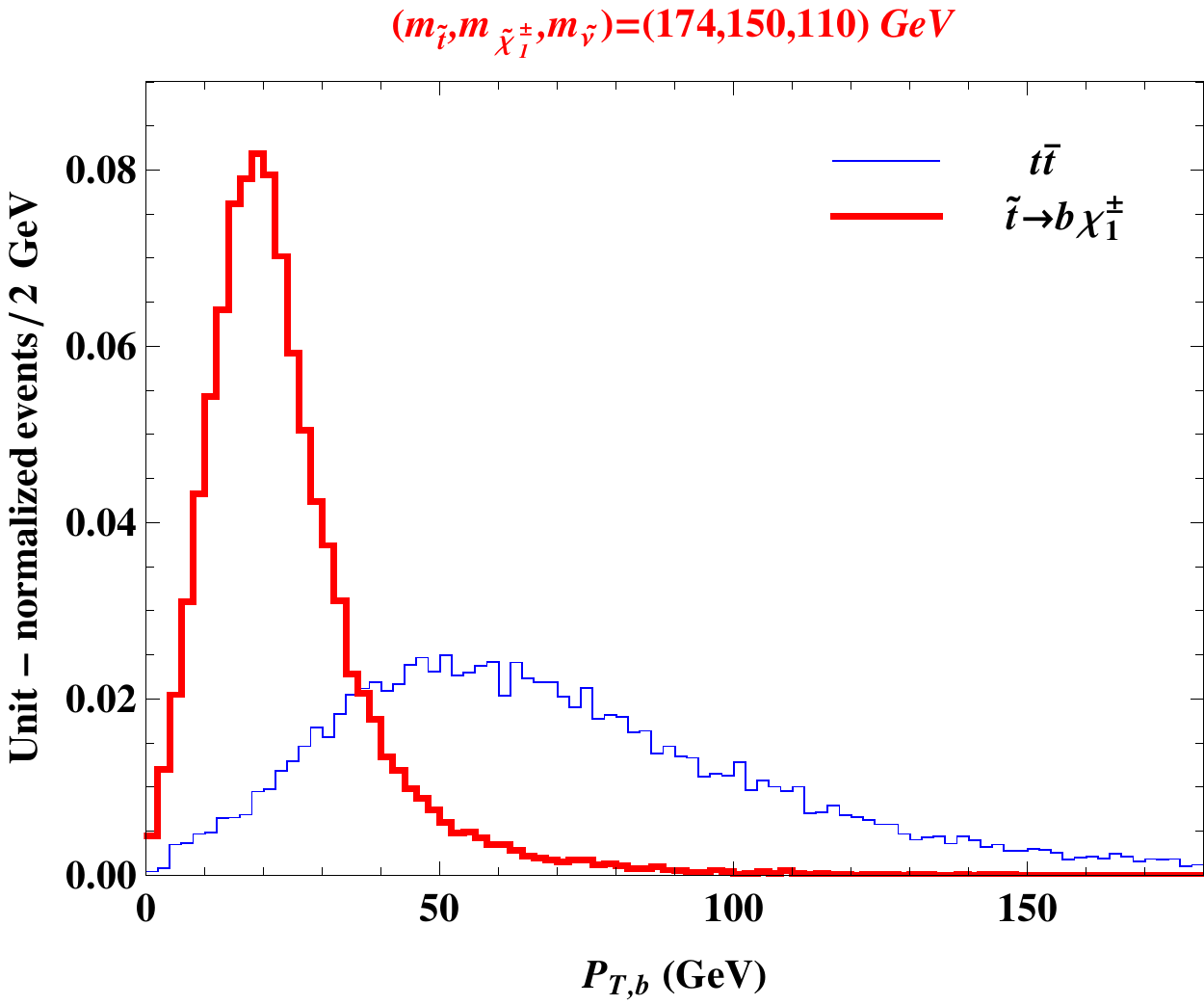}
\caption{\label{fig:cuts4} 
The same as Fig.~\ref{fig:cuts3}, but for signal events for
study point 4 ($m_{\tilde t}=174$ GeV,  $m_{\tilde \chi^\pm}=150$ GeV and $m_{\tilde \nu}=110$ GeV).} 
\end{figure}

Our final example for the signal event topology of
Fig.~\ref{fig:DecayTopologies}(b) is a study point in the most
challenging region, \textcolor{region6}{vi}, where no endpoint
violations should occur. The mass spectrum for study point 4 is given
in Table~\ref{tab:study-points}, and the resulting signal and
background distributions are displayed in Figs.~\ref{fig:MT2cuts4} and
\ref{fig:cuts4}.
Discovery is clearly challenging in this scenario, as the
$b$-jets will be quite soft, while the lepton $p_T$ and $MET$
distributions for the signal are very similar to those for the
background.  The invariant mass variables in Fig.~\ref{fig:MT2cuts4}
are not particularly helpful either, since the kinematic endpoints of the
signal distributions are always below those of the background. Whether
stops can be discovered at study point 4 thus remains an open
question, which we shall revisit in
Sec.~\ref{sec:mixed-events}.

\section{$M_2$ endpoint study for Topology 2}
\label{sec:M2_endpoint_study_for_topology_2}

In this section we shall focus on the other symmetric signal event
topology in Fig.~\ref{fig:DecayTopologies}(c), when both stops decay
according to Topology 2 in Fig.~\ref{fig:process}(b).  The on-shell
constrained invariant mass variables discussed in the previous section
will be useful here as well, since they were constructed with the {\em
  background} topology in mind, which has not changed.  Even though
the signal event topology is now more complicated (there are two
invisible particles in each decay chain), the signal distributions
still exhibit kinematic endpoints.   We find that the $M_{T2}$
endpoints are given by (see, e.g., \cite{Mahbubani:2012kx})
\bea
M_{T2}^{max}(ab;\tilde m=0) & = &  2\, C_{+}(m_t, m_\nu), 
\label{susymtab2}\\ [2mm]
M_{T2}^{max}(a;\tilde m=m_W) & = &
C_{+}(m_{t},m_{W})+\sqrt{C^2_{+}(m_{t},m_{W})+m^2_W},
\label{susymta2} \\ [2mm]
M_{T2}^{max}(b;\tilde m=0) & = & 2\, \sqrt{C_{+}(m_{W},m_{\nu})\, 
\left[ C_{+}(m_{W},m_{\nu})- C_{-}(m_{t},m_{W}) \right]},
\label{susymwb2}
\eea
where
\bea
C_{\pm}(x,y) &\equiv&  \frac{x^2-y^2}{4\,m_{\tilde t}\,x^2}\, 
\left\{m^2_{\tilde t}+ m^2_{t} -m^2_{\tilde\chi^0}
\pm \sqrt{\lambda(m^2_{\tilde t}, m^2_{t}, m^2_{\tilde\chi^0})}\right\},
\\[2mm]
\lambda(x,y,z)&\equiv& x^2+y^2+z^2 - 2xy - 2yz - 2zx.
\eea
Upon careful examination of Eqs.~(\ref{susymtab2}-\ref{susymwb2}), one
can show that these kinematic endpoints are always above the
corresponding background endpoints (\ref{mtab}-\ref{mwb}), as long as
the channel $\tilde t \to t \tilde\chi^0$ is open (i.e., the decay is
kinematically allowed).  As we move close
to the threshold for $\tilde t \to t \tilde\chi^0$, the signal
kinematic endpoints (\ref{susymtab2}-\ref{susymwb2}) converge to the
corresponding SM values (\ref{mtab}-\ref{mwb}), and discovery becomes
very challenging.  In this section, therefore, we shall consider two
study points: one above this threshold and one at threshold. The mass
spectra for those study points  are listed in
Table~\ref{tab:study-points-topo2}.

\begin{table}[t]
\centering
\begin{tabular}{| c | c | c | c | c |}
\hline
Study Point & Stop Mass & Neutralino Mass \\
\hline \hline
5  & $300$ GeV & $100$ GeV  \\
6  & $174$   GeV & $0$ GeV  \\
\hline
\end{tabular}
\caption{\label{tab:study-points-topo2} The stop and neutralino masses
  for the two study points considered in this section.}
\end{table}

\subsection{Study point 5: a case above the $\tilde t \to t
  \tilde\chi^0$ threshold}
\label{subsec:point-5}

\begin{figure}[t]
\centering
\includegraphics[width=7.0cm]{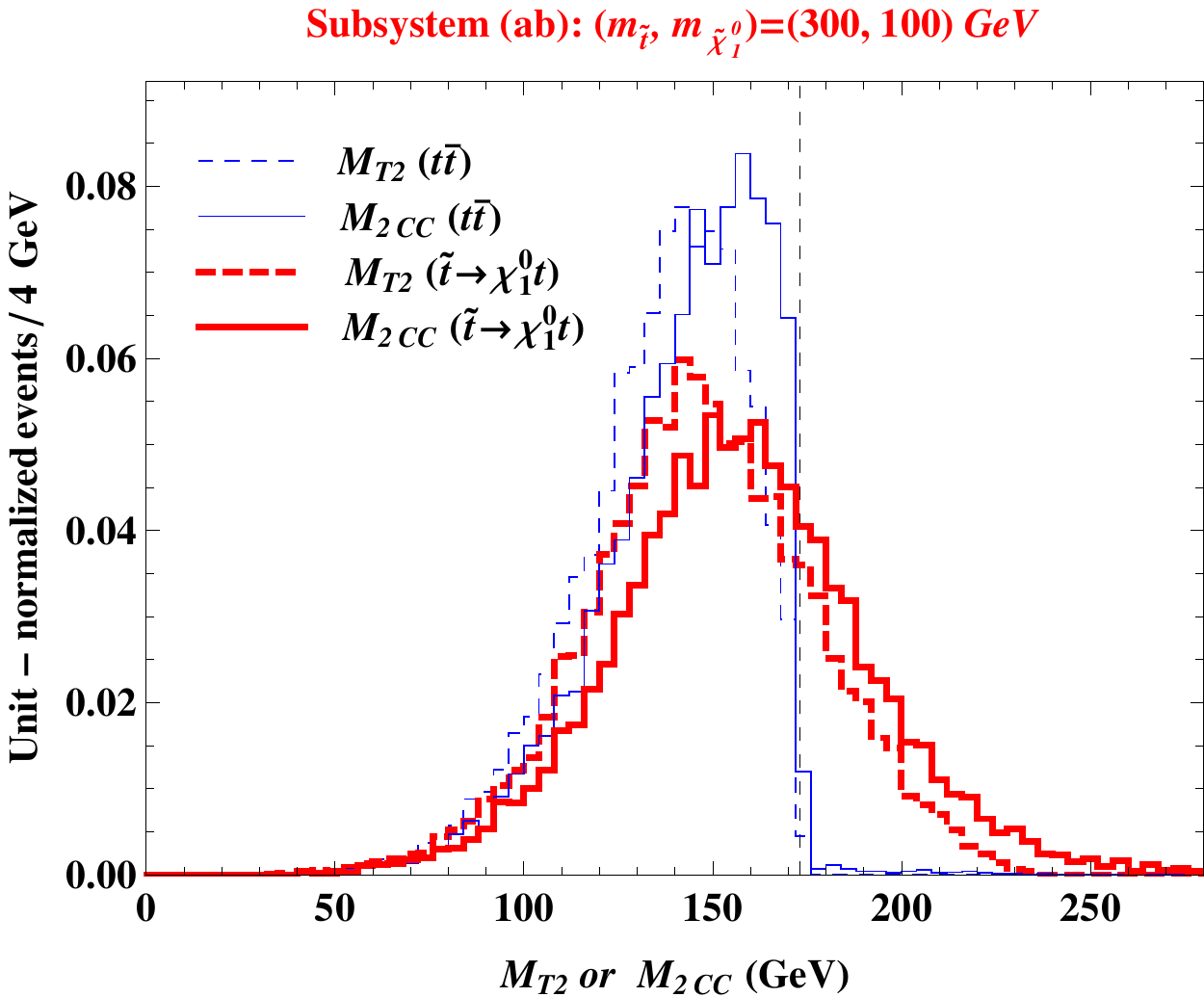}
\includegraphics[width=7.0cm]{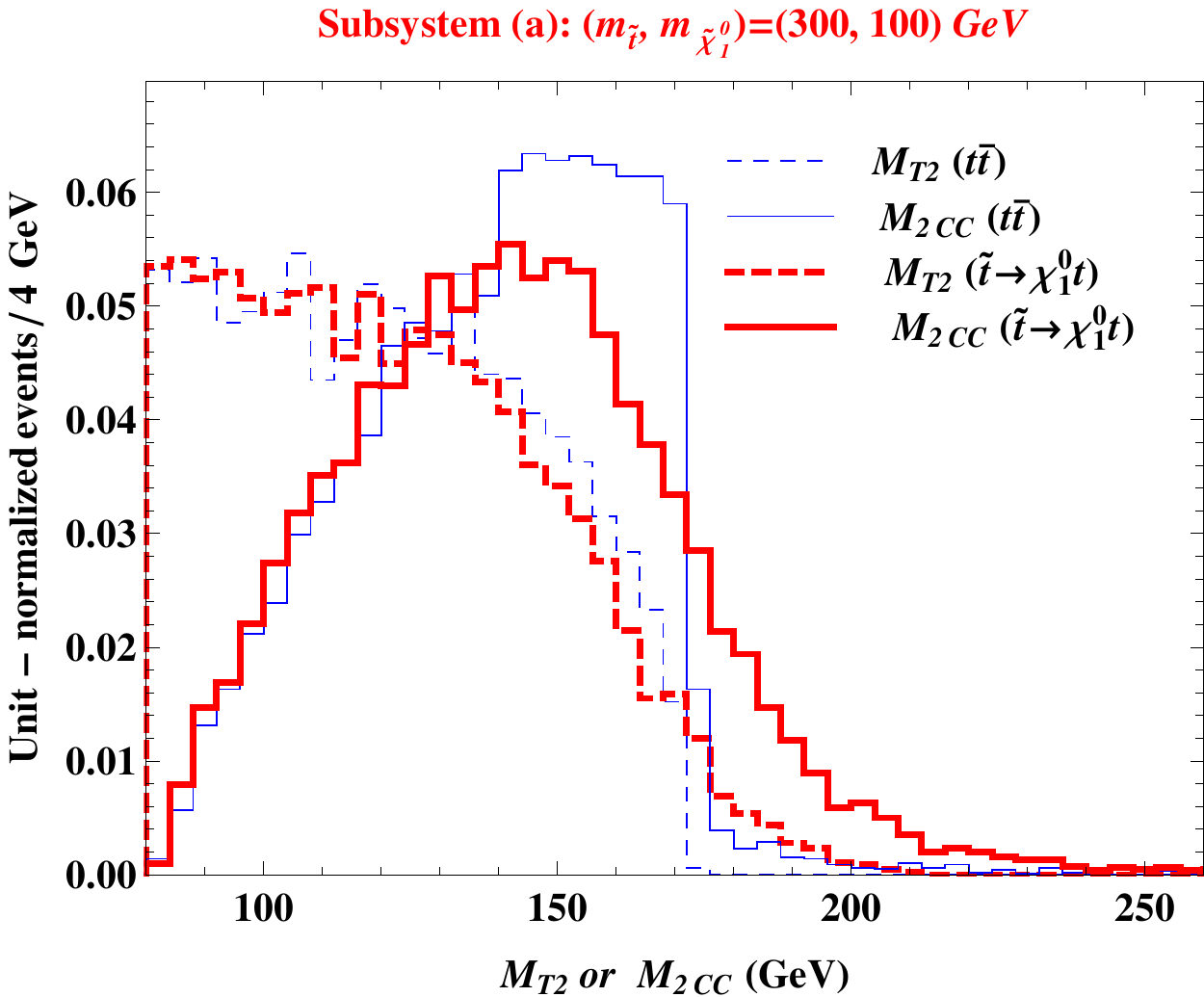}\\
\includegraphics[width=7.0cm]{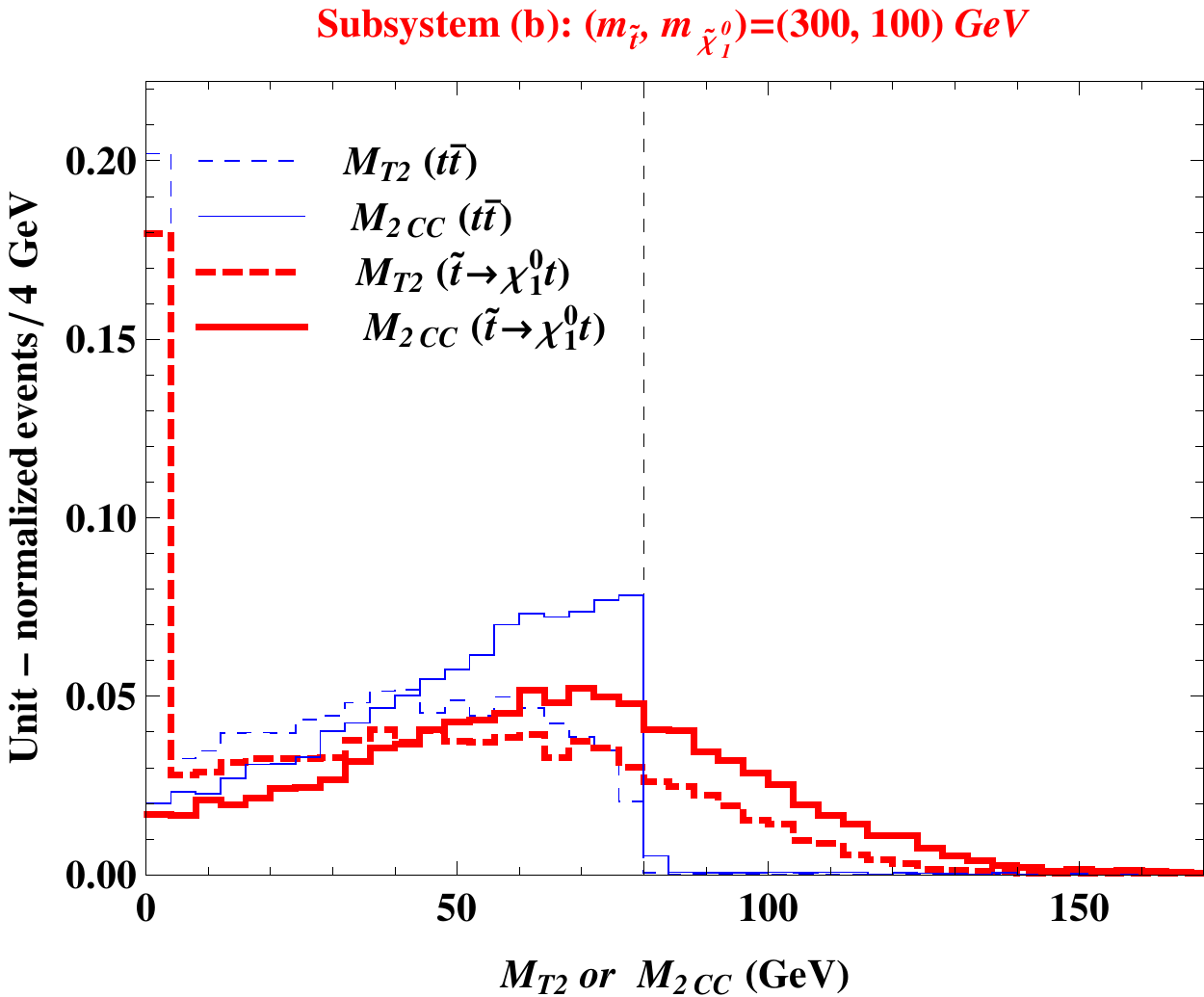}
\includegraphics[width=7.0cm]{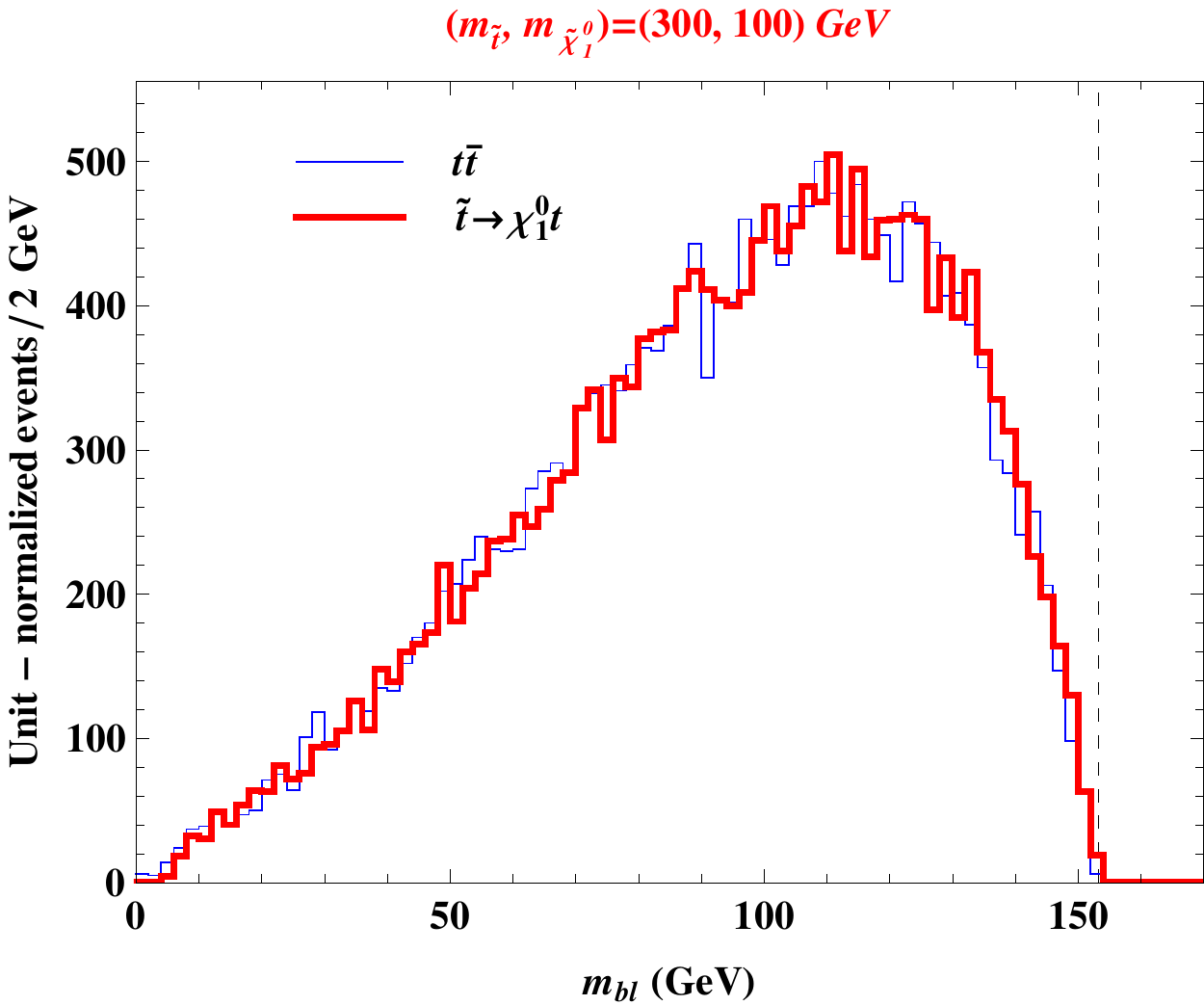}
\caption{\label{fig:MT2cuts5} 
The same as Fig.~\ref{fig:MT2cuts3}, but for signal events with the event topology of Fig.~\ref{fig:DecayTopologies}(c)
for study point 5 ($m_{\tilde t}=300$ GeV and $m_{\tilde \chi^0}=100$ GeV). } 
\end{figure}

\begin{figure}[h]
\centering
\includegraphics[width=4.7cm]{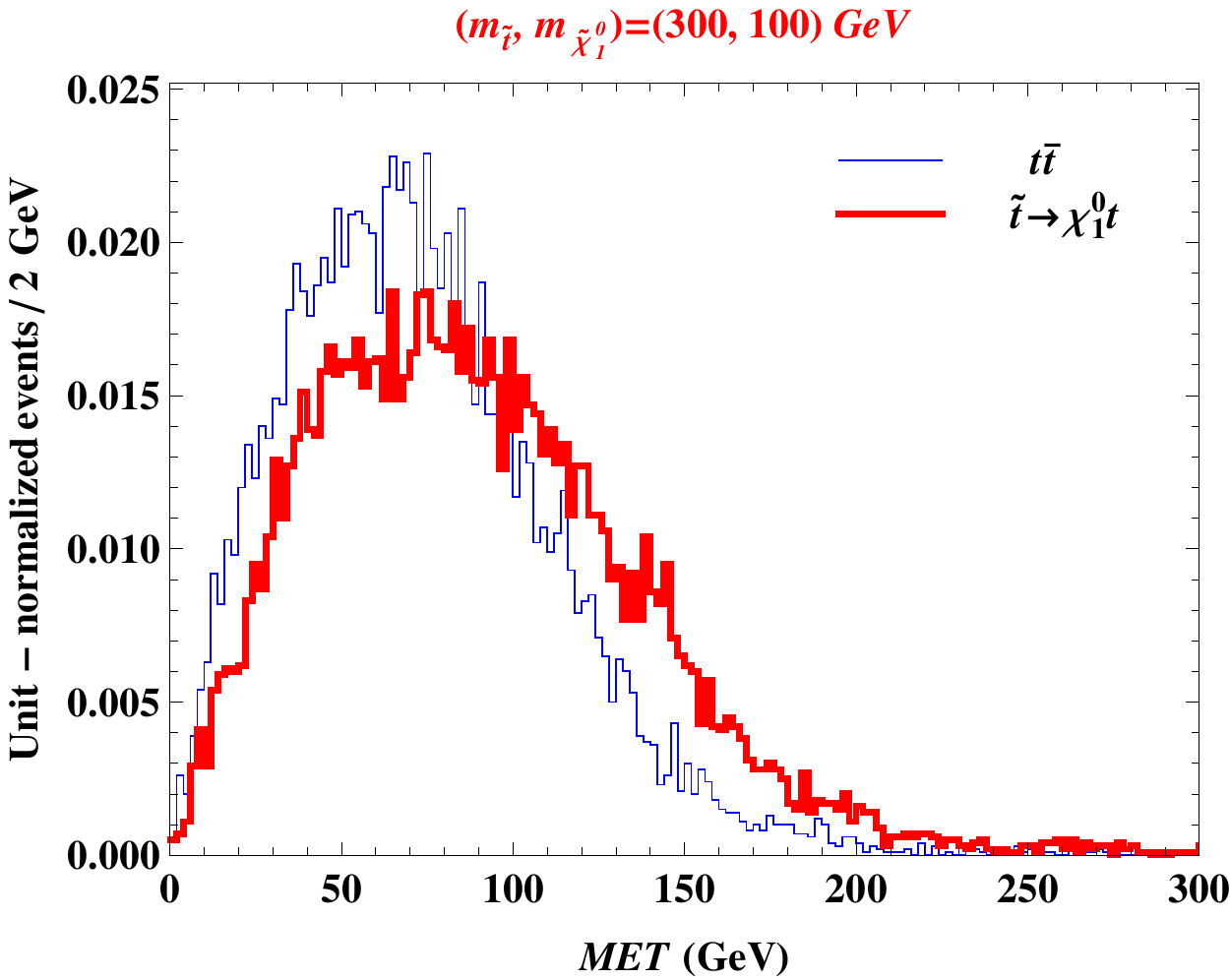}
\includegraphics[width=4.7cm]{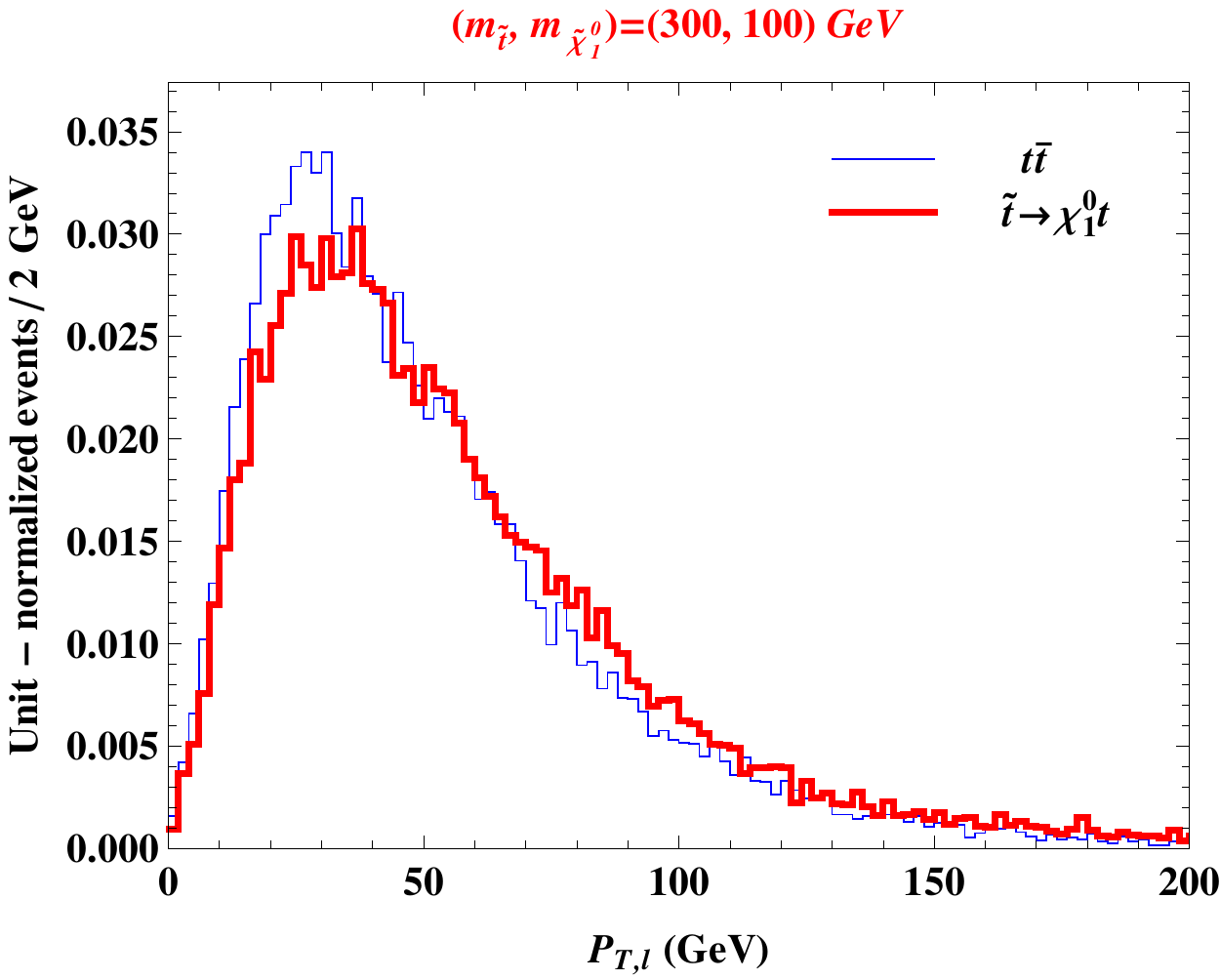}
\includegraphics[width=4.7cm]{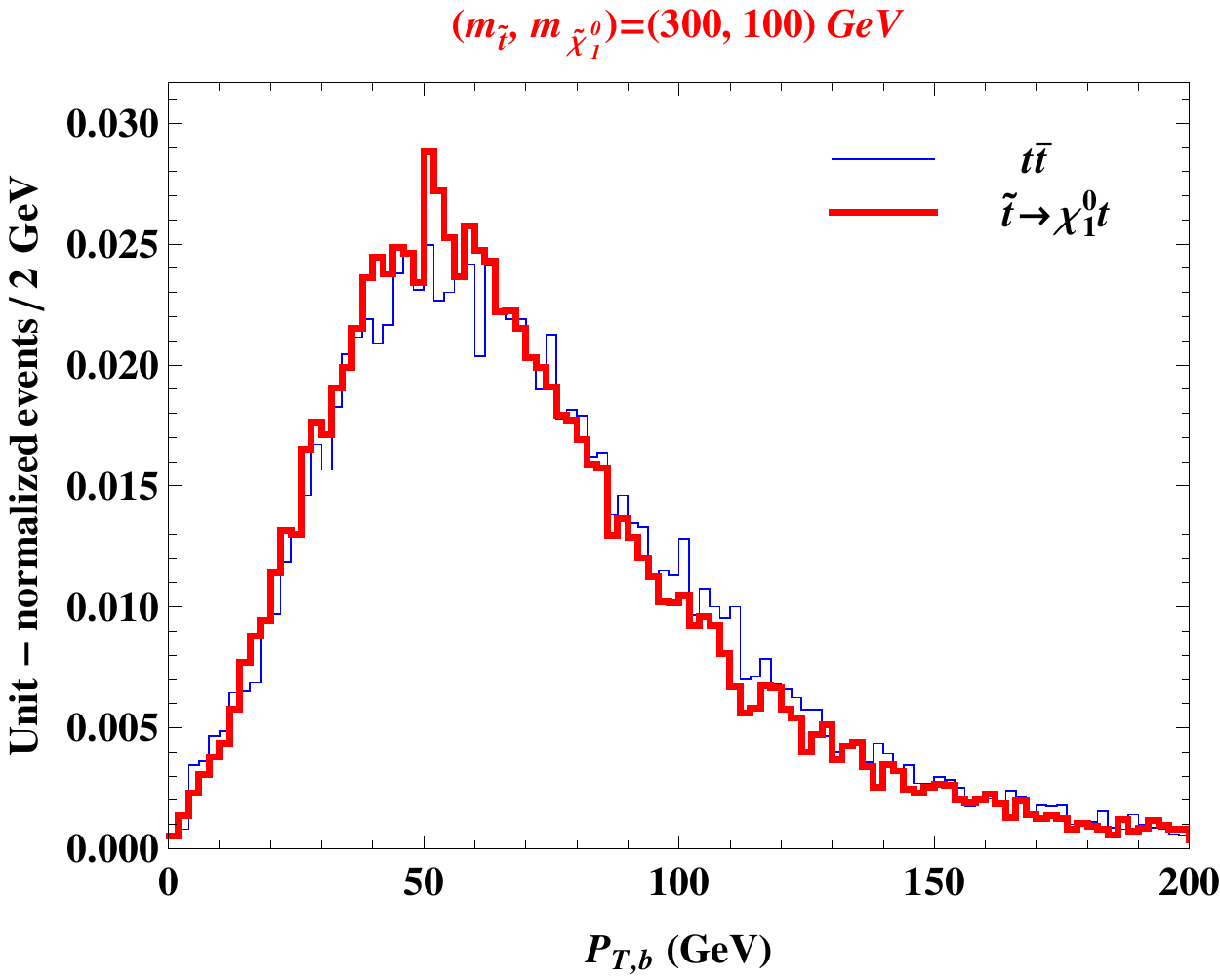}
\caption{\label{fig:cuts5} 
The same as Fig.~\ref{fig:cuts3}, but for signal events with the event topology of Fig.~\ref{fig:DecayTopologies}(c)
for study point 5 ($m_{\tilde t}=300$ GeV and $m_{\tilde \chi^0}=100$ GeV). } 
\end{figure}

We first discuss study point 5, where the mass splitting is large
enough that the decay $\tilde t \to t \tilde\chi^0$ is open, and the
resulting top quark is on-shell.  The corresponding kinematic
distributions are shown in Figs.~\ref{fig:MT2cuts5} and \ref{fig:cuts5}.
It is clear that this is already a challenging case --- the signal and
background distributions for the ``conventional" variables in
Fig.~\ref{fig:cuts5} are rather similar. The jet and lepton $p_T$
spectra are governed by the known mass differences between the {\em SM}
  particles $t$, $W^\pm$, and $\nu$, thus, there is very little
distinction between the signal and background $p_T$
distributions. Similarly, the $m_{b\ell}$ distribution in
Fig.~\ref{fig:MT2cuts5} is the same for signal and
background. The $MET$ distribution in Fig.~\ref{fig:cuts5} is slightly
harder for the signal, due to the presence of two
additional invisible particles. However, the effect is very small and
hence unlikely to be useful in practice.

This motivates the use of the $M_{T2}$ and $M_{2\sqcup\sqcup}$
variables whose distributions are shown in the first three panels of
Fig.~\ref{fig:MT2cuts5}. As anticipated from
Eqs.~(\ref{susymtab2}-\ref{susymwb2}), for all three subsystems
$(ab)$, $(a)$, and $(b)$, the signal distributions for the $M_{T2}$
variable have a tail which extends beyond the background
endpoints. This effect is most pronounced for subsystem $(ab)$ and
less so for subsystem $(a)$.

Note how the situation improves if one were to use the on-shell
constrained variable $M_{2CC}$ (solid lines) instead of $M_{T2}$
(dashed lines). For background events, $M_{2CC}$ is computed by
applying the correct kinematic constraints; therefore, the
kinematic endpoints (\ref{mtab}-\ref{mwb}) are still obeyed. For
signal events, we get a somewhat different story --- a much larger
fraction of signal events now violate these endpoints, leading
to an improvement in the signal efficiency.  The largest benefit is
observed in the case of subsystem $(a)$, for which previously the
$M_{T2}$ variable was the least helpful.  There are two separate reasons
why $M_{2CC}$ separates signal from background better than $M_{T2}$:
\begin{enumerate}
\item Due to the hierarchy (\ref{eq:inequality}), the $M_{2CC}$
  distributions are harder than the $M_{T2}$ distributions, thus more
  signal events are expected to migrate above the background
  endpoint.  The shape difference between the $M_{2CC}$ and $M_{T2}$
  distributions is especially noticeable in the case of subsystems
  $(a)$ and $(b)$ in Fig.~\ref{fig:MT2cuts5}. Notice, in particular,
  the completely different shapes of the $M_{T2}(a)$ and $M_{2CC}(a)$
  distributions, as well as the disappearance of the big spike at
  $M_{T2}(b)=0$. 
\item For signal events, the $M_{2CC}$ kinematic endpoints
  themselves are even higher\footnote{We have not attempted to obtain
    analytical formulas analogous to (\ref{susymtab2}-\ref{susymwb2})
    for the $M_{2CC}$ kinematic endpoints, but our numerical studies
    clearly showed that the bounds (\ref{susymtab2}-\ref{susymwb2})
    themselves are violated in the case of the $M_{2CC}$ variable. }
  than the $M_{T2}$ kinematic endpoints given in
  Eqs.~(\ref{susymtab2}-\ref{susymwb2}).
This can be readily observed in Fig.~\ref{fig:MT2cuts5}, where the
signal $M_{2CC}$ distributions (red solid lines) extend to higher
values than the signal $M_{T2}$ distributions (red dashed lines).
Contrast this situation with the examples considered in 
Sec.~\ref{sec:M2_endpoint_study_for_topology_1}, when $M_{T2}$ and
$M_{2CC}$ always shared the same kinematic endpoint. There, the
signal event topology (Fig.~\ref{fig:DecayTopologies}(b)) was the same
as the background event topology
(Fig.~\ref{fig:DecayTopologies}(a)). As a result, the kinematic
constraints being imposed in the calculation of $M_{2CC}$ did
correspond to the actual physics of the signal events.  Now, in the
case of study point 5, the signal event topology of
Fig.~\ref{fig:DecayTopologies}(c) is completely different --- in a
sense, one is applying ``the wrong" constraints when calculating
$M_{2CC}$. Somewhat paradoxically then, Fig.~\ref{fig:MT2cuts5}
teaches us that one obtains a beneficial result, despite
applying ``the wrong" constraints. 
\end{enumerate}

\subsection{Study point 6: a case at the $\tilde t \to t \tilde\chi^0$ threshold}
\label{subsec:point-6}

\begin{figure}[t]
\centering
\includegraphics[width=7.0cm]{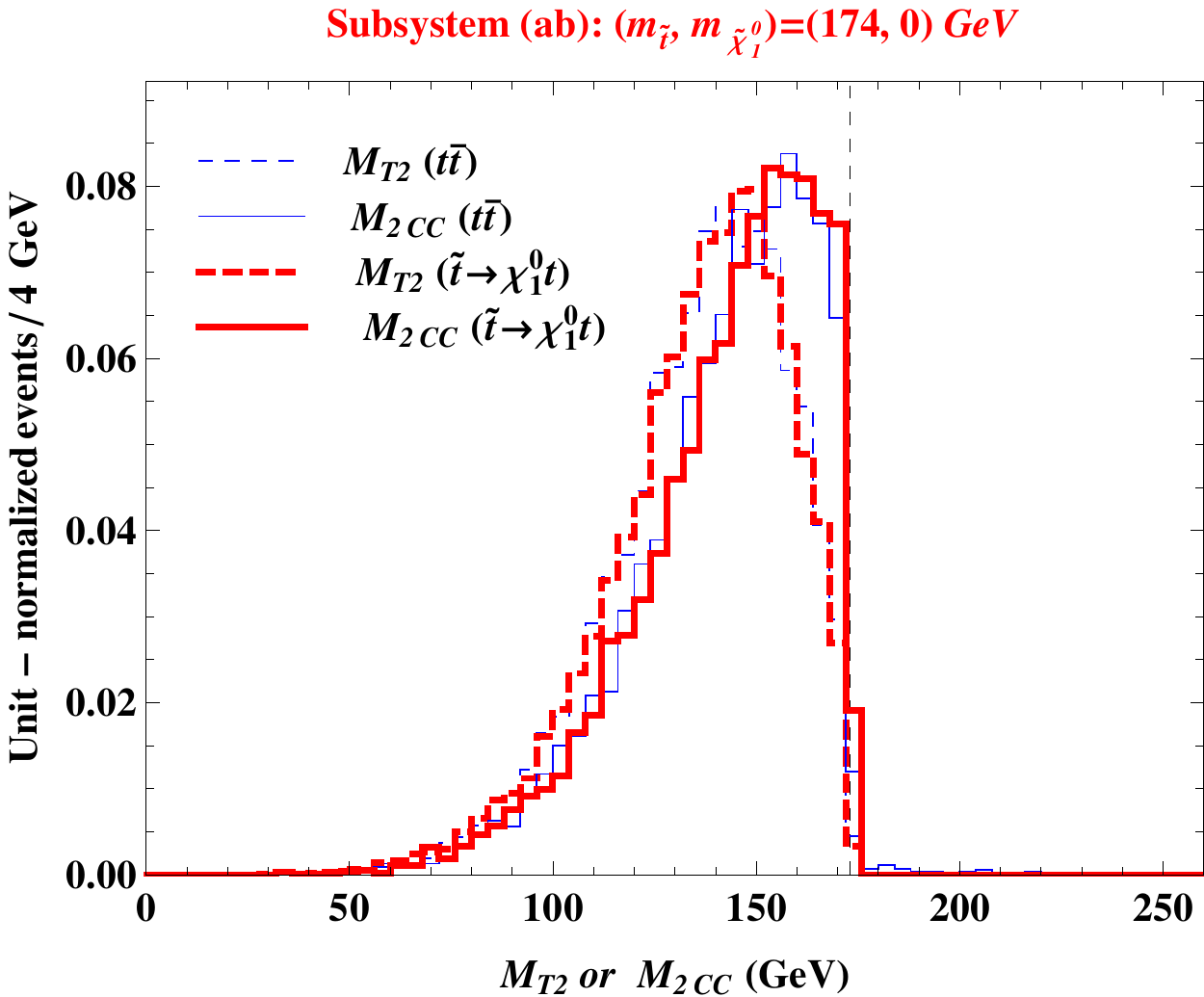}
\includegraphics[width=7.0cm]{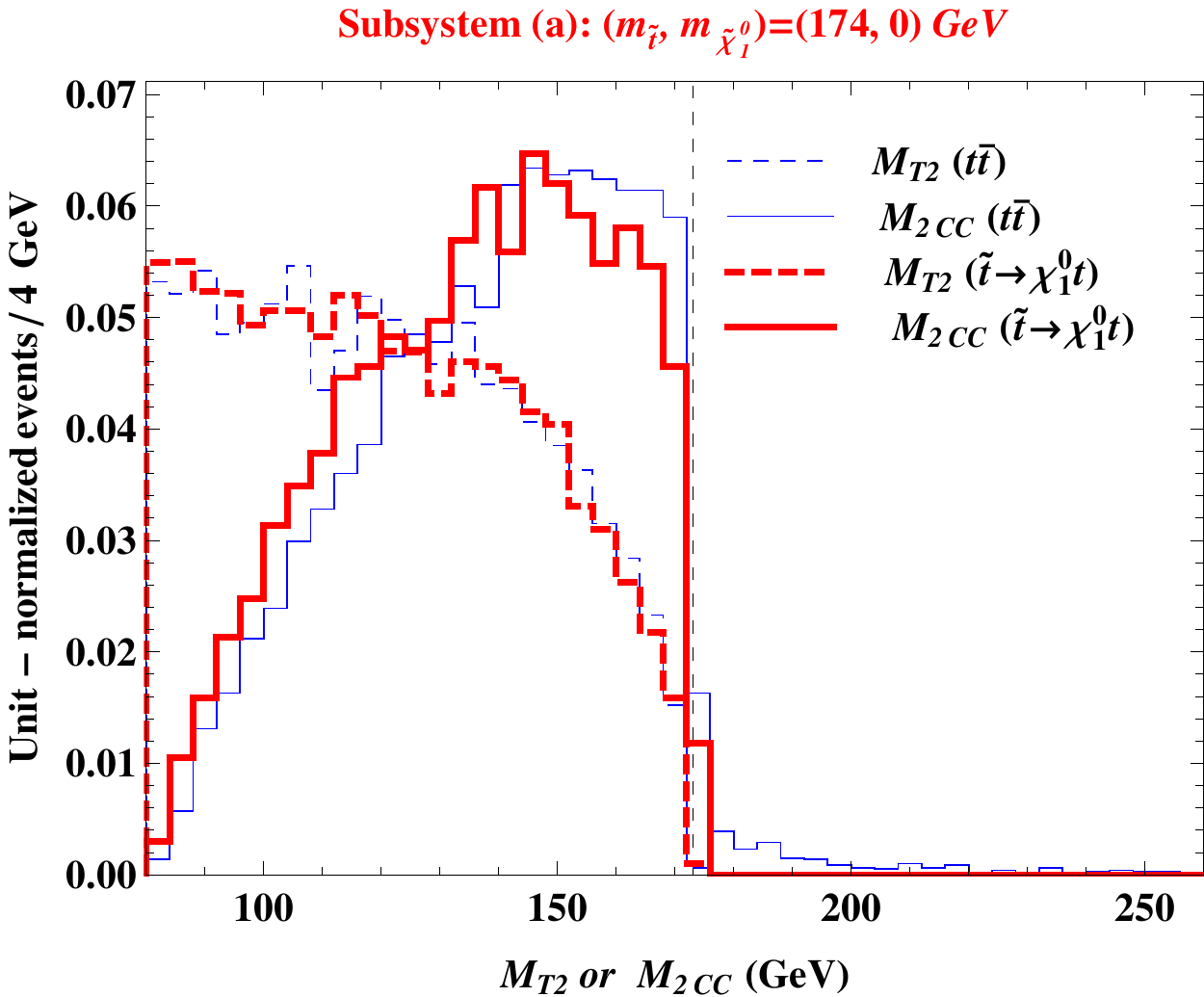}
\includegraphics[width=7.0cm]{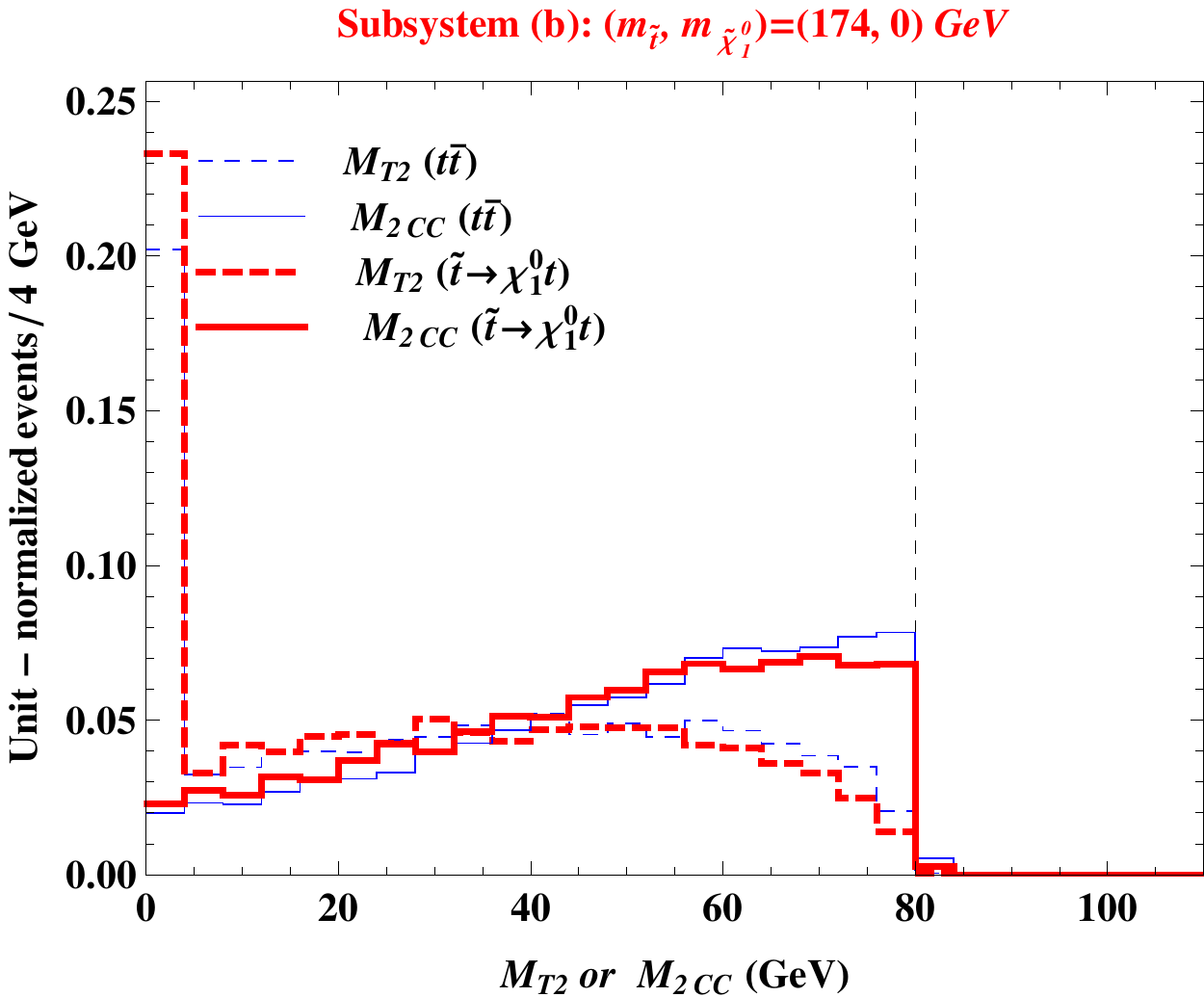}
\includegraphics[width=7.0cm]{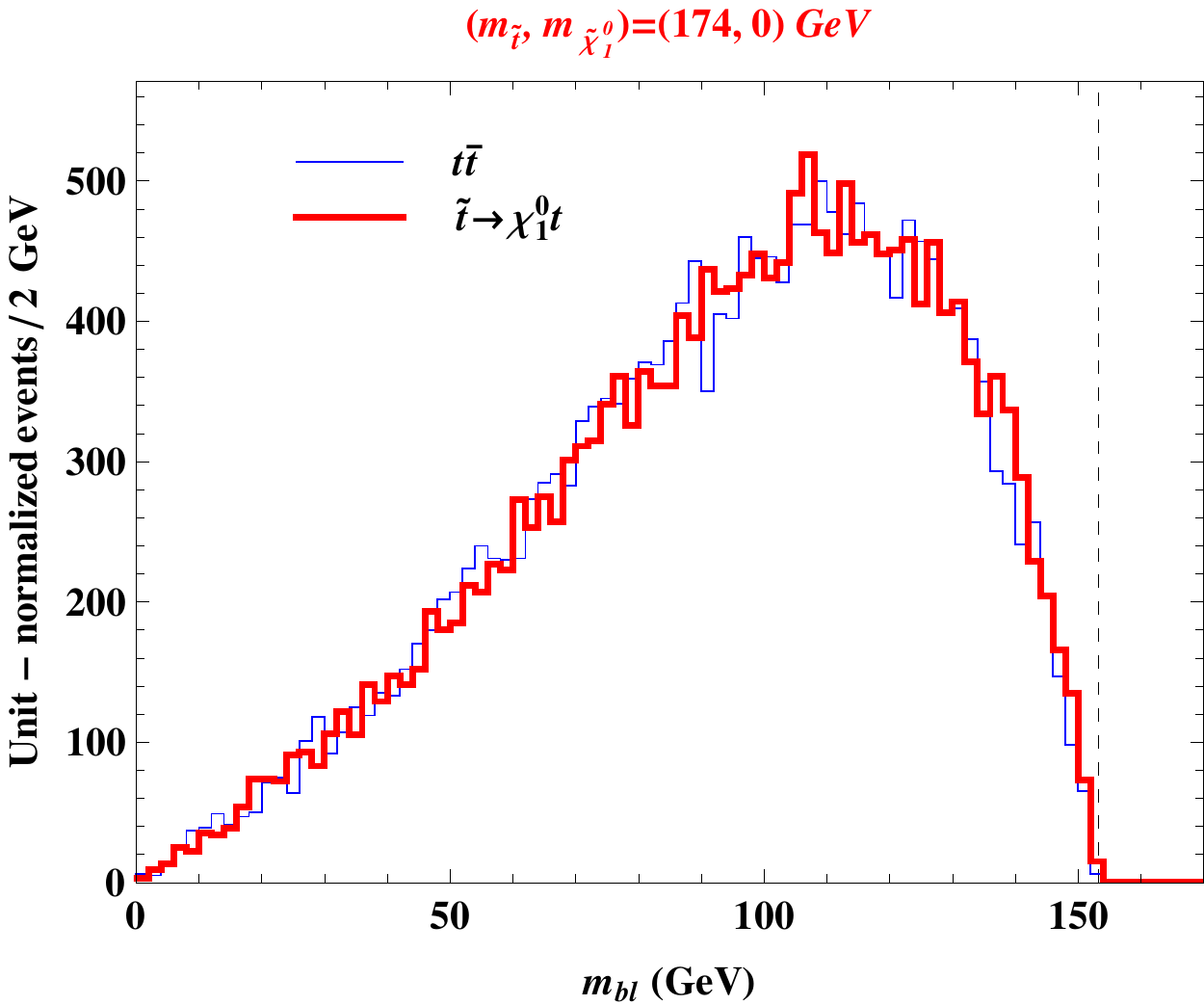}
\caption{\label{fig:MT2cuts6} 
The same as Fig.~\ref{fig:MT2cuts3}, but for signal events with the event topology of Fig.~\ref{fig:DecayTopologies}(c)
for study point 6 ($m_{\tilde t}=174$ GeV and $m_{\tilde \chi^0}=0$ GeV). } 
\end{figure}

\begin{figure}[t]
\centering
\includegraphics[width=4.7cm]{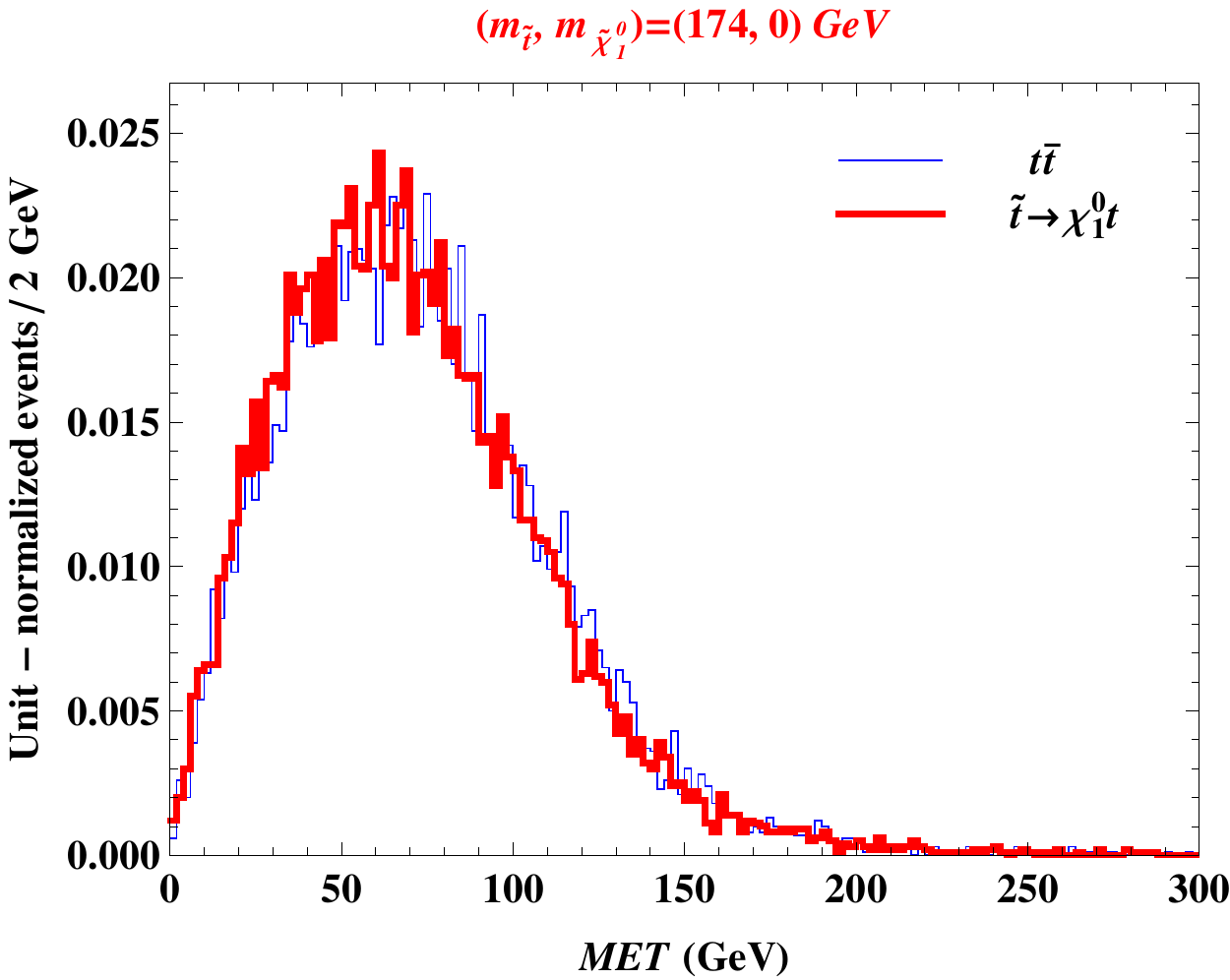}
\includegraphics[width=4.7cm]{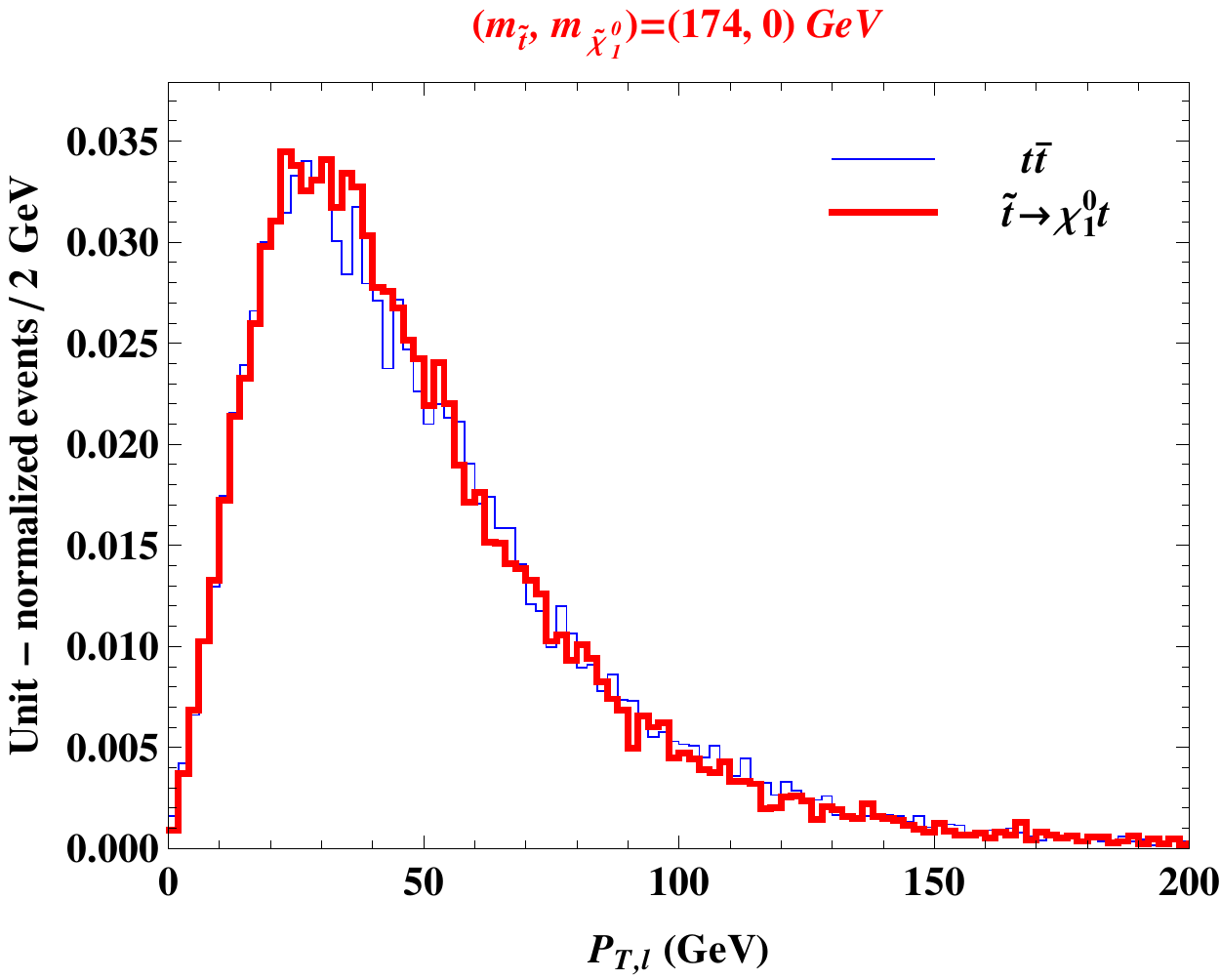}
\includegraphics[width=4.7cm]{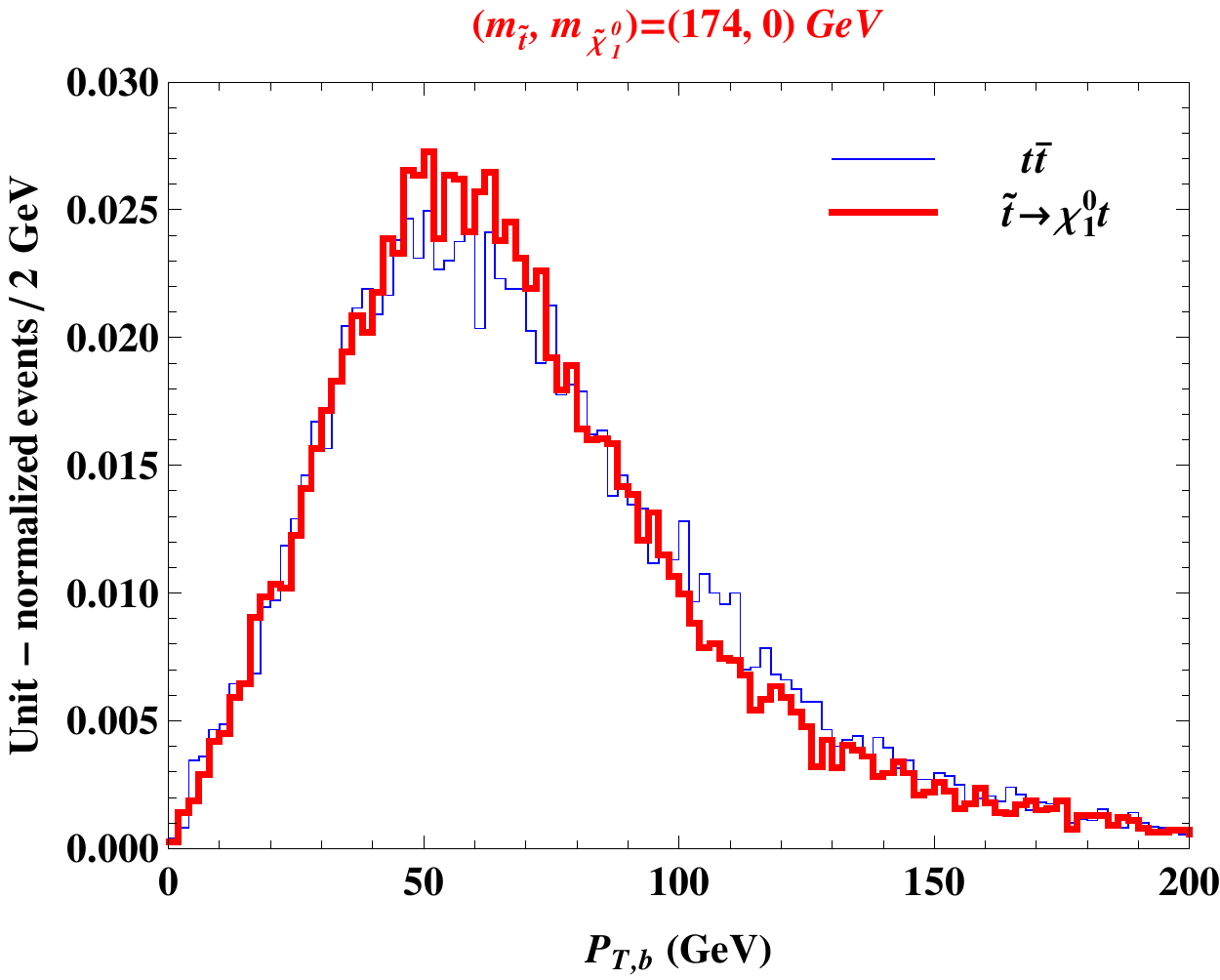}
\caption{\label{fig:cuts6} 
The same as Fig.~\ref{fig:cuts3}, but for signal events with the event topology of Fig.~\ref{fig:DecayTopologies}(c)
for study point 6 ($m_{\tilde t}=174$ GeV and $m_{\tilde \chi^0}=0$ GeV). } 
\end{figure}

Our last example is a very difficult one: study point 6 in Table
\ref{tab:study-points-topo2}.  Here the new physics mass spectrum is
such that the decay $\tilde t \to t \tilde\chi^0$ occurs exactly at
threshold. As a result, the (massless) neutralinos carry away a
negligible amount of momentum, and the signal events look very
top-like. This is illustrated in Figs.~\ref{fig:MT2cuts6} and
\ref{fig:cuts6}, where we compare our standard set of kinematic
distributions for signal and background. 

Fig.~\ref{fig:cuts6} shows that the $p_T$ distributions and the $MET$
distribution are almost identical for signal and background. The
invariant mass distributions from Fig.~\ref{fig:MT2cuts6} are also
very similar; there are slight differences in the shapes due to the
top quarks in the signal being more likely to be off-shell, but  the
kinematic endpoints are the same.  Thus, barring a shape-based
analysis, there are no obvious cuts which could discriminate signal
from background. Therefore, just like study point 4, this would be a
very difficult, and most likely impossible, scenario for discovery
using these methods. As before, we shall leave this as an open
question to be revisited in Sec.~\ref{sec:mixed-events}.

\section{$M_2$ endpoint study for mixed events }
\label{sec:mixed-events}

In this section, we shall consider signal events with the mixed event topology of 
Fig.~\ref{fig:DecayTopologies}(d).  In doing so, we are motivated by two factors:
\begin{itemize}
\item In any realistic model, the stop is likely to have several
  relevant decay modes.  (Here we consider the simplest scenario with
  only the two decay modes from Fig.~\ref{fig:process}.)  Since the
  stops are pair-produced, the number of signal events in each
  symmetric channel is proportional to the corresponding branching
  ratio squared.  For mixed events, where the two stops decay {\em
    differently}, the number of signal events benefits from an
  additional combinatorial factor of 2.
\item In the course of our study of the symmetric event topologies from 
Fig.~\ref{fig:DecayTopologies}(b) (in
Sec.~\ref{sec:M2_endpoint_study_for_topology_1}) and
Fig.~\ref{fig:DecayTopologies}(c) (in
Sec.~\ref{sec:M2_endpoint_study_for_topology_2}), we determined
that there are ``blind spots" in the mass parameter space, where the
signal resembles the background, and discovery is very
challenging. Study points 4 and 6 are examples of such difficult
cases. In this section, therefore, we shall investigate the question
of whether one can recover some sensitivity by considering mixed
events constructed from precisely those two difficult cases. In other
words, we consider events with the event topology of
Fig.~\ref{fig:DecayTopologies}(d), where the upper (lower) decay chain
corresponds to study point 4 (study point 6). (Study point 4 gives the
stop mass ($174$ GeV) and the neutralino mass ($0$ GeV); study point 6
uses the same stop mass, a chargino mass of $150$ GeV and a sneutrino
mass of $110$ GeV.)
We shall assume that the two stop decays occur with equal
branching fractions.
\end{itemize}

The idea to use mixed stop events was previously discussed in
Ref.~\cite{Graesser:2012qy},
which suggested a new variable, ``topness", that quantifies how well an
event can be reconstructed under the top background hypothesis. In
order to calculate the ``topness" of an event, one minimizes the total
$\sqrt{s}$ of the event, making the reasonable ansatz that the
momentum configuration thus obtained provides a good approximation to the
true kinematics of the event~\cite{Graesser:2012qy,Konar:2008ei}. Our
approach is similar to the extent that the
 on-shell constrained invariant mass variables, like $M_{2CC}$, are also
 found by minimization, though not of the total $\sqrt{s}$ but of the parent
 mass in the respective subsystem.
 By imposing the symmetry constraints (\ref{eq:parents-equal}) and
 (\ref{eq:relatives-equal}), we focus on the one key difference
 between the signal and background events: 
 the signal event topology is asymmetric while the background event
 topology is symmetric.
 We can therefore expect that the constraints (\ref{eq:parents-equal})
 and (\ref{eq:relatives-equal})
 will affect signal and background events {\em differently}.

\begin{figure}[t]
\centering
\includegraphics[width=7.0cm]{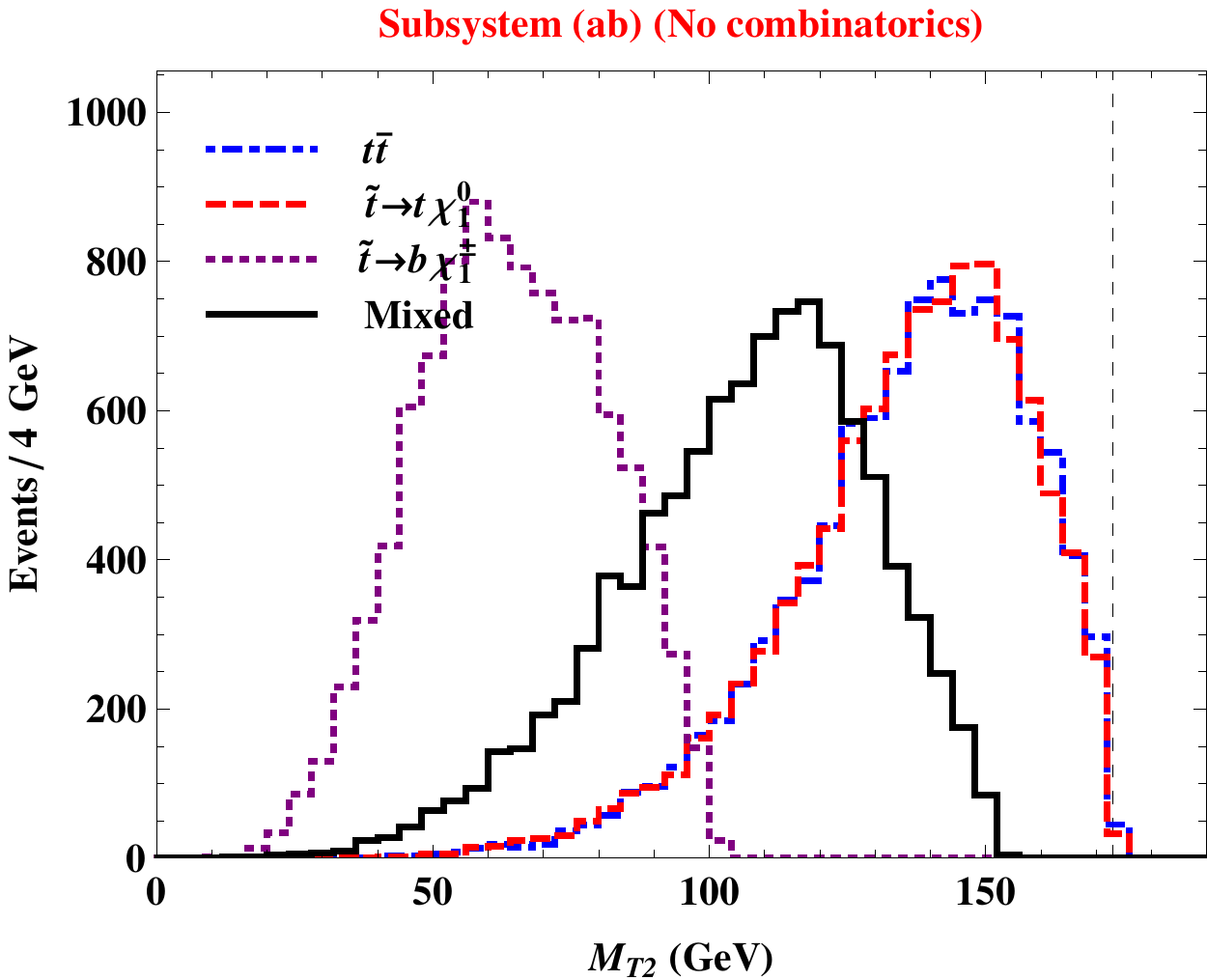}
\includegraphics[width=7.0cm]{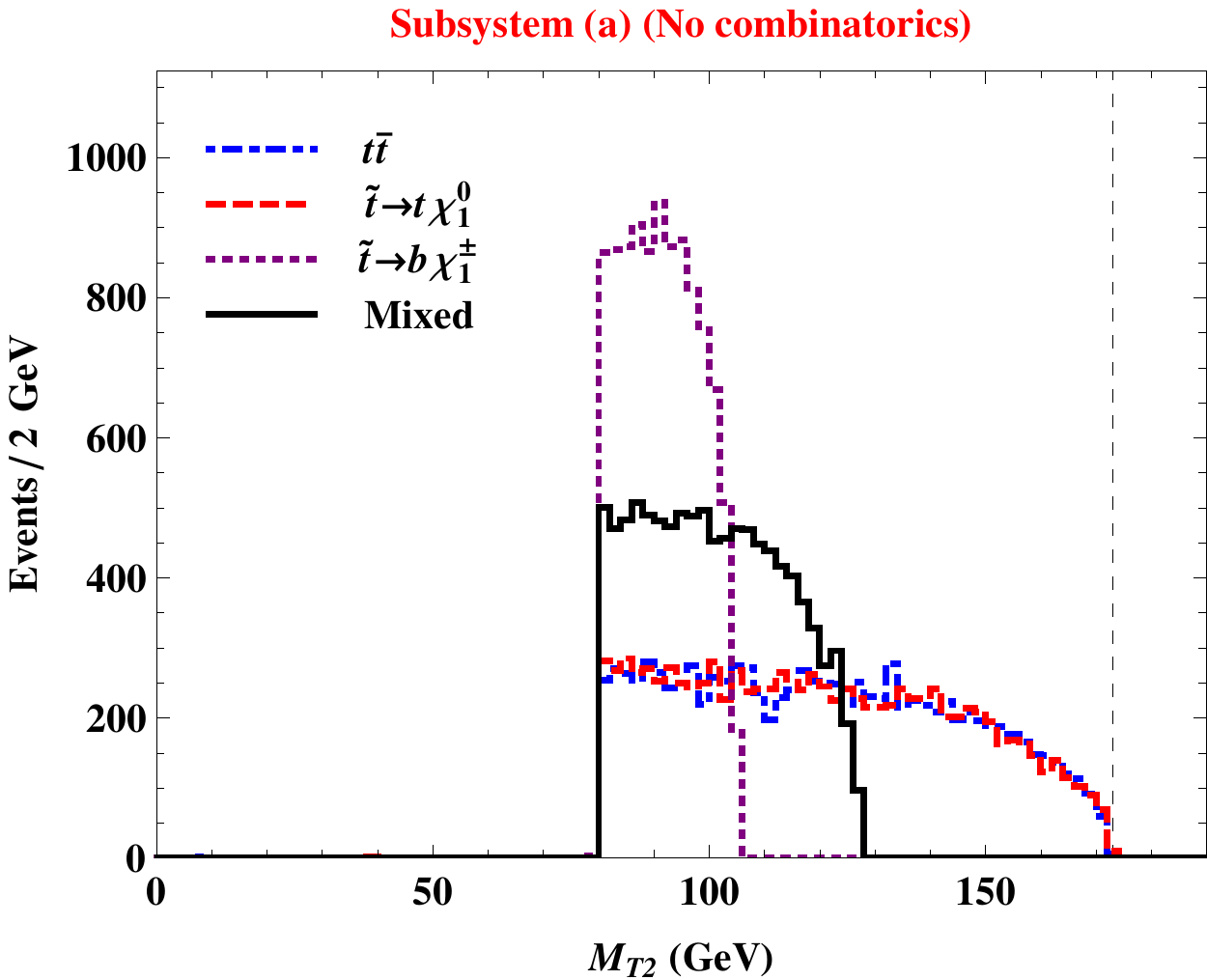}\\
\includegraphics[width=7.0cm]{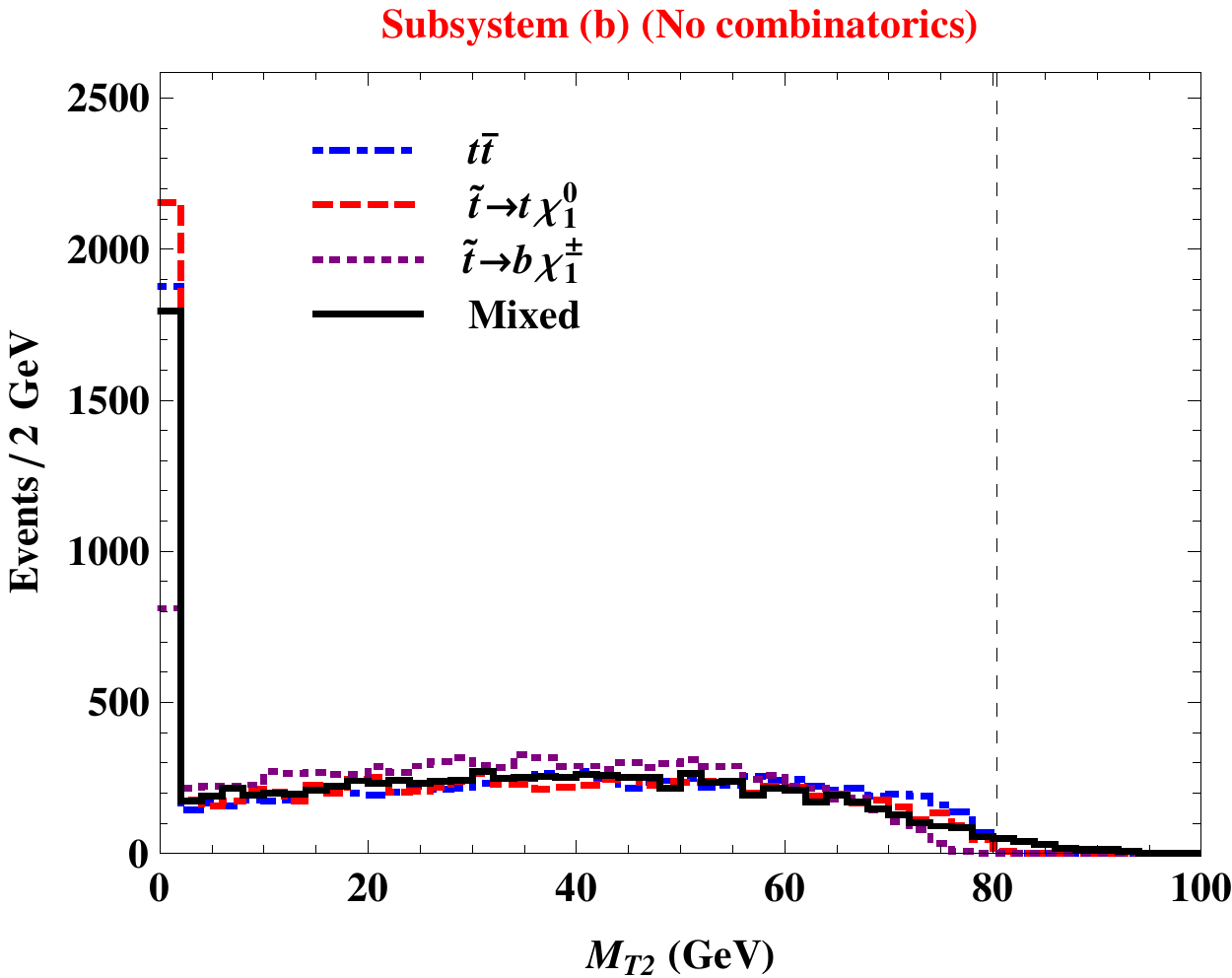}
\includegraphics[width=7.0cm]{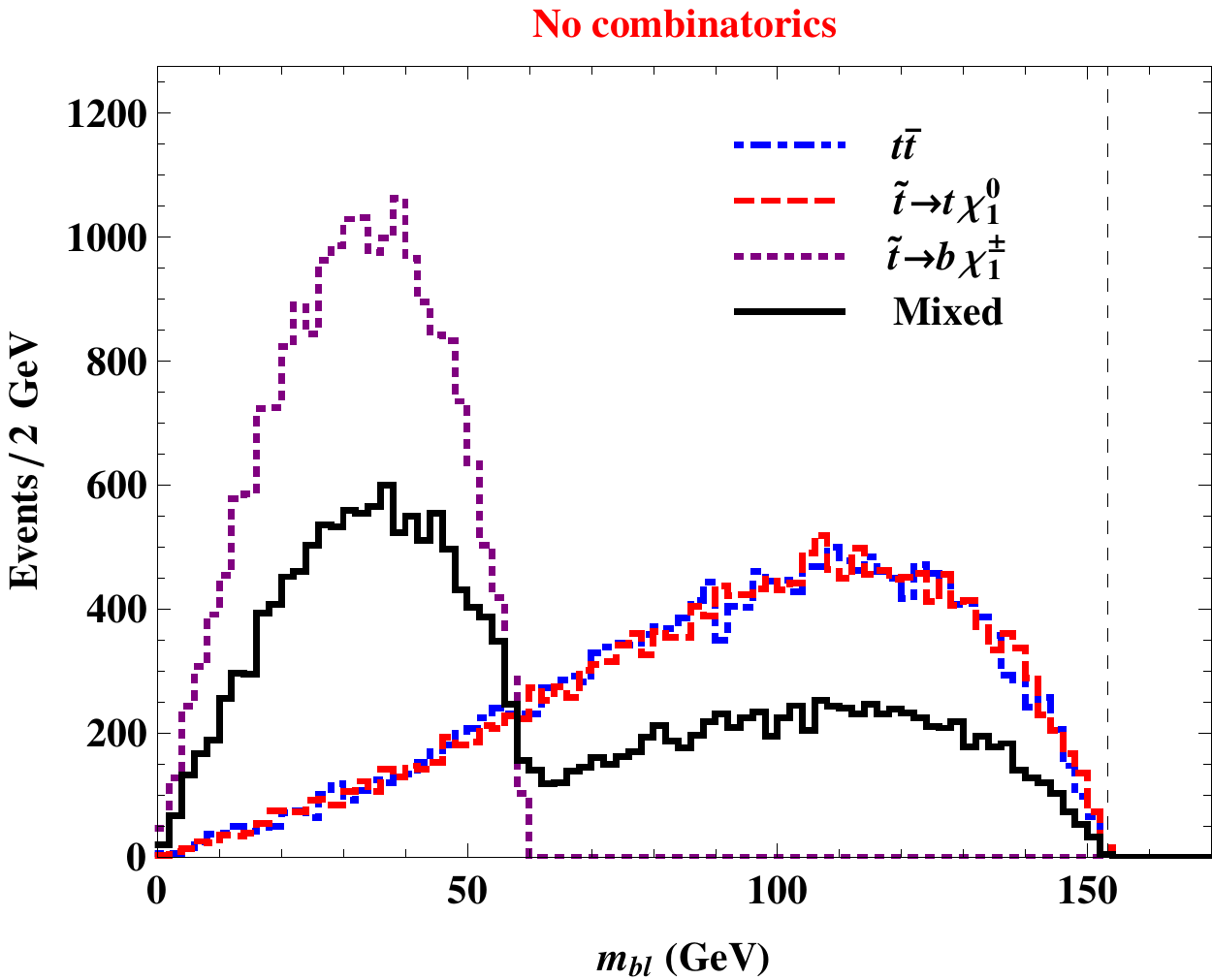}
\caption{\label{fig:MT2cuts5again} 
The same as Figs.~\ref{fig:MT2cuts4} and \ref{fig:MT2cuts6}, 
comparing distributions of the three $M_{T2}$ subsystem variables
and the invariant mass $m_{b\ell}$ for the four types of events 
in Fig.~\ref{fig:DecayTopologies}. The mass spectrum corresponds to study points 4 and 6: 
$m_{\tilde t}=174$ GeV,  $m_{\tilde \chi^\pm}=150$ GeV, $m_{\tilde \nu}=110$ GeV, and 
$m_{\tilde \chi^0}=0$ GeV. } 
\end{figure}

\begin{figure}[t]
\centering
\includegraphics[width=7.0cm]{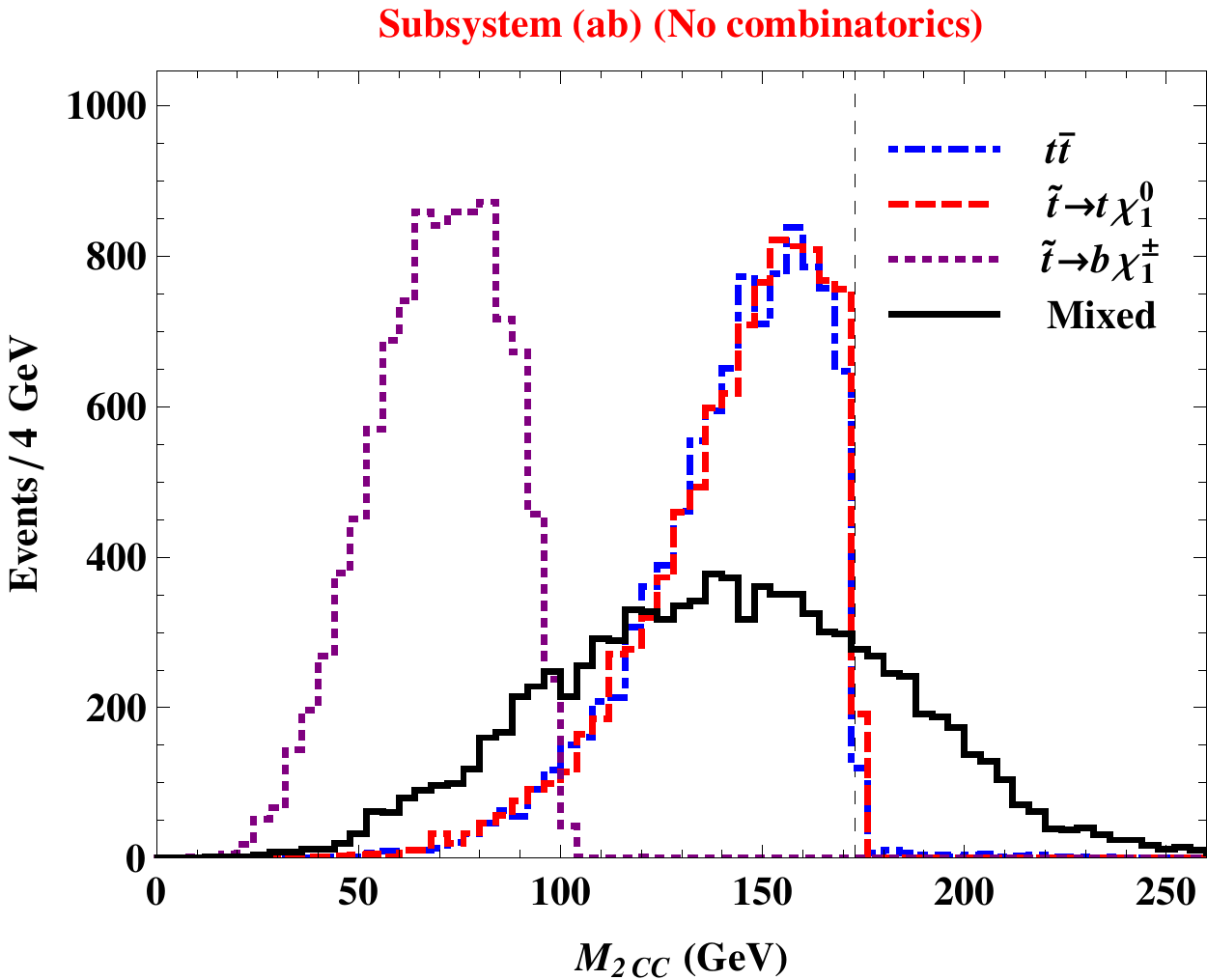}
\includegraphics[width=7.0cm]{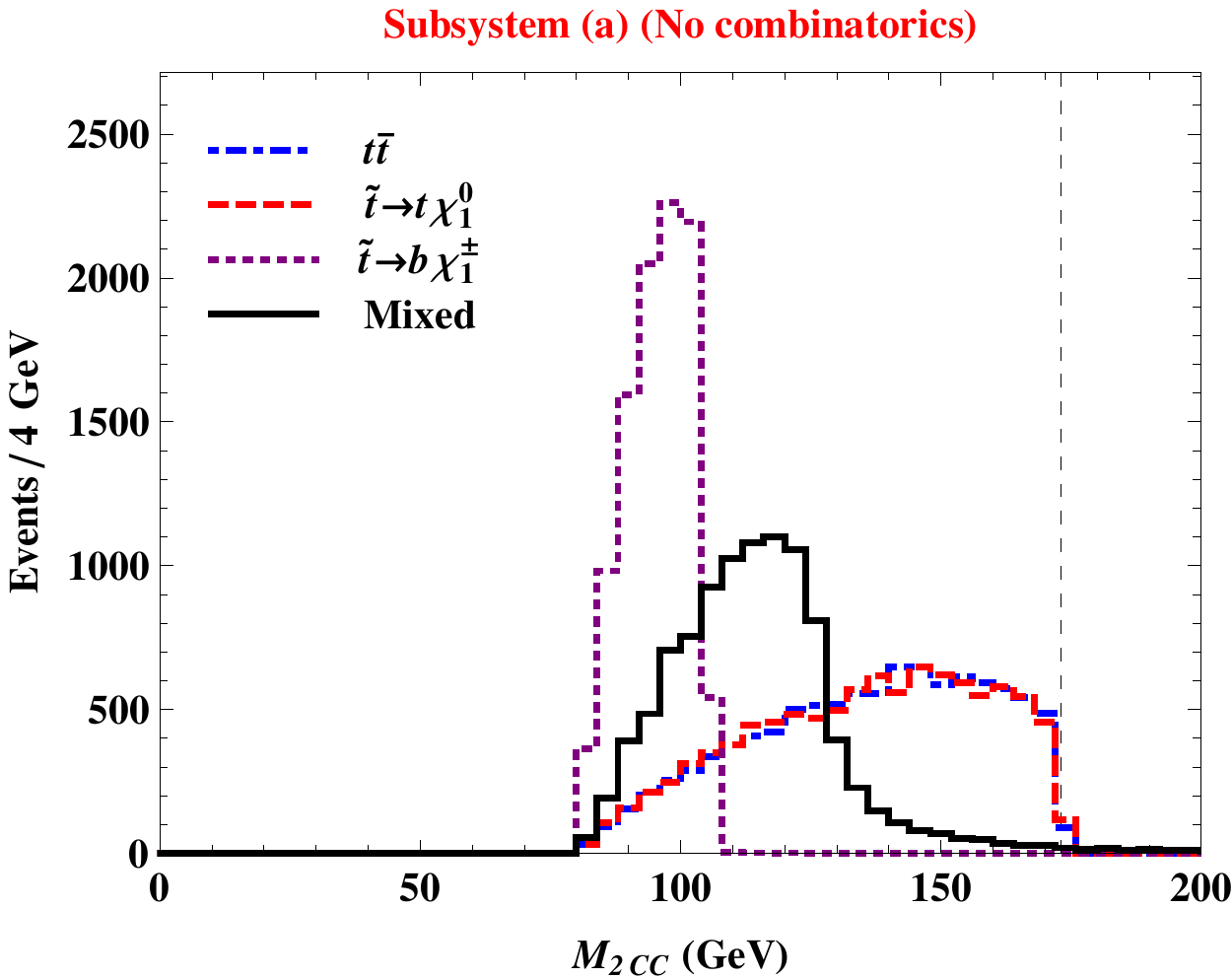}\\
\includegraphics[width=7.0cm]{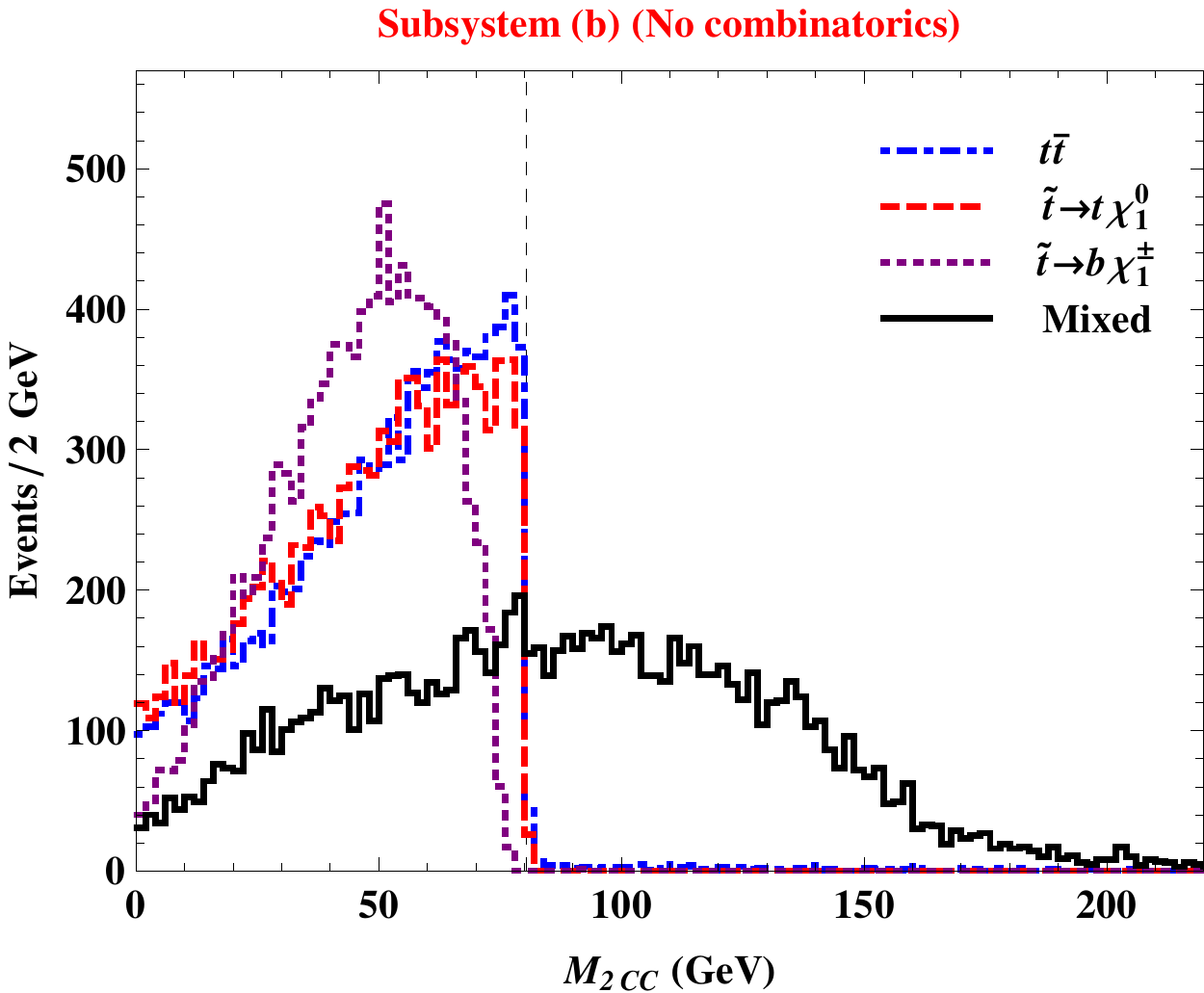}
\caption{\label{fig:M2CCcuts5} 
The same as Fig.~\ref{fig:MT2cuts5again}, but for the on-shell constrained variable $M_{2CC}$ 
in the three possible subsystems $(ab)$, $(a)$, and $(b)$. } 
\end{figure}

As a point of reference, we begin by showing distributions of
variables for which no improvement can be expected in the case of
mixed events in Fig.~\ref{fig:MT2cuts5again}. The four variables
depicted in the figure are the three subsystem $M_{T2}$ variables and
the invariant mass $m_{b\ell}$.  All four variables are calculated
here using the assignment of the leptons and b-quarks to the correct
decay chains.
Each panel contains four distributions, one for each event type from
Fig.~\ref{fig:DecayTopologies}: the $t\bar{t}$ background from
Fig.~\ref{fig:DecayTopologies}(a) (blue dot-dashed lines); the
symmetric $\tilde t\to b\tilde\chi^+_1$ signal events from
Fig.~\ref{fig:DecayTopologies}(b) (magenta dotted lines); the
symmetric $\tilde t\to t\tilde\chi^0_1$ signal events from
Fig.~\ref{fig:DecayTopologies}(c) 
(red dashed lines); and the asymmetric signal events from
Fig.~\ref{fig:DecayTopologies}(d) (black solid lines).
The vertical black dashed line in each panel marks the location of the 
upper kinematic endpoint of the $t\bar{t}$ background distribution.
We see that for all three types of signal events (symmetric or asymmetric), 
the respective distributions do not violate the background kinematic
endpoints, thus discovery appears to be just as difficult with mixed
events as it was with the symmetric events considered earlier in
Secs.~\ref{subsec:point-4} and \ref{subsec:point-6}. 
This conclusion is easy to understand; the $M_{T2}$ variable is a
variable defined on the transverse plane, where it is impossible to
impose a 3+1-dimensional mass constraint like
Eq.~(\ref{eq:parents-equal}) or (\ref{eq:relatives-equal}).

\begin{figure}[t]
\centering
\includegraphics[width=7.0cm]{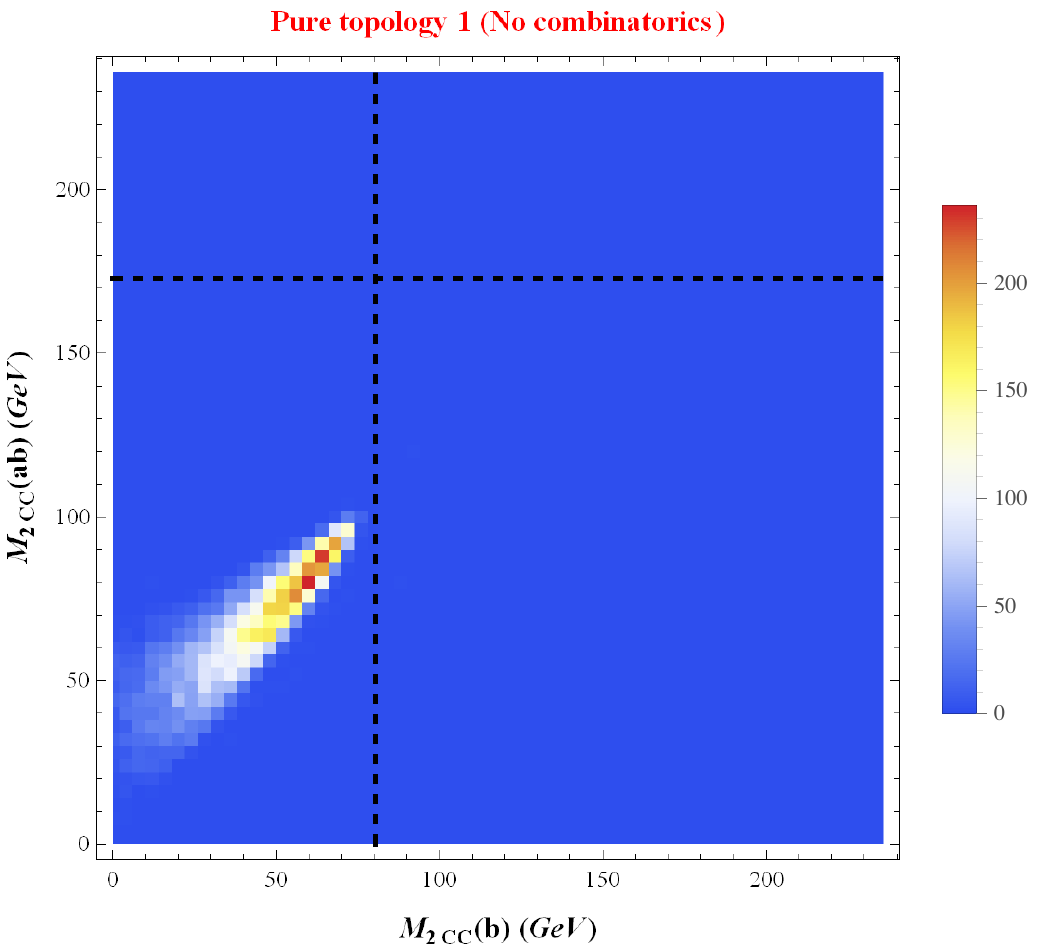}
\includegraphics[width=7.0cm]{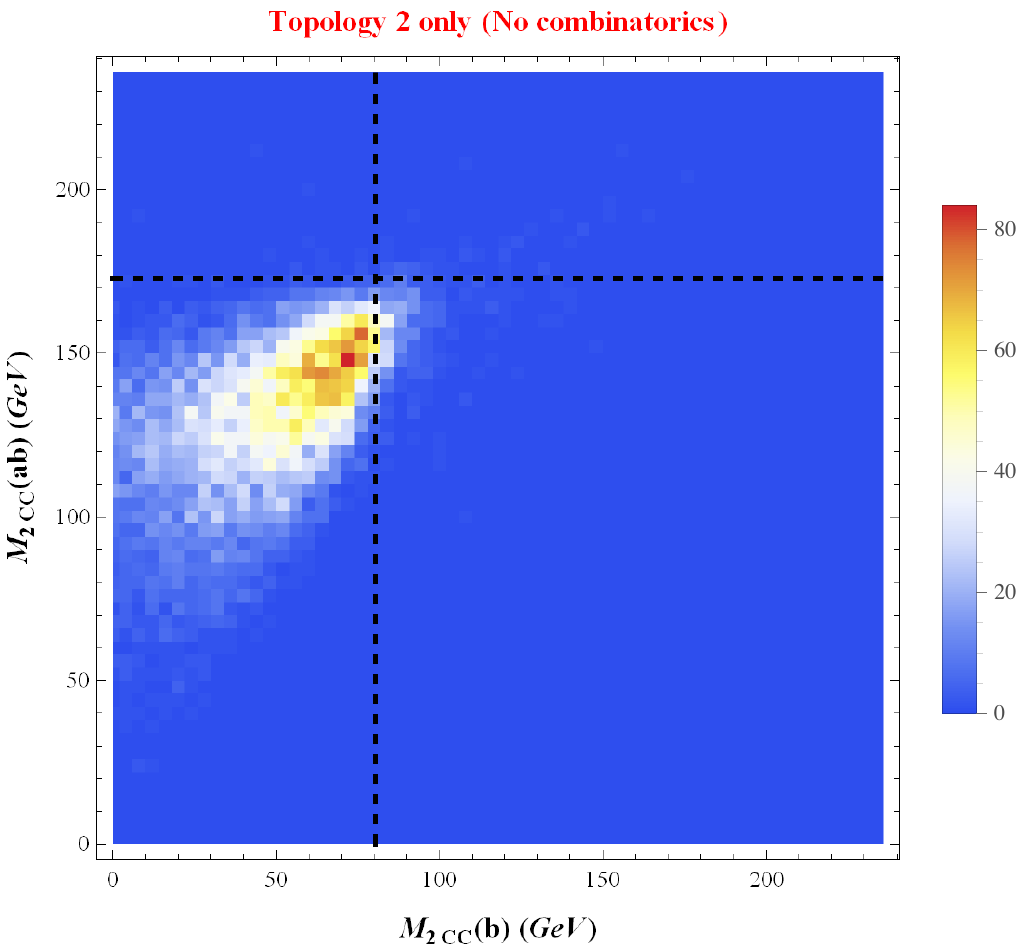}\\
\includegraphics[width=7.0cm]{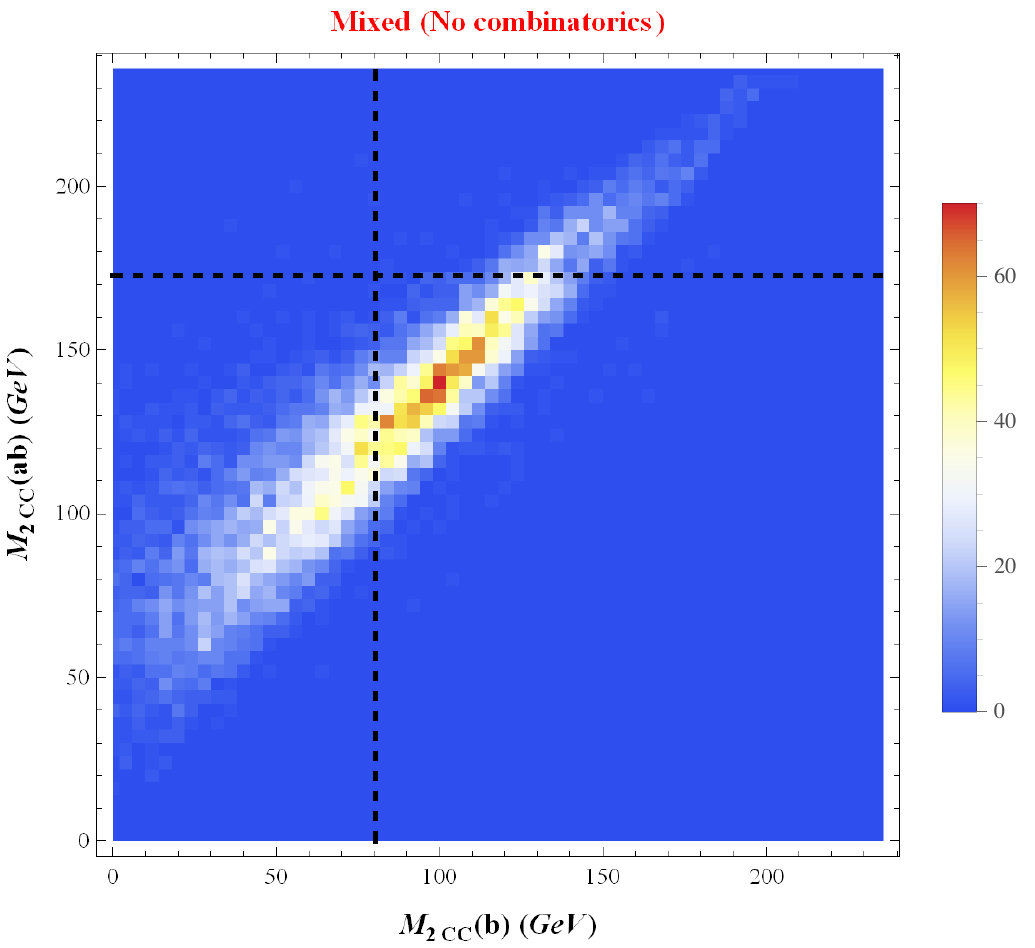}
\includegraphics[width=7.0cm]{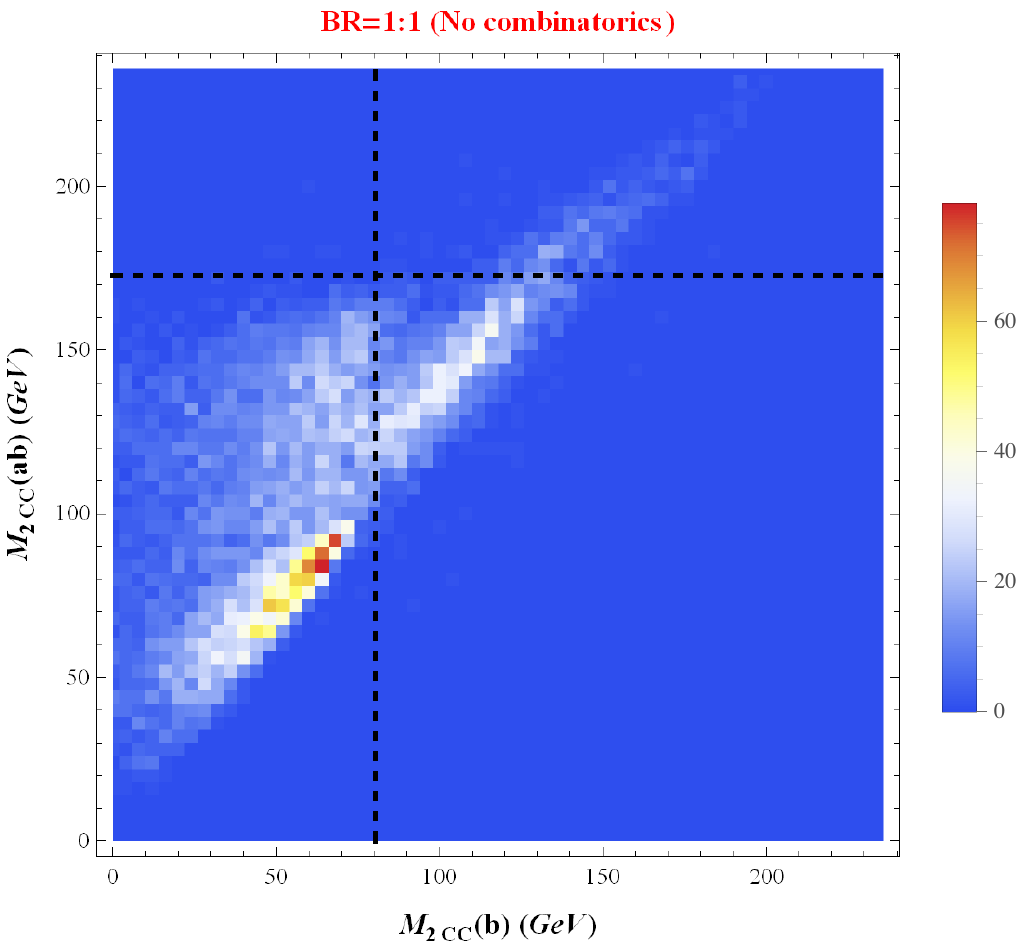}
\caption{\label{fig:nocomb} 
The correlation between the two best performing variables from
Fig.~\ref{fig:M2CCcuts5},
for the three types of signal events:
pure Topology 1 from Fig.~\ref{fig:DecayTopologies}(b) (upper left),
pure Topology 2 from Fig.~\ref{fig:DecayTopologies}(c) (upper right),
mixed topology from Fig.~\ref{fig:DecayTopologies}(d) (lower left),
and combined total (lower right). } 
\end{figure}

The situation is quite different when we consider distributions of
on-shell constrained variables, $M_{2CC}$, for which the constraints 
of Eqs.~(\ref{eq:parents-equal}) or (\ref{eq:relatives-equal}) are imposed. 
As seen in Fig.~\ref{fig:M2CCcuts5},
the signal distributions for mixed events may now exhibit endpoint violation, 
even when the signal distributions for symmetric events do not.
The effect is most pronounced in the case of the subsystem variable $M_{2CC}(b)$;
for the subsystem variable $M_{2CC}(ab)$ it is less noticeable, while for
$M_{2CC}(a)$ it is absent altogether. Fig.~\ref{fig:M2CCcuts5} showcases the main result 
of this section: that with the help of an appropriately chosen 
on-shell constrained variable (in this case $M_{2CC}(b)$), one can obtain 
a relatively good separation of signal from background {\em for mixed events}. 
It is worth emphasizing that this separation was achieved for a very unfavorable 
choice of mass parameters, as the study points 4 and 6 were not
observable using events where both decay chains had the same topology. 

\begin{figure}[t]
\centering
\includegraphics[width=7.0cm]{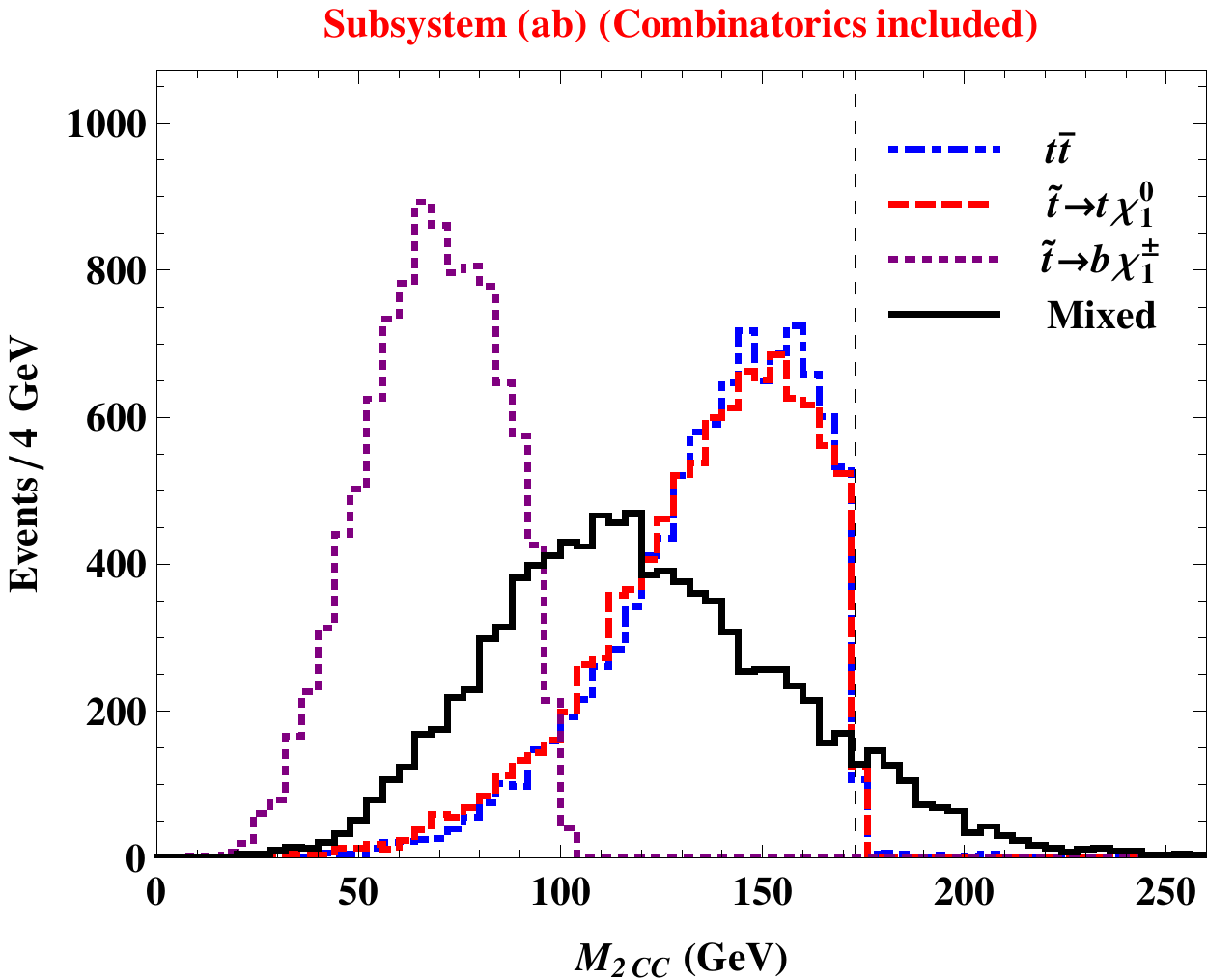}
\includegraphics[width=7.0cm]{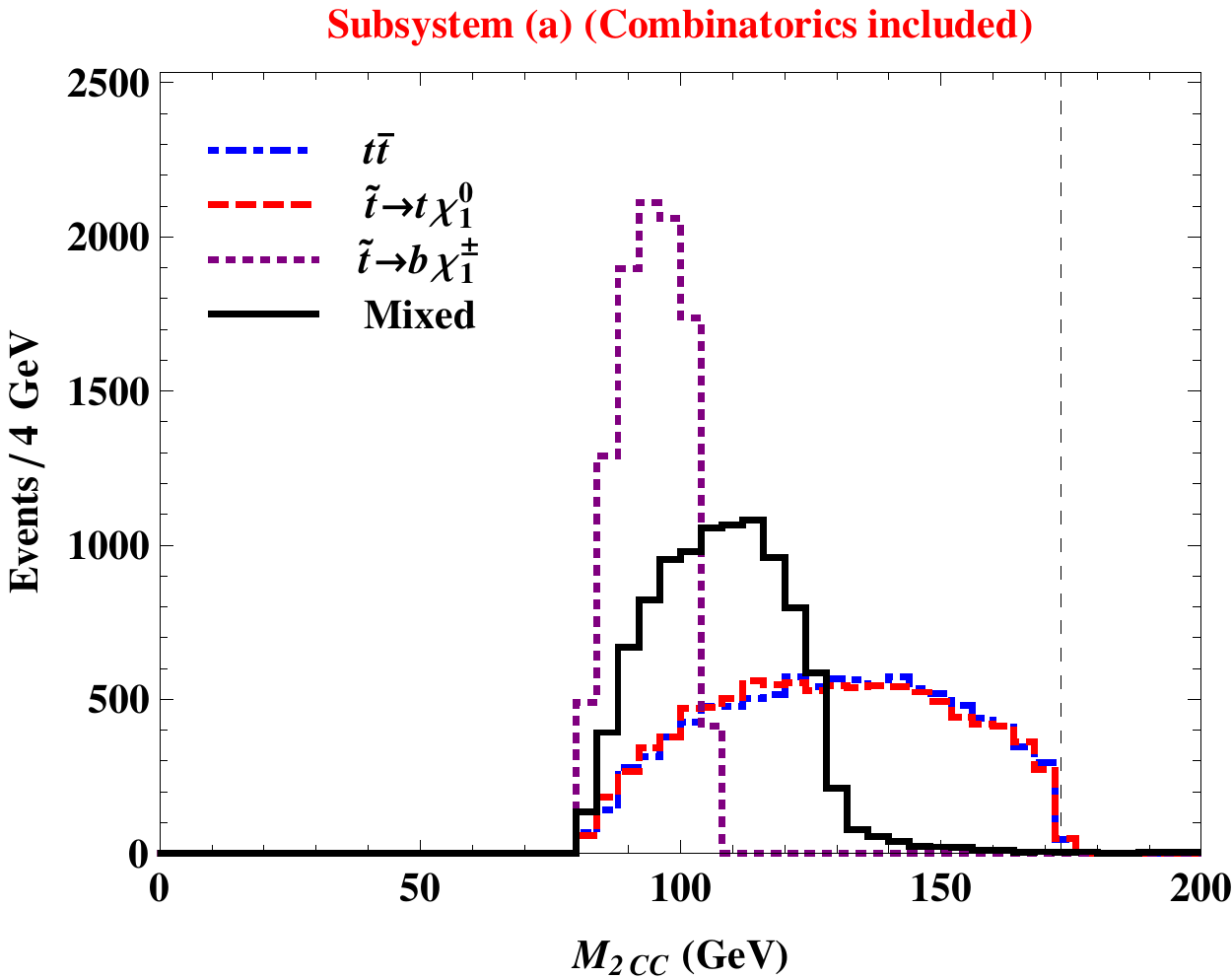}
\includegraphics[width=7.0cm]{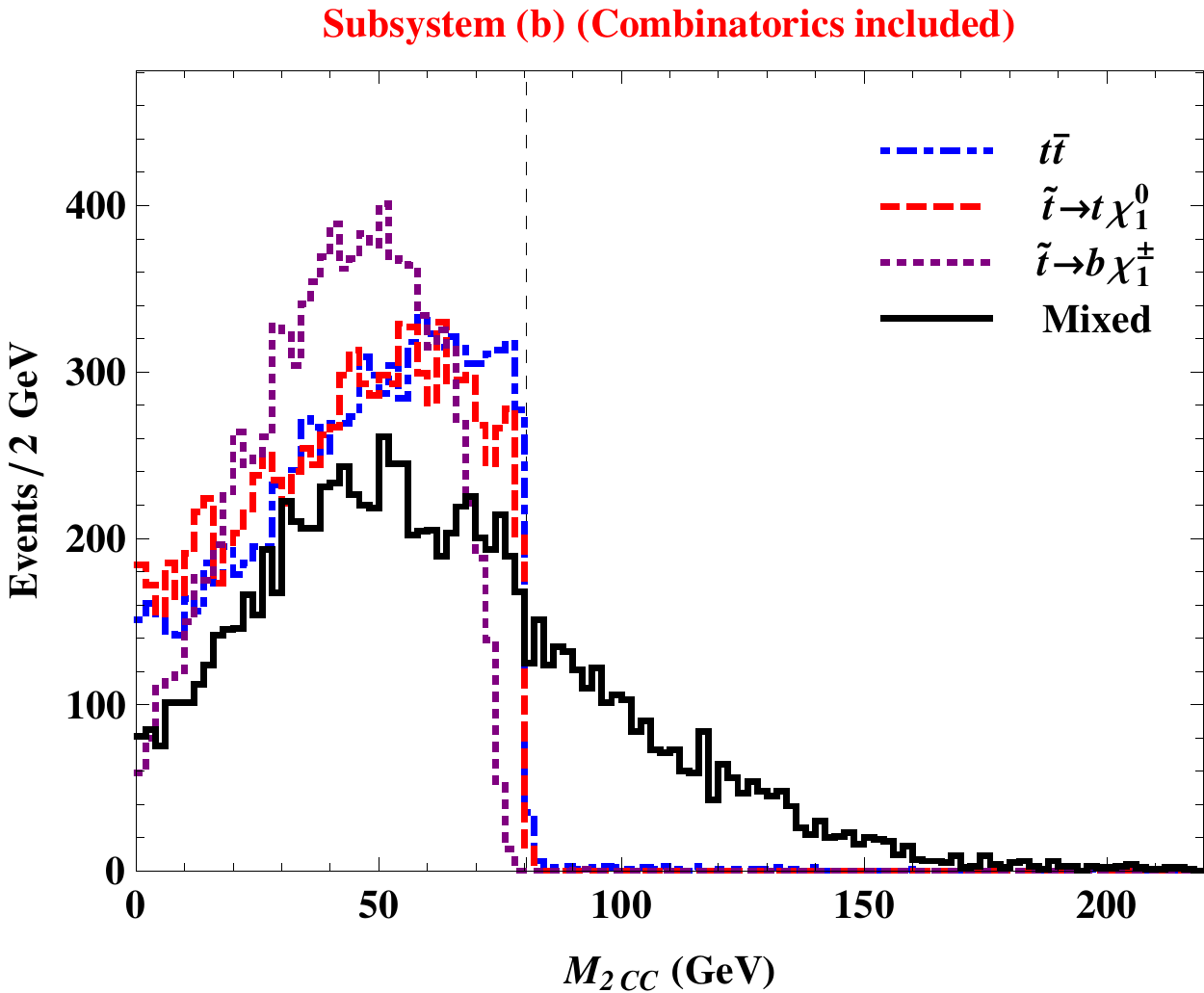}
\caption{\label{fig:M2CCcuts5comb} 
The same as Fig.~\ref{fig:M2CCcuts5}, but including the effects of
combinatorics: for each event, we try both possible assignments of the
lepton-$b$pairs, and plot the smaller of the two resulting $M_{2CC}$
values. } 
\end{figure}

Given that endpoint violation was observed for both $M_{2CC}(b)$
and $M_{2CC}(ab)$, it is worth investigating the possible correlation
between those two variables.  In Fig.~\ref{fig:nocomb} we show
two-dimensional plots exhibiting those correlations.  We consider
separately the three types of signal events: pure Topology 1 from
Fig.~\ref{fig:DecayTopologies}(b) (upper left), pure Topology 2 from
Fig.~\ref{fig:DecayTopologies}(c) (upper right), and the mixed
topology from Fig.~\ref{fig:DecayTopologies}(d) (lower left).
Finally, the lower right panel in Fig.~\ref{fig:nocomb} shows the
result for the full signal sample, with equal branching fractions for
Topology 1 and Topology 2.   The black dashed lines in
Fig.~\ref{fig:nocomb} mark the locations of the expected upper
kinematic endpoints for background events, following Eqs.~(\ref{mwb})
and
(\ref{mtab}).  Any events which appear to the right of the vertical
black dashed lines and/or above the horizontal black dashed lines in
Fig.~\ref{fig:nocomb} are expected to be signal-like. In agreement
with Fig.~\ref{fig:M2CCcuts5}, we see that for symmetric signal event
topologies (the upper two panels in Fig.~\ref{fig:nocomb}), the signal
events are contained within the ``background-like'' rectangular region
adjacent to the origin.  On the other hand, for the asymmetric event
topology of Fig.~\ref{fig:DecayTopologies}(d) (the lower left panel),
many signal events leak out of the background-like box.  The figure
also reveals a linear correlation between $M_{2CC}(b)$ and
$M_{2CC}(ab)$.  Furthermore, the slope is such that if an event
violates the background $M_{2CC}(ab)$ endpoint (\ref{mtab}), it also
necessarily violates the background $M_{2CC}(b)$ endpoint (\ref{mwb}),
while the reverse is not true. We therefore conclude that the
$M_{2CC}(b)$ distribution alone is sufficient in separating signal
from background in this scenario with mixed events.

\begin{figure}[t]
\centering
\includegraphics[width=7.0cm]{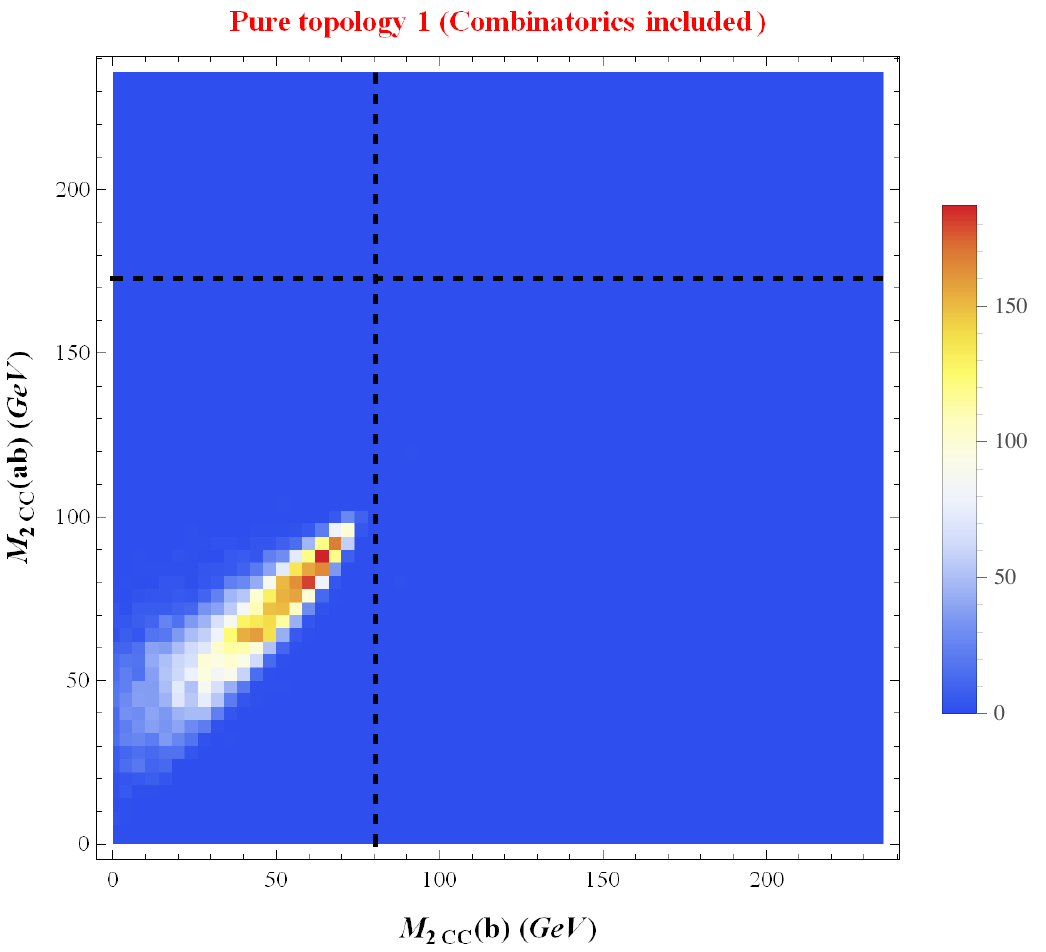}
\includegraphics[width=7.0cm]{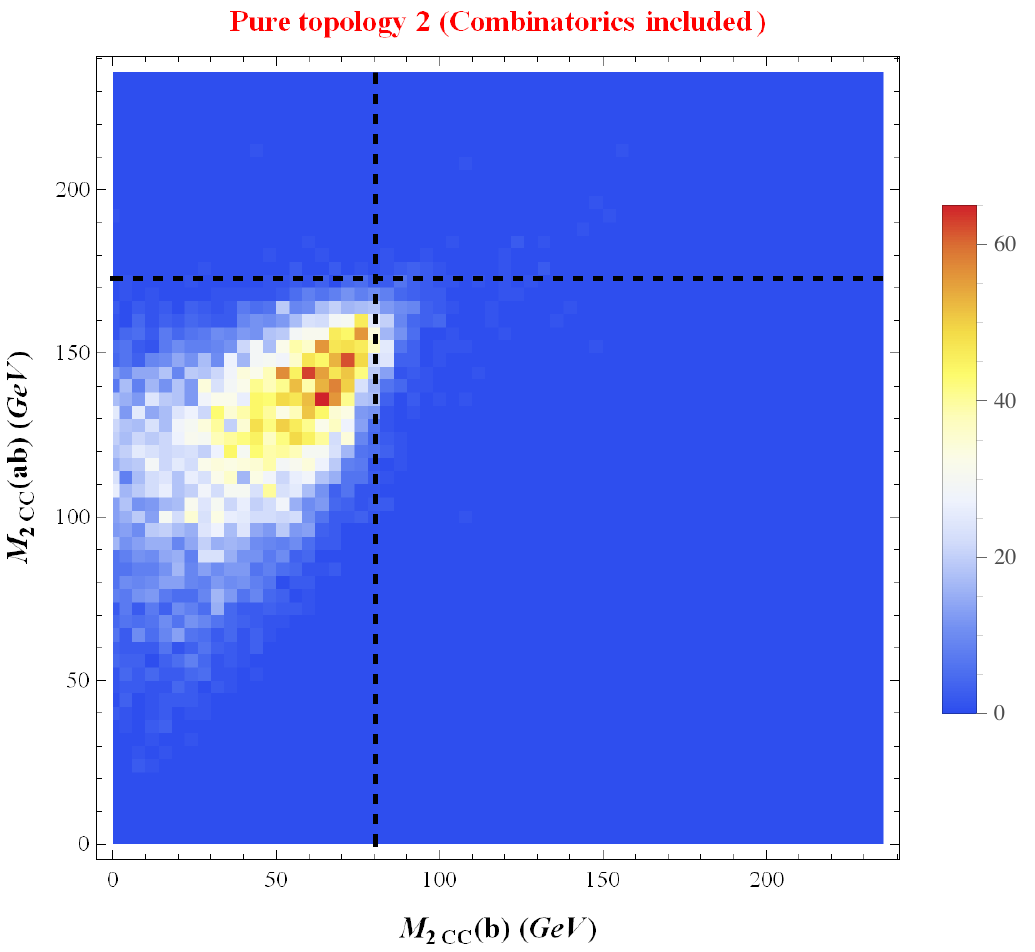}\\
\includegraphics[width=7.0cm]{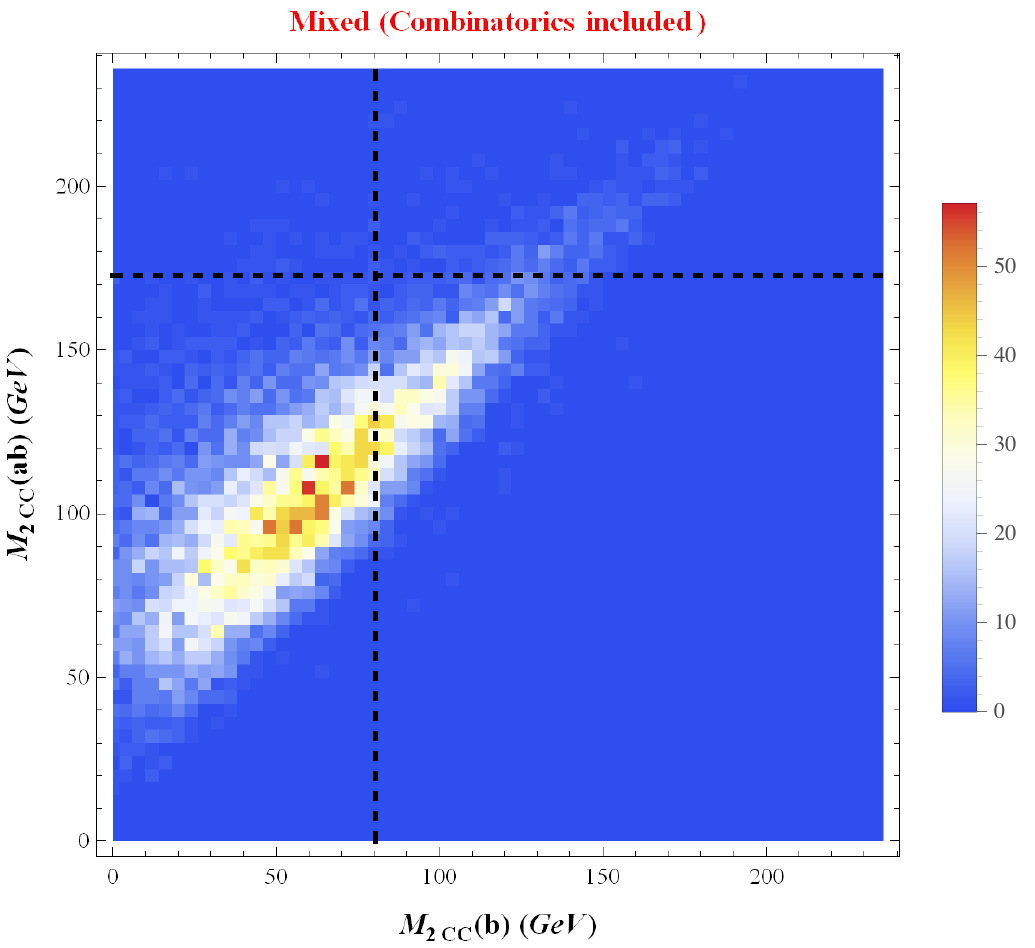}
\includegraphics[width=7.0cm]{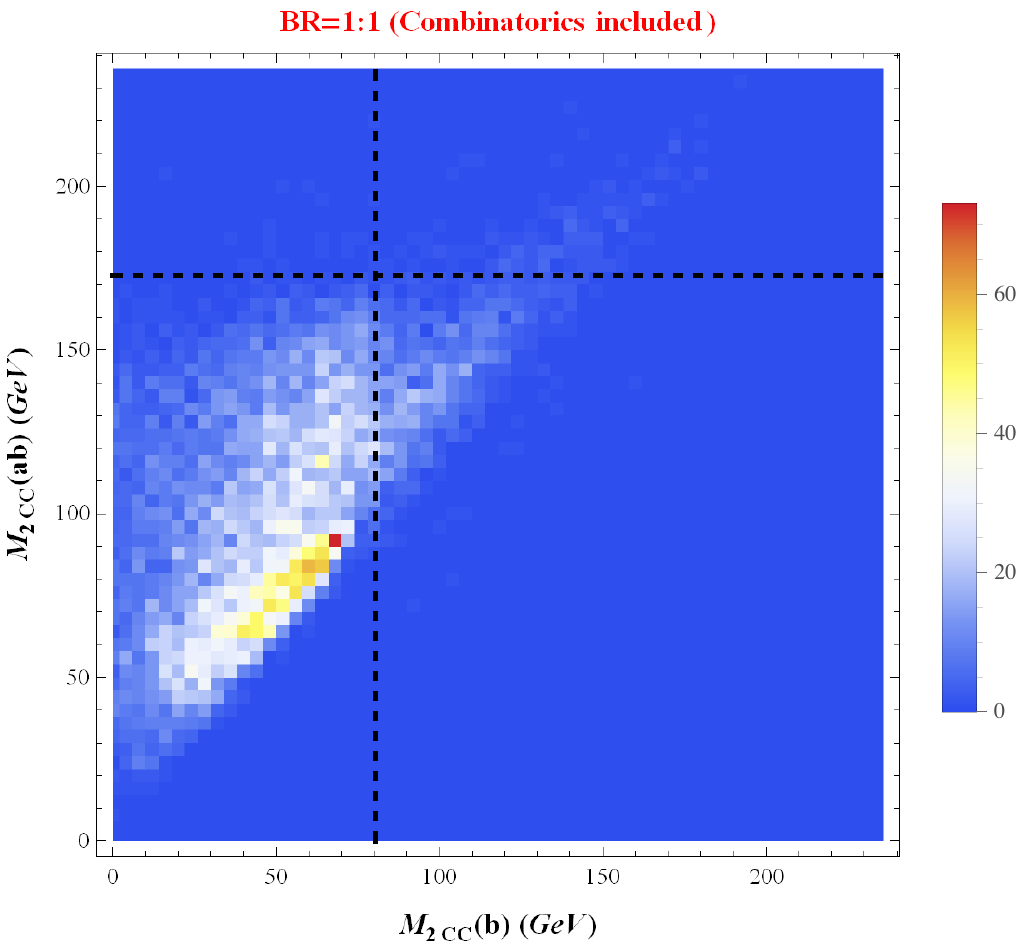}
\caption{\label{fig:comb} 
The same as Fig.~\ref{fig:nocomb}, but including the effects of
combinatorics as in Fig.~\ref{fig:M2CCcuts5comb}. } 
\end{figure}

In our discussion so far in this section, we have been ignoring the
combinatorial problem arising when we try to pair up the two $b$-jets
with the two leptons. Since the $b$-quark charge  is not measured, we
have two possible pairings, each resulting in a candidate value for
the kinematic variable. Since we are interested in upper kinematic
endpoints, the simplest solution is to consider both pairings and then
pick the one which gives the {\em smaller} value for the kinematic
variable. This approach has been followed  in recreating
Figs.~\ref{fig:M2CCcuts5} and \ref{fig:nocomb} as
Figs.~\ref{fig:M2CCcuts5comb} and \ref{fig:comb}, respectively.  As
expected, this procedure tends to shift all distributions towards
lower values, thus the number of signal events which violate the
background endpoints is fewer than before; compare, e.g., the
$M_{2CC}(b)$ distributions for mixed events in
Figs.~\ref{fig:M2CCcuts5} and \ref{fig:M2CCcuts5comb}.  Nevertheless,
the effect is still present, offering hope that difficult cases like
study points 4 and 6 could perhaps best be looked for in such mixed
event topologies instead.

\section{Results with realistic detector simulation}
\label{sec:detector_resolution_etc}

In the previous three sections we saw that the $M_{T2}$ and $M_{2CC}$ variables
allow us to identify signal events as tails which extend beyond the
upper kinematic endpoint for background events. However, in a
realistic  experiment, the background distributions themselves may
acquire high tails, for a variety of reasons.  This is why it is
necessary to test our previous observations, which were made at
parton level, with realistic simulation, including the effects of
detector resolution, initial and final state radiation, jet
reconstruction, cuts, etc.  It is clear that our positive conclusions
drawn for fortuitous cases of new physics like study point 1 will
survive all these complications, therefore in this section we shall
only focus on the difficult scenario discussed in 
Sec.~\ref{sec:mixed-events}, i.e., the mixed events which were a
hybrid between the difficult study points 4 from
Sec.~\ref{subsec:point-4} and 6 from Sec.~\ref{subsec:point-6}.

\subsection{Event simulation details}
\label{sec:simulation}

As before, the parton-level event generation is done 
by \texttt{MadGraph\_aMC@NLO}~\cite{Alwall:2014hca}, where by default
the parton distributions are evaluated by \texttt{NNPDF23}~\cite{Ball:2012cx}.
The relevant output is then piped through \texttt{Pythia 6.4}~\cite{Sjostrand:2006za} 
and \texttt{Delphes3}~\cite{deFavereau:2013fsa}. For both signal and 
$t\bar{t}$ background, the decays of top quarks are handled by \texttt{Pythia 6.4}, 
while Topology 1 is explicitly generated by \texttt{MadGraph\_aMC@NLO} 
without any prior cuts. All simulations are performed at leading order
for a $pp$ collider of $\sqrt{s}=14$ TeV. 

For the signal process, we assume that the branching ratio of
Topology 1 relative to Topology 2 is $1:1$. 
In addition, in Topology 1, the chargino is forced to decay
exclusively into a sneutrino (which may further decay invisibly), and
a lepton (i.e., electron and muon only). In Topology 2, the stop
decays to the lightest neutralino and a top quark, which subsequently
decays with the relevant branching ratios predicted in the SM. For our
purposes, we only consider the dilepton final state, in which
both top quarks decay leptonically.  The input top quark mass is set
to $173$ GeV, while the $W^\pm$ gauge boson mass is $80$ GeV. 
Jets are reconstructed with the anti-$k_t$ algorithm~\cite{Cacciari:2008gp}, 
using a radius parameter $R=0.5$. The $b$-tagging efficiency is taken
to be $70$\%, while light quark jets are mis-tagged at the rate of $1$\%. 

Given the final state $2b+2\ell+\met$, in principle there are several
sources of SM background that need to be taken into account. In order
to suppress the reducible SM backgrounds, we apply the following
pre-selection cuts:
\begin{itemize}
\item The event must contain exactly two opposite sign leptons with 
$p_T>10$ GeV and $|\eta|<2.5$ ($|\eta|<2.4$) for electrons (muons).
\item In order to reduce background from low mass resonances, 
in the $ee$ and $\mu\mu$ channels, we demand $m_{ee/\mu\mu}>20$
GeV. Furthermore, to reduce the $Z+$jets
background, events with dilepton masses 
within the $Z$-mass window are vetoed by requiring
$|m_{ee/\mu\mu}-m_Z| > 15$ GeV.
\item To further suppress Drell-Yan, for the $ee$ and $\mu\mu$
  channels, we apply a missing transverse energy cut of $\met>40$ GeV.
 \item The event is required to have $\geq 2$ jets with $p_T > 30$ GeV and 
$|\eta|<2.4$. It is also required that exactly two of these jets are $b$-tagged. 
\end{itemize}
After these cuts, we are left with $t\bar{t}$ as the dominant (irreducible) 
background, and this will be the only background process we will consider here.
The posterior cuts using $M_2$ variables are imposed for
events already passing the above set of pre-selection cuts. 

\subsection{Results for $M_{T2}$ and $M_{2CC}$}
\label{sec:MT2}

We first revisit our results from Sec.~\ref{sec:mixed-events}, this
time including the effects of detector simulation and
combinatorics. As before, we consider both pairings of the two tagged
$b$-jets and the two leptons, and use the {\em smaller} of the two
resulting values for the kinematic variable under consideration.
However, unlike the plots in Sec.~\ref{sec:mixed-events}, here we
do not separate the three types of signal events (Topology 1,
Topology 2 and mixed), since in the real experiment there is no way to
tell which is which.

\begin{figure}[t]
\centering
\includegraphics[width=7.0cm]{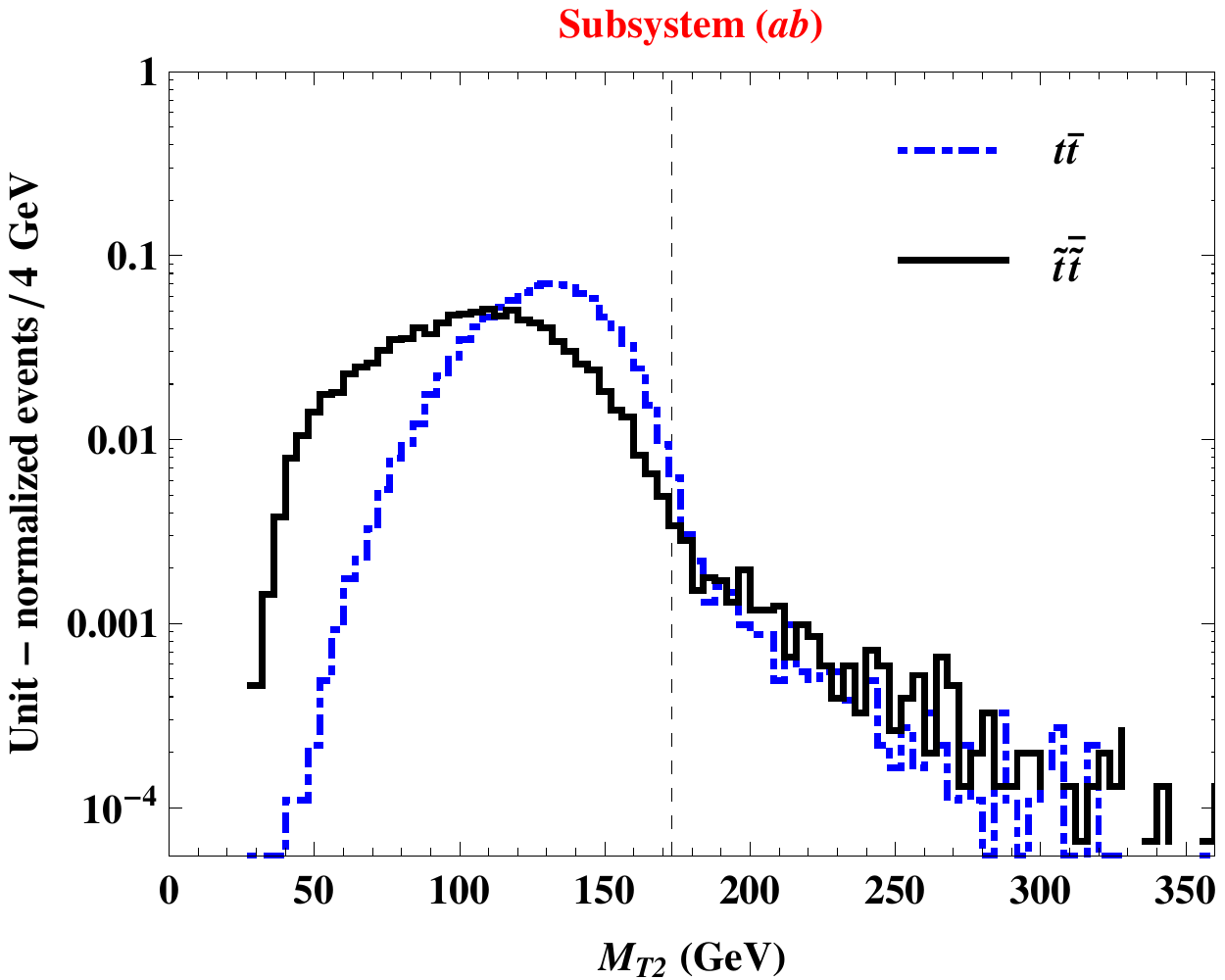}
\includegraphics[width=7.0cm]{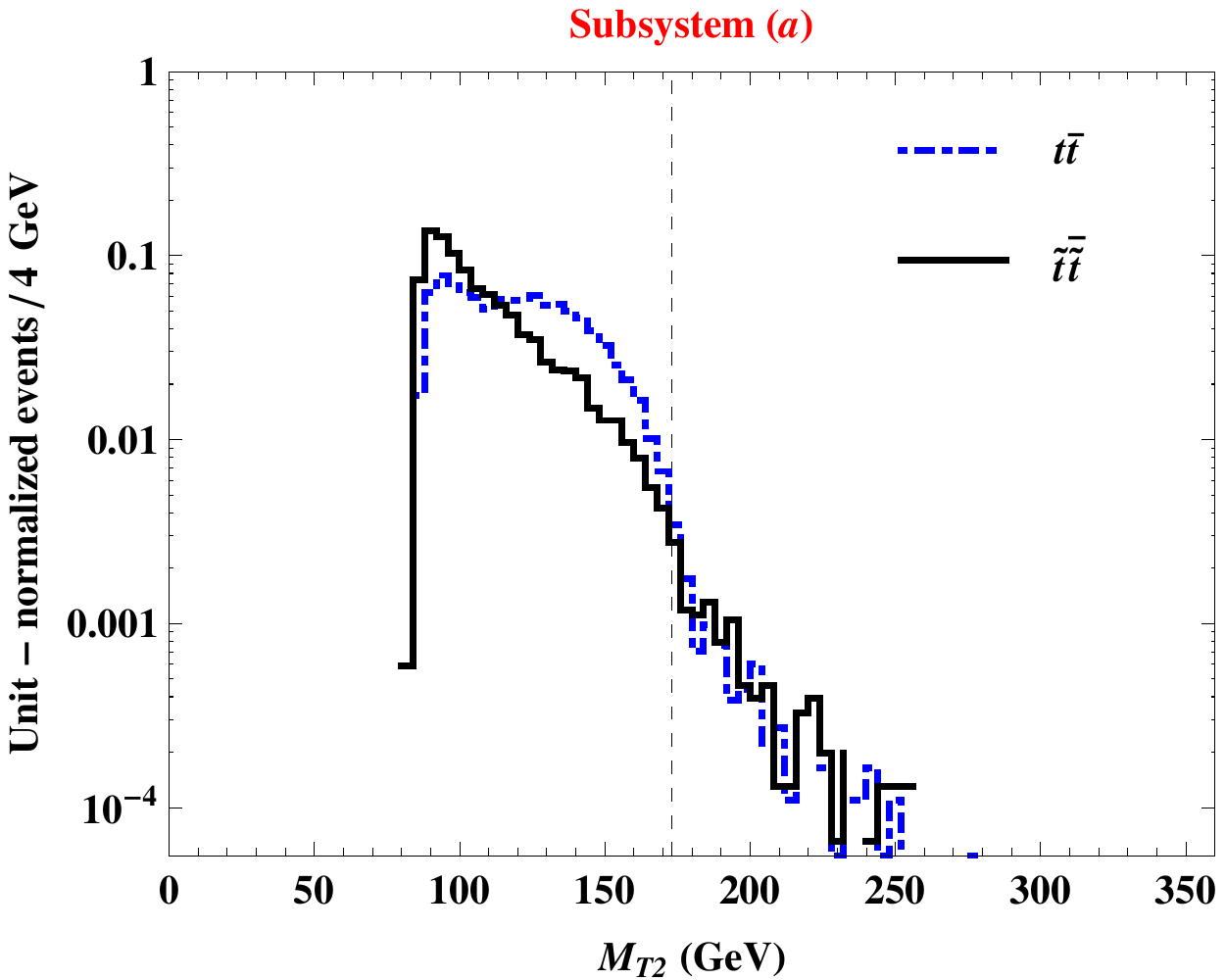}
\includegraphics[width=7.0cm]{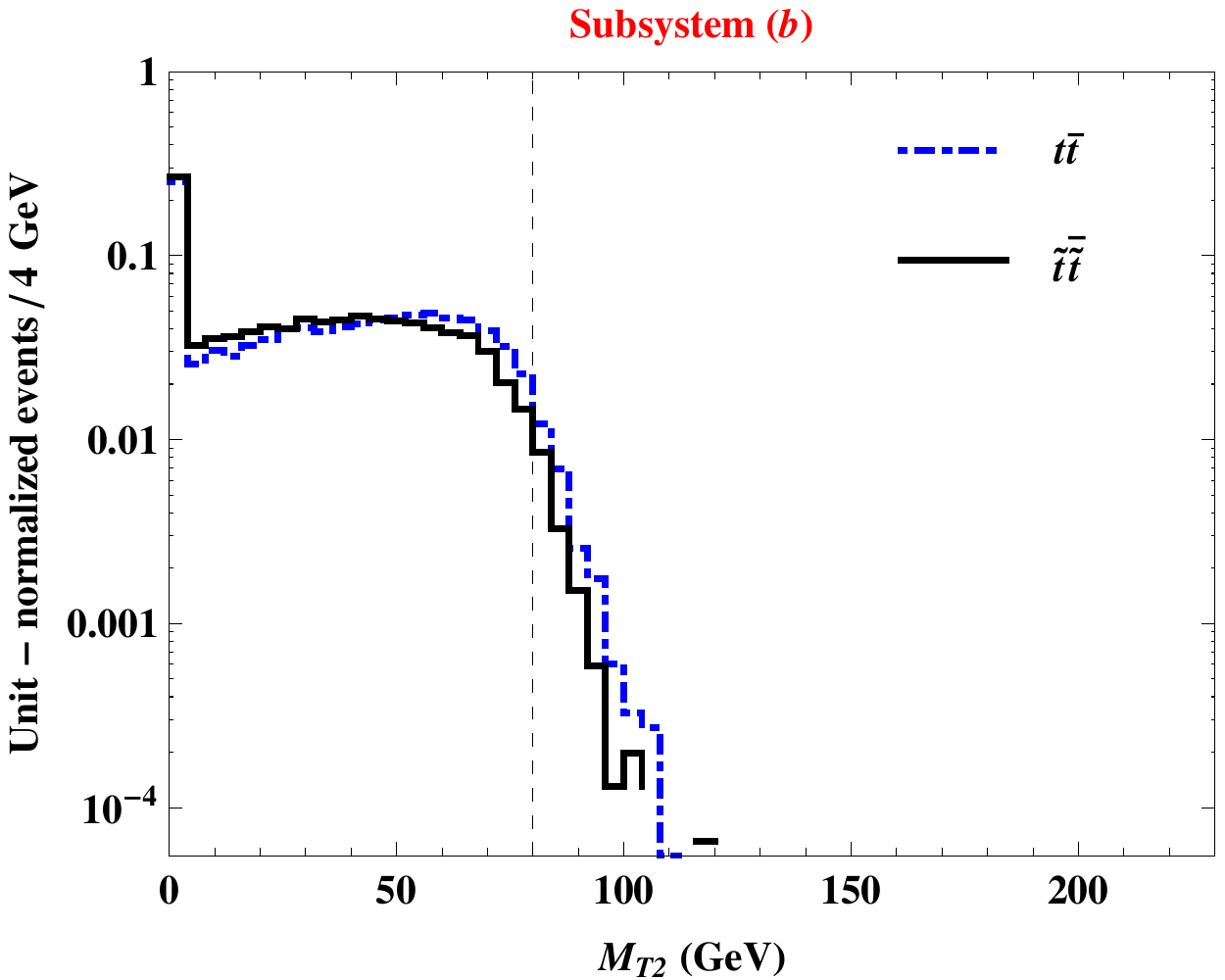}
\caption{\label{fig:MT2detector} The same as Fig.~\ref{fig:MT2cuts5again}, 
but including the effects of combinatorics and detector simulation.
The vertical black dashed lines denote the expected $M_{T2}$ endpoints
of the $t\bar{t}$ background in each subsystem. }
\end{figure}

Fig.~\ref{fig:MT2detector} compares the signal and $t\bar{t}$
background distributions for the three different $M_{T2}$ subsystem variables.
As expected from the parton-level results in
Sec.~\ref{sec:mixed-events} (see Fig.~\ref{fig:MT2cuts5again}),
the discrimination power in the high tail region is relatively poor, since
the signal and the background events obey the same kinematic endpoints.

\begin{figure}[t]
\centering
\includegraphics[width=7.0cm]{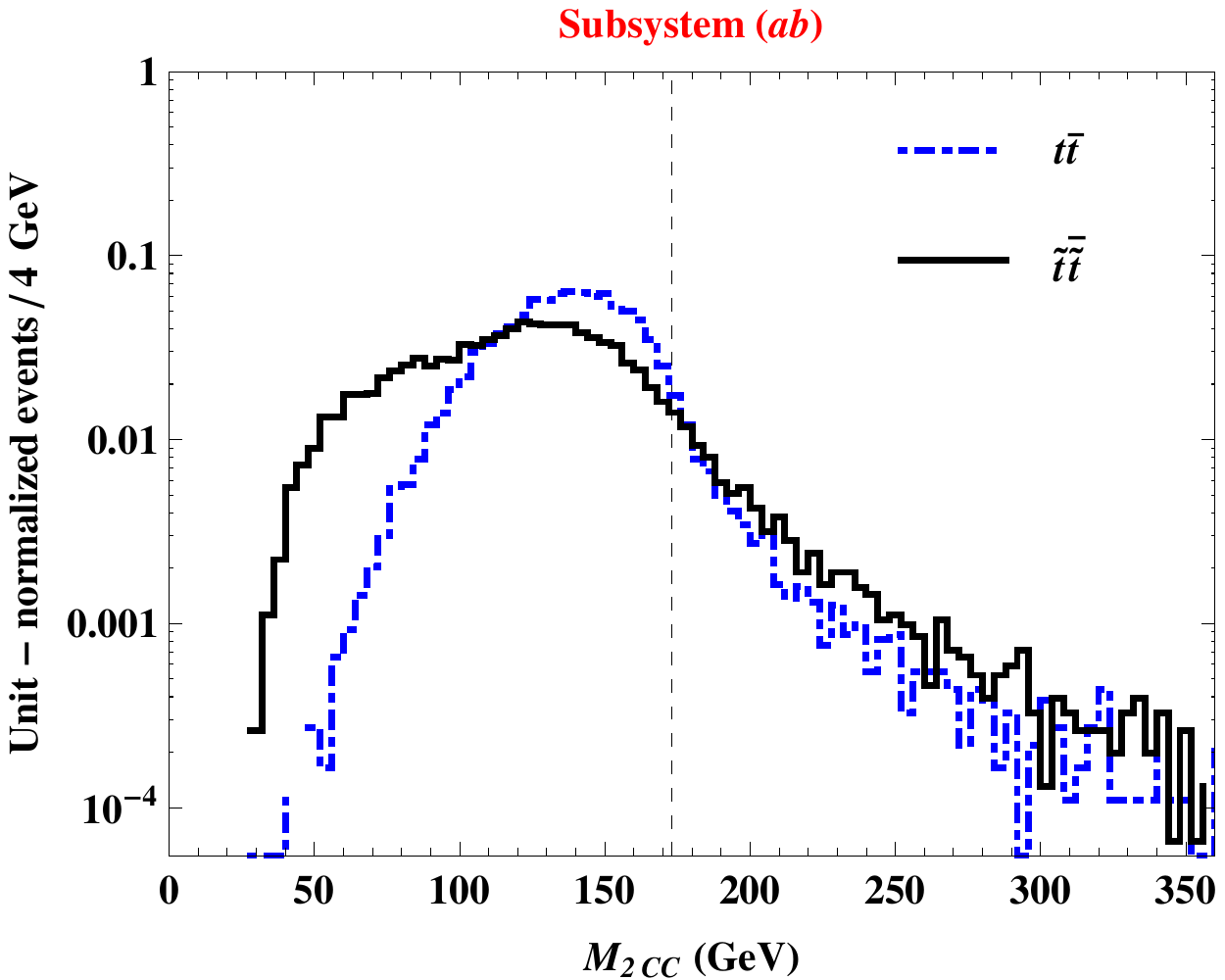}
\includegraphics[width=7.0cm]{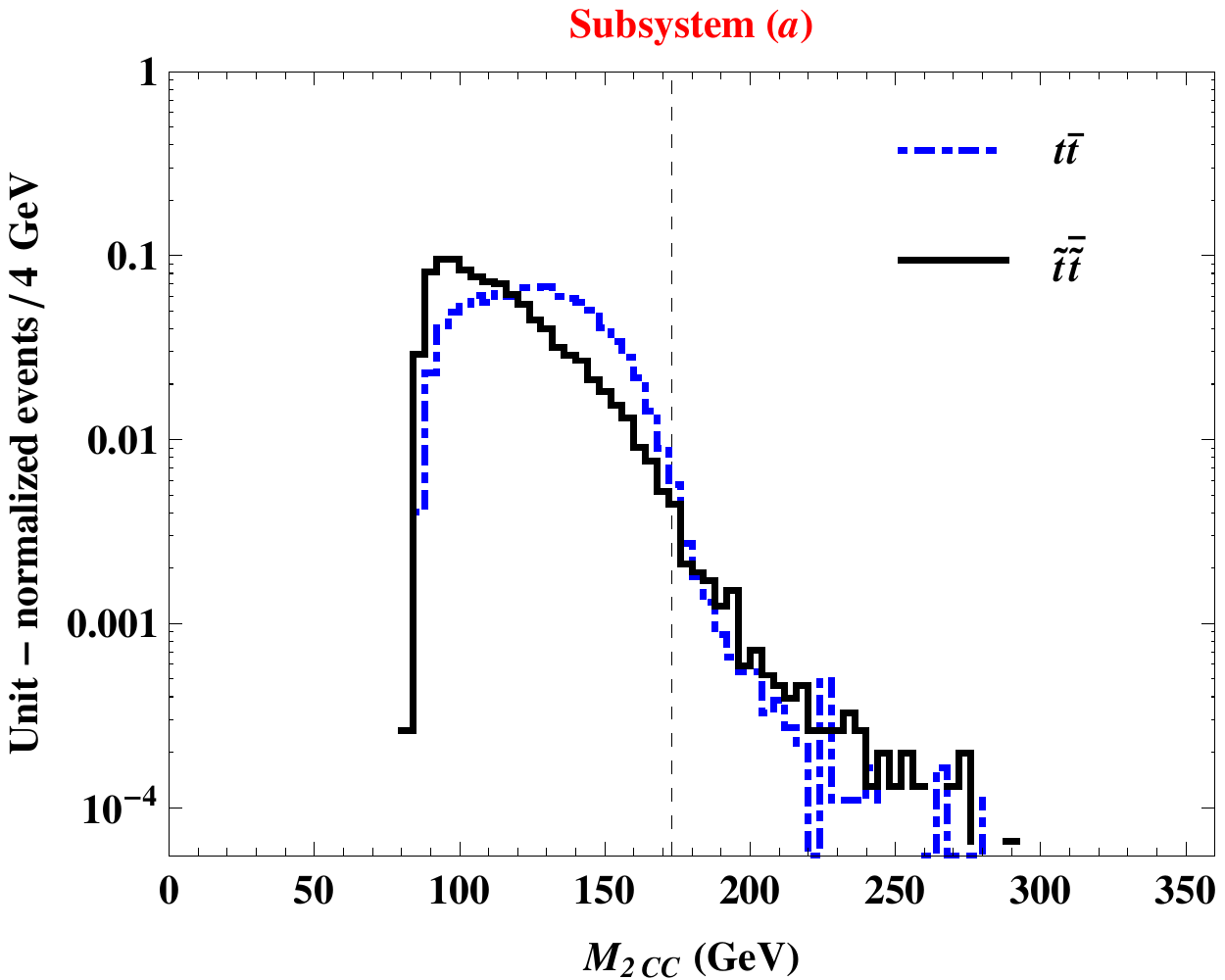}
\includegraphics[width=7.0cm]{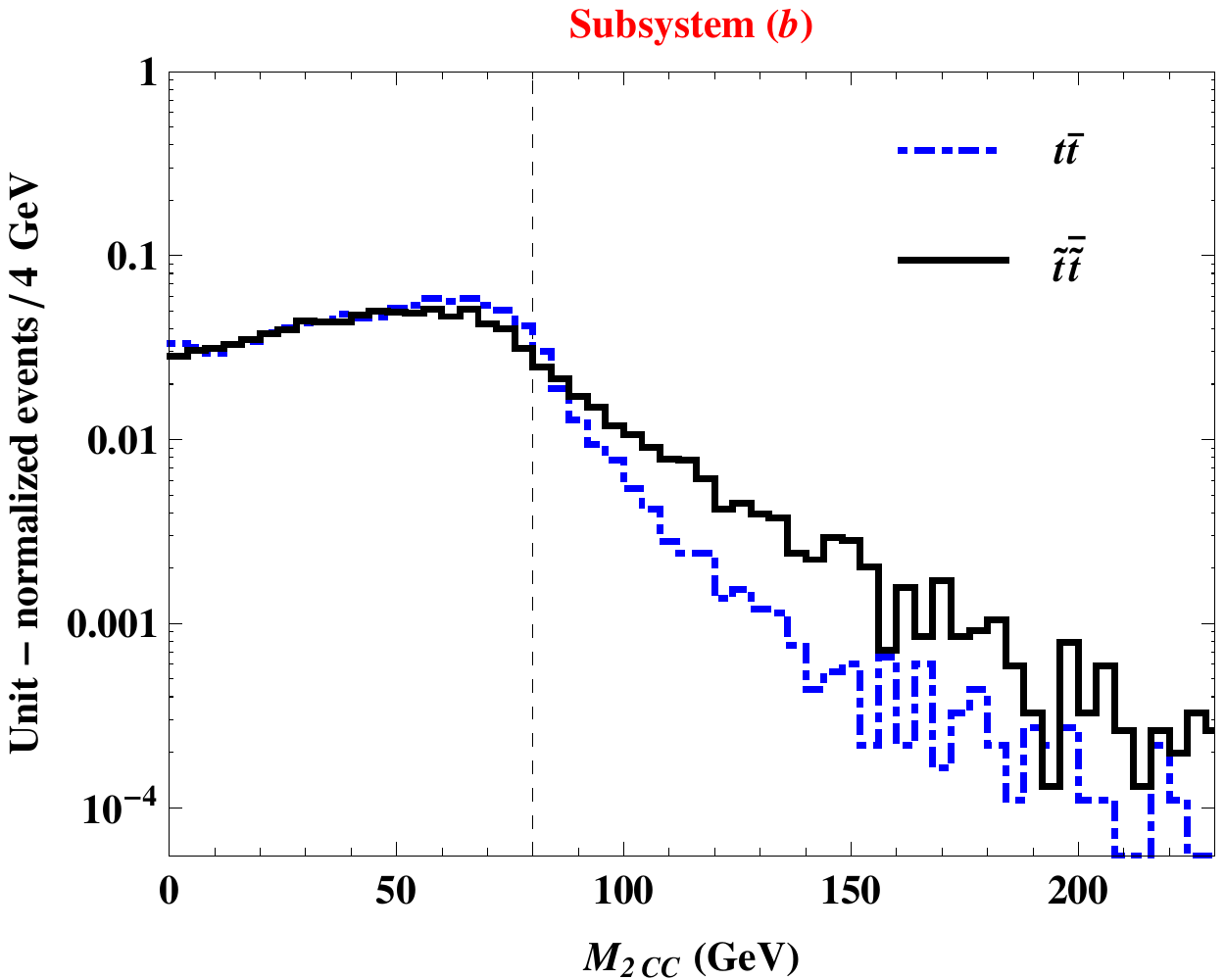}
\caption{\label{fig:M2CCdetector} 
The same as Fig.~\ref{fig:MT2detector}, but for the corresponding
$M_{2CC}$ variables.}
\end{figure}

Fig.~\ref{fig:M2CCdetector} shows the corresponding $M_{2CC}$
distributions for signal and $t\bar{t}$ background events in the three
subsystems.  As anticipated from the parton-level result in
Figs.~\ref{fig:M2CCcuts5} and \ref{fig:M2CCcuts5comb}, there is a
noticeable improvement in the $(b)$ subsystem (for which the visible
particle is a lepton) as seen in the lower panel
and a slight improvement in the $(ab)$ subsystem as well. 
Therefore, one would expect that a minimum $M_{2CC}$ cut would be beneficial.
The optimal value of the cut would depend on the expected signal
cross-section, and on the assumed systematic uncertainty on the
background normalization in the high tail region.

A careful comparison of the parton-level results in
Figs.~\ref{fig:MT2cuts5again},~\ref{fig:M2CCcuts5}
and~\ref{fig:M2CCcuts5comb} versus the detector-level results in
Figs.~\ref{fig:MT2detector} and~\ref{fig:M2CCdetector} reveals that at
the detector level the background distributions develop high tails
which, unless properly understood, could be confused with a signal.
We have checked that in the majority of cases, background events
populate the high tail due to imperfect $b$-tagging. A typical event
looks as follows: one of the two $b$-jets is either too soft to pass
the jet ID cuts, or is not tagged as a  $b$-jet.  (Recall that the
$b$-jet tagging inefficiency is $30$\%.)  Instead, a gluon from initial
state radiation (ISR) forms a hard jet which is subsequently mistagged
as a $b$-jet. Thus in computing the $M_{T2}$ and $M_{2CC}$ variables
one is using the wrong $b$-jet object, which leads to the endpoint
violation.
An improvement in the $b$-tagging algorithm, especially one which
lowers the mistag rate for ordinary QCD jets, would help alleviate
this problem.

\subsection{Results for the relative shift from $M_{T2}$ to $M_{2CC}$}
\label{sec:shift}

In all of our analysis so far, we have relied on the existence of the background 
kinematic endpoints (\ref{mtab}-\ref{mwb}) and focused on the high tails
{\em above} those endpoints. Ideally, this kinematic region should be
populated only by signal events, even when one accounts for the
two-fold combinatorial ambiguity in pairing the leptons and the
$b$-jets. Unfortunately, as we have already seen, this straightforward
approach has two drawbacks:
\begin{itemize}
\item {\em Presence of high tails in the background distributions.}
While in theory the background distributions are not supposed to
extend beyond their kinematic endpoints, in practice this is not
always the case. Such tails were readily observed in
Figs.~\ref{fig:MT2detector} and \ref{fig:M2CCdetector}, which were
obtained using realistic detector simulation. 
\item {\em Low signal efficiency.} Unless we are dealing with a new
  physics model with a widely split spectrum (see related discussion
  in Sec.~\ref{subsec:regions-and-study-points}), a significant
  fraction of the signal events will also lie below the background
  kinematic endpoints, thus by cutting at or near the endpoint, we
  will be removing a large chunk of signal events as well.
This was very evident in the ``worst case" scenarios like study points
4 and 6, or the mixed event case discussed in
Secs.~\ref{sec:mixed-events} and \ref{sec:MT2}.
\end{itemize}
These two problems suggest that we should reexamine the region {\em below} 
the background kinematic endpoints and search for a good
discriminating variable which would be applicable to that region as
well. As in Sec.~\ref{sec:mixed-events}, our goal will be to target
signal events with the mixed event topology of
Fig.~\ref{fig:DecayTopologies}(d).

To begin with, recall the main difference between the background
events described by Fig.~\ref{fig:DecayTopologies}(a) and the signal
described by Fig.~\ref{fig:DecayTopologies}(d): the background events
are symmetric
while the signal events are asymmetric. The on-shell constrained
variables $M_{2CX}$, $M_{2XC}$, and $M_{2CC}$ are obtained by applying
the additional constraints of Eqs.~(\ref{eq:parents-equal}) and
(\ref{eq:relatives-equal}),
which assume that the events are symmetric. Enforcing these
constraints leads to the hierarchy (\ref{eq:inequality}) which is
simply due to the fact that a constrained minimum is larger than an
unconstrained minimum. Since the background events are symmetric, the
constraints (\ref{eq:parents-equal}) and (\ref{eq:relatives-equal})
will be satisfied for the {\em true} values of the invisible momenta,
and, as long as the global minimum is not too far away (in momentum
space), one can expect a relatively mild hierarchy
(\ref{eq:inequality}). Conversely, for signal events with the mixed
event topology, the {\em true} values of the invisible momenta in
general do not satisfy the constraints (\ref{eq:parents-equal}) and
(\ref{eq:relatives-equal}). Thus one could expect that the effect of
imposing the constraints would be larger, leading to a larger
hierarchy (\ref{eq:inequality}).

These intuitive considerations suggest that we look at the shift of
the on-shell constrained invariant mass variable which is caused by
the constraint itself. Keeping in mind the identity
$M_{T2}=M_{2XX}$~\cite{Cho:2014naa}, we can take the usual stransverse 
mass $M_{T2}$ as our benchmark variable in the absence of any constraints.
Then, we can ``measure" the effect of the constraints by comparing
$M_{T2}$ to 
$M_{2CC}$, where both (\ref{eq:parents-equal}) and (\ref{eq:relatives-equal}) 
have been applied. This motivates the consideration of a new
variable\footnote{Since 
the shift $M_{2CC}- M_{T2}$ is relatively small compared to the individual 
values of $M_{2CC}$ or $M_{T2}$, in Eq.~(\ref{eq:deltaM}) we prefer to define
$\Delta M$ in terms of the difference of the {\em squared} masses.
The square root is then used merely to lower the mass dimension of $\Delta M$ back to GeV.} 
\beq
\Delta M(S;\tilde m) \equiv \sqrt{M^2_{2CC}(S;\tilde m) - M^2_{T2}(S; \tilde m)}.
\label{eq:deltaM}
\eeq 
As indicated in (\ref{eq:deltaM}), this variable can be computed for
any of the three subsystems.
We have checked that in our example here, the $(ab)$ subsystem shows the
best discrimination between the signal and $t\bar{t}$ background. 
\begin{figure}[t]
\centering
\includegraphics[width=6.0cm]{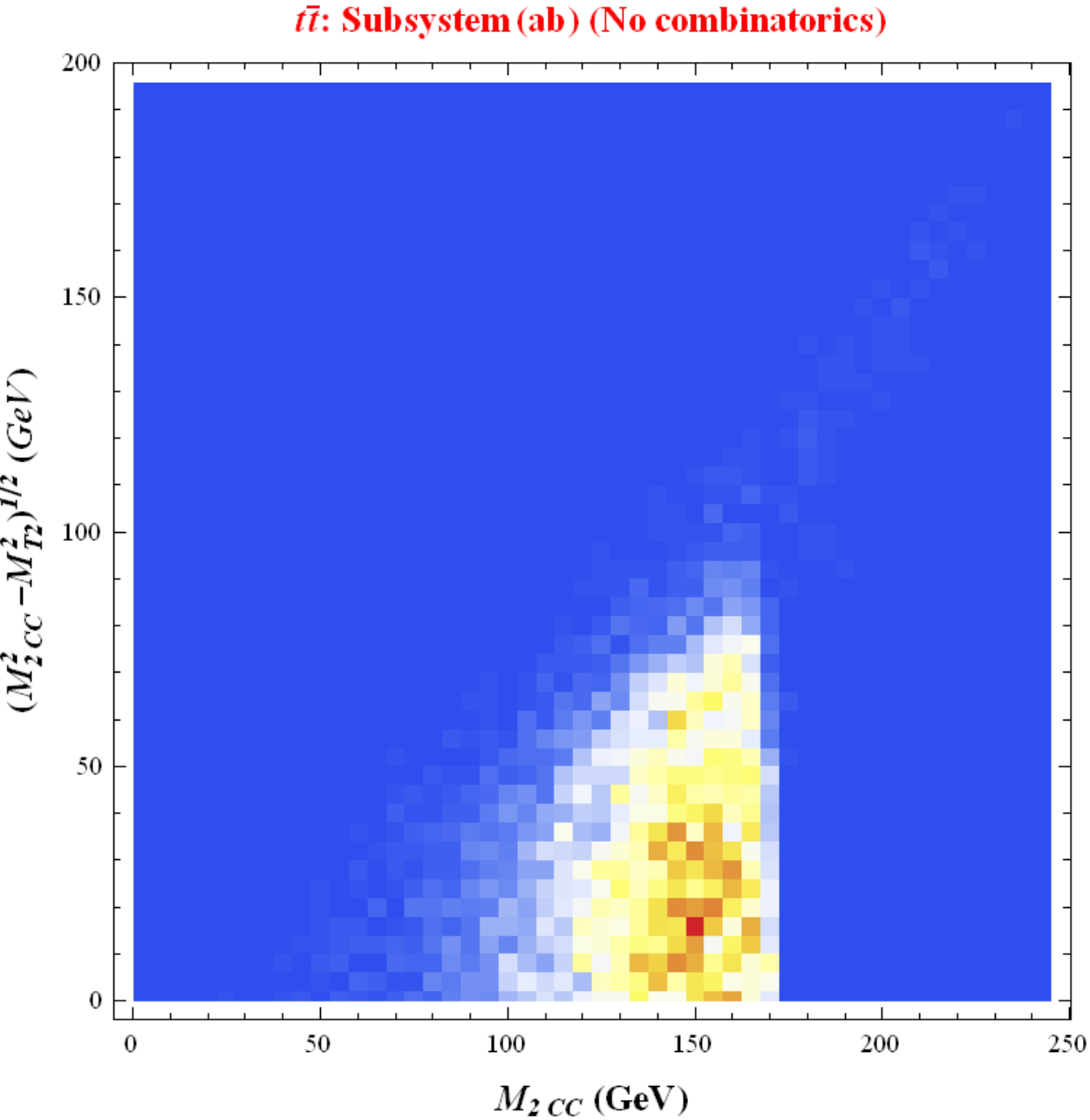}
\includegraphics[width=6.0cm]{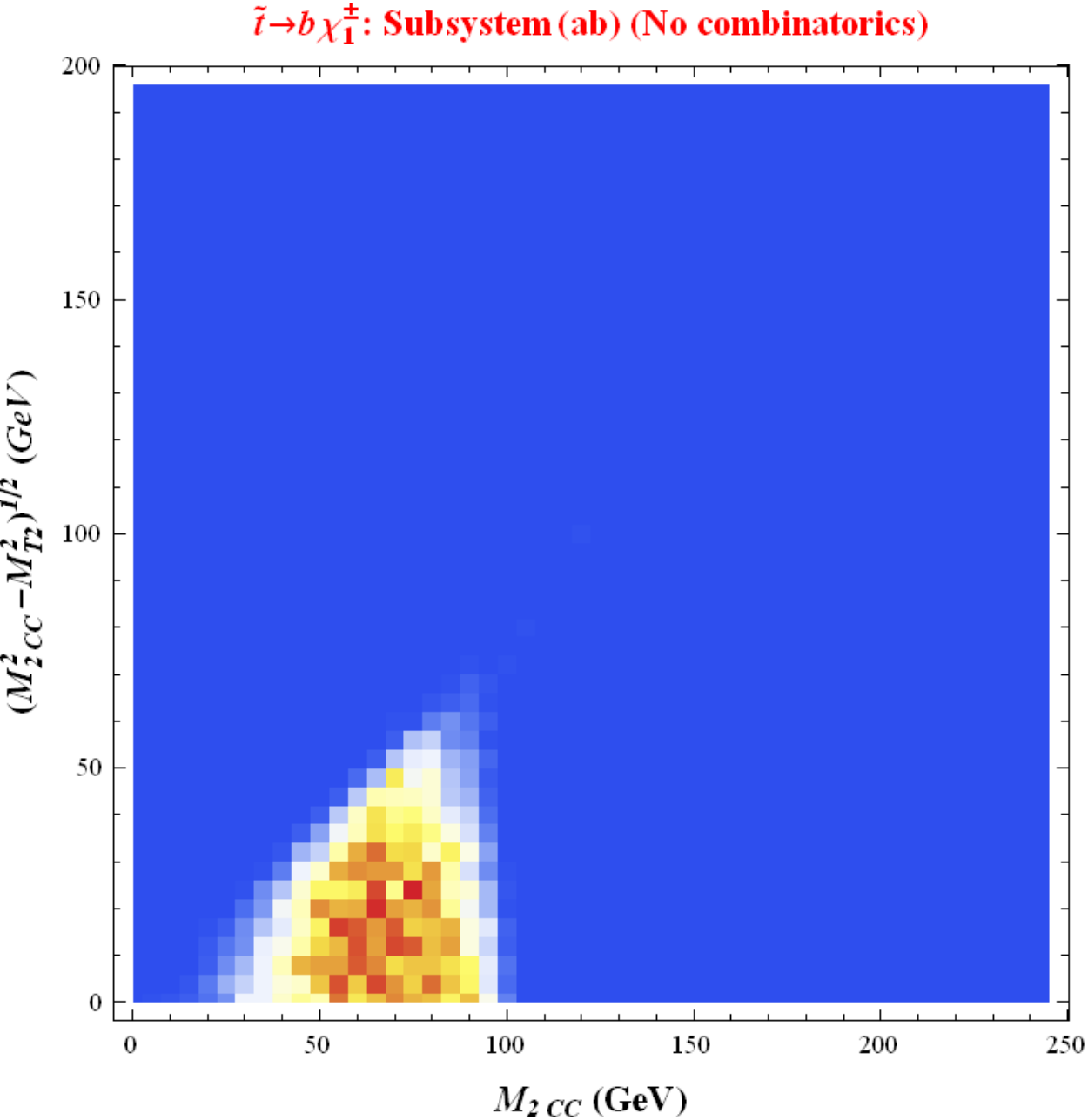}\\
\includegraphics[width=6.0cm]{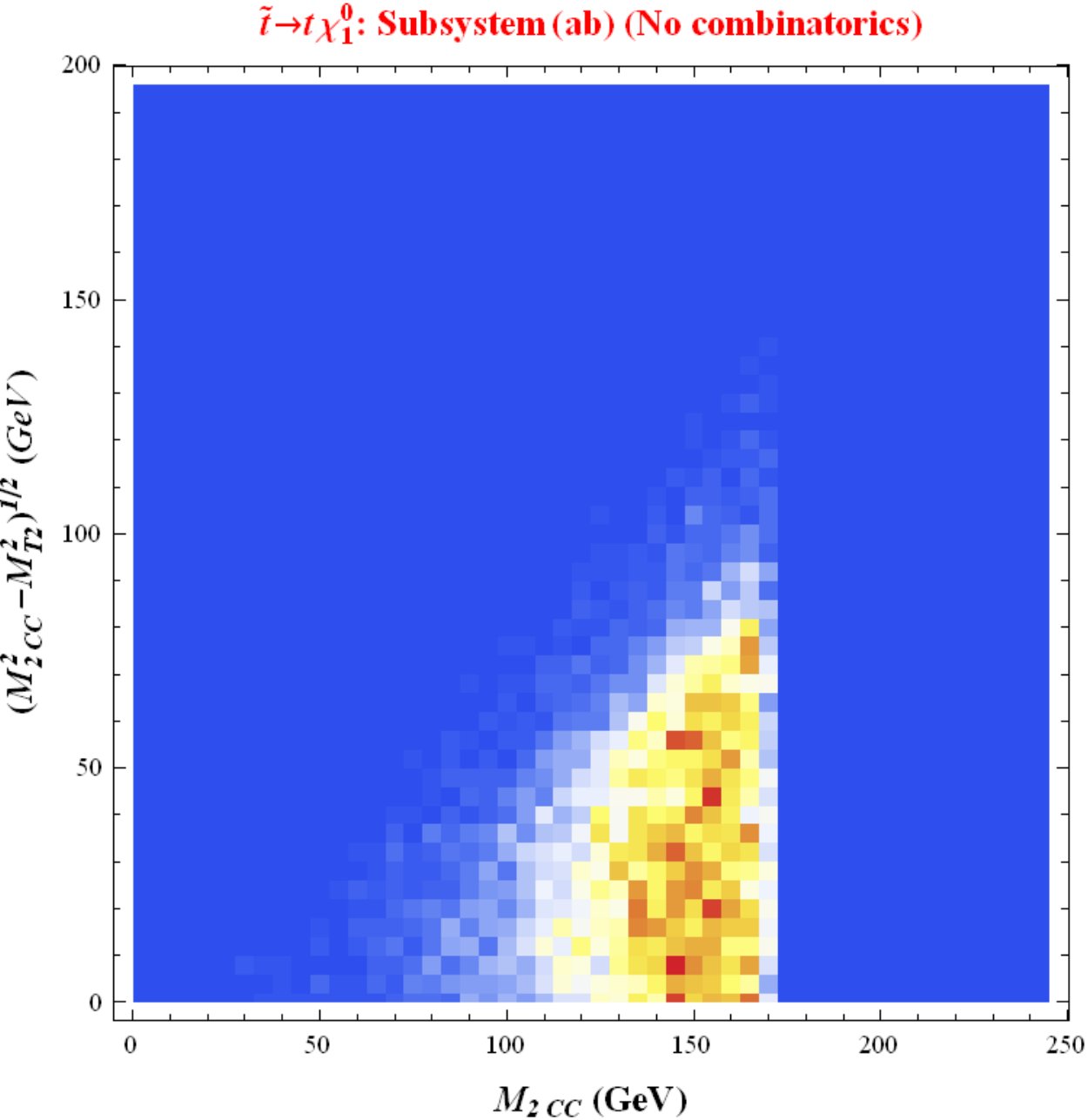} 
\includegraphics[width=6.0cm]{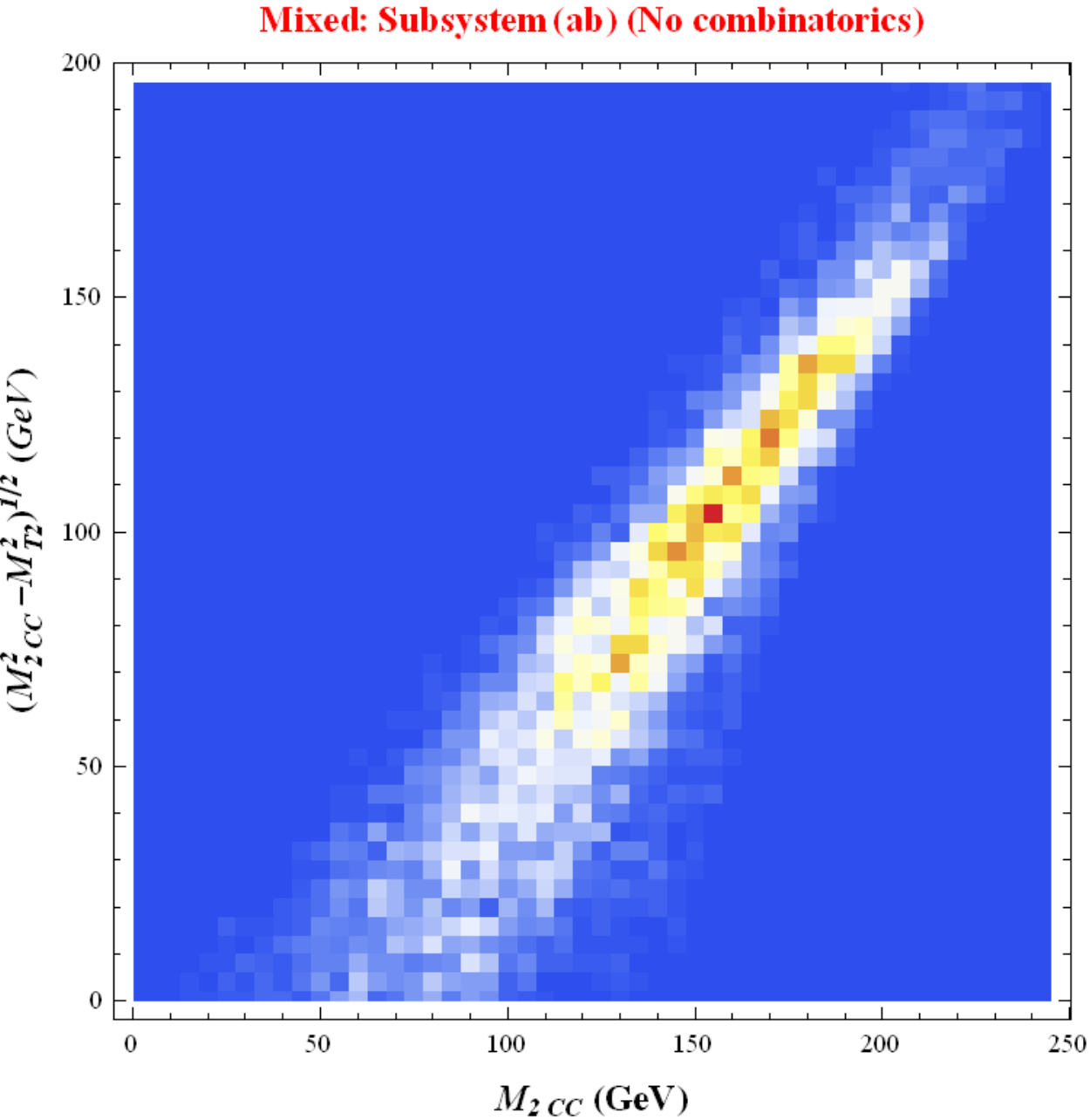}
\caption{\label{fig:DM} Density plots exhibiting the correlations between the
variables $\Delta M(ab)$ and $M_{T2}(ab)$ for background events (upper left)
and for the three types of signal events:
with pure Topology 1 from Fig.~\ref{fig:DecayTopologies}(b) (upper right),
pure Topology 2 from Fig.~\ref{fig:DecayTopologies}(c) (lower left),
and the mixed topology from Fig.~\ref{fig:DecayTopologies}(d) (lower right). }
\end{figure}
Therefore, in Fig.~\ref{fig:DM}, we contrast the new variable $\Delta M(ab)$ defined
in (\ref{eq:deltaM}) with the variable $M_{2CC}(ab)$ advocated 
above in Sec.~\ref{sec:mixed-events}. Each panel in Fig.~\ref{fig:DM} 
shows a specific type of events at parton level:
$t\bar{t}$ background events (upper left panel),
signal events with pure Topology 1 from
Fig.~\ref{fig:DecayTopologies}(b) (upper right panel),
signal events with pure Topology 2 from
Fig.~\ref{fig:DecayTopologies}(c) (lower left panel), and
signal events with mixed topology from
Fig.~\ref{fig:DecayTopologies}(d) (lower right panel). 
We see that, as already observed in Fig.~\ref{fig:M2CCcuts5},
a certain number of signal events in the mixed channel exceed the
background endpoint for $M_{2CC}$.
More importantly, the figure also shows that there are many more
signal events which do {\em not}
exceed the background endpoint, yet their value for $\Delta M$ is
significantly larger than 
that for a typical background event. The situation does not change much if we 
account for the two-fold combinatorial ambiguity, as demonstrated by
Fig.~\ref{fig:DMcomb}.
We conclude that $\Delta M$ possesses additional 
discriminating power, and therefore, for an optimal analysis, one should 
use {\em both} $\Delta M$ and $M_{2CC}$ as discriminating variables.

\begin{figure}[t]
\centering
\includegraphics[width=6.0cm]{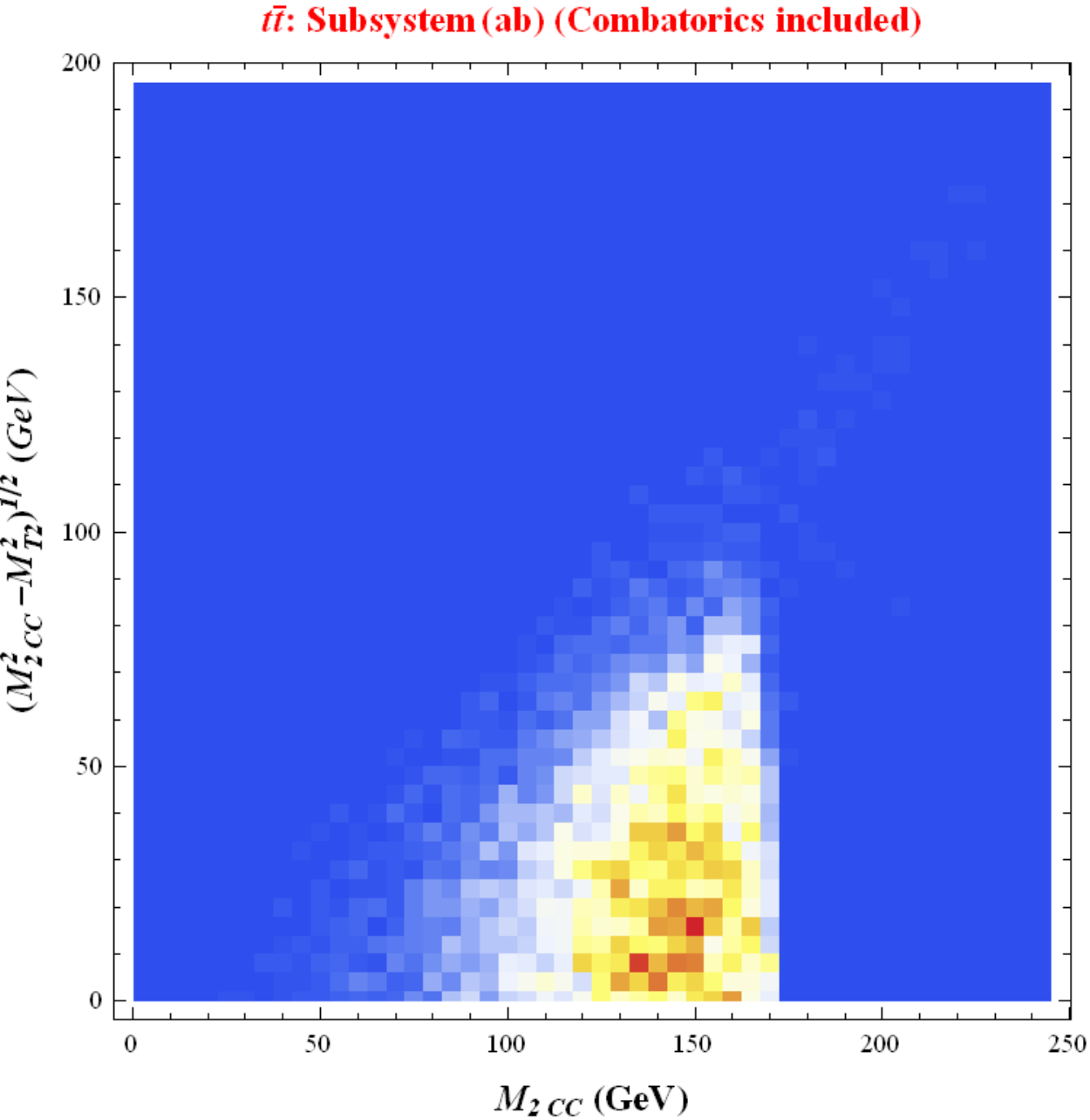}
\includegraphics[width=6.0cm]{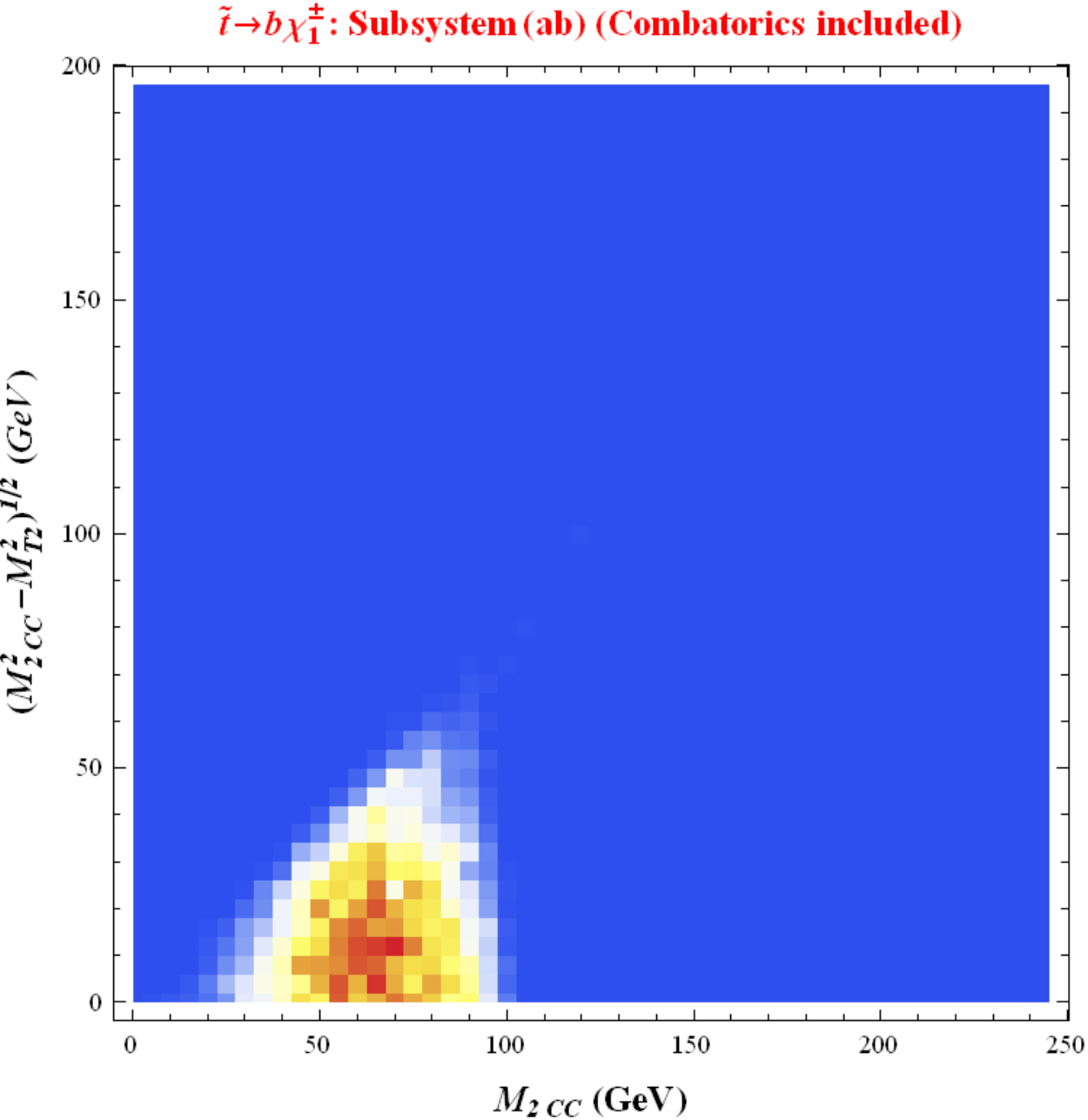}\\
\includegraphics[width=6.0cm]{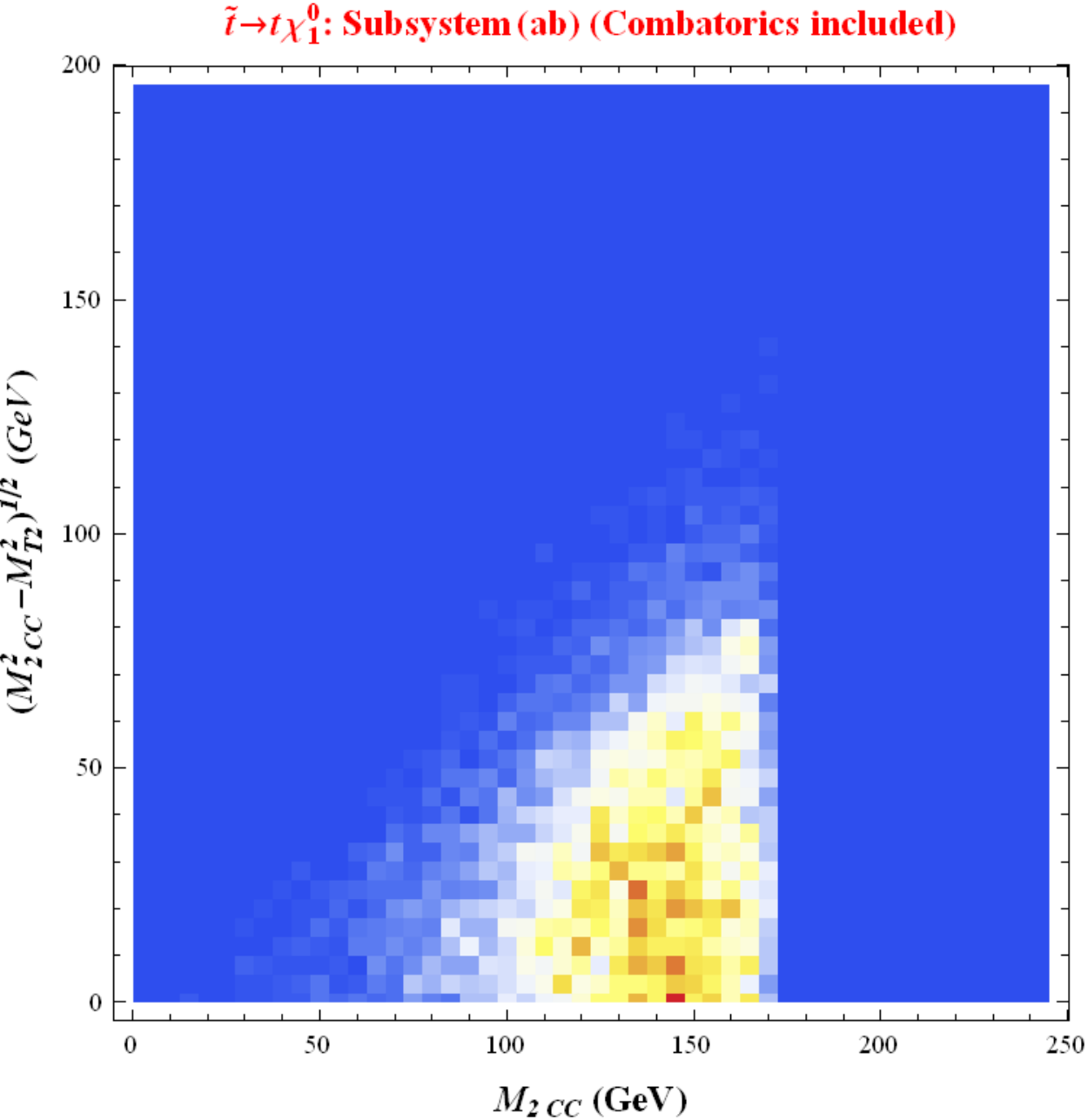} 
\includegraphics[width=6.0cm]{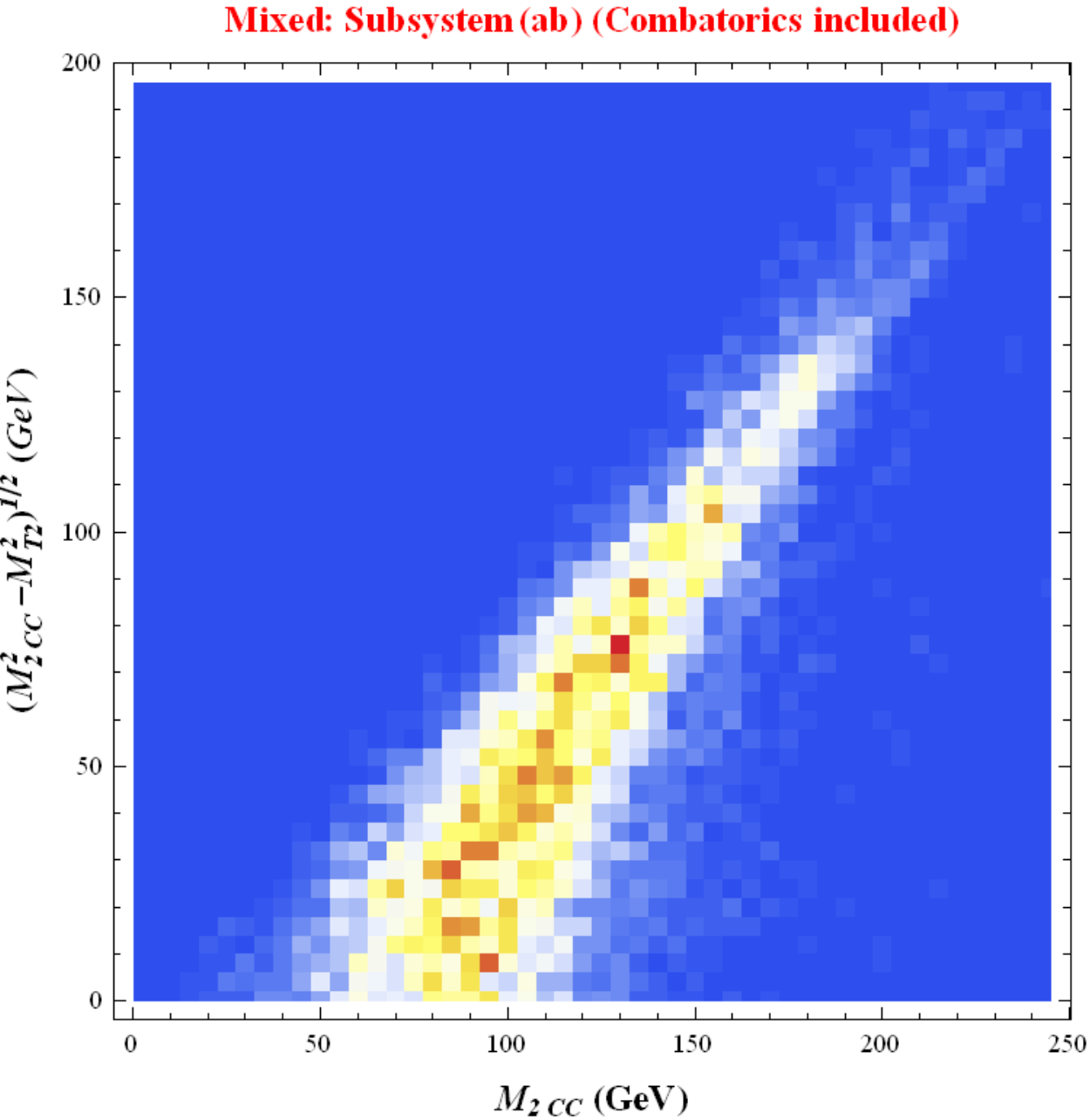}
\caption{\label{fig:DMcomb} The same as Fig.~\ref{fig:DM}, but including combinatoric effects:
each variable $M_{T2}$ and $M_{2CC}$ is separately calculated with the $b$-$\ell$ pairing 
which gives the lower value for the corresponding variable.
 }
\end{figure}

\begin{figure}[t]
\centering
\includegraphics[width=7.0cm]{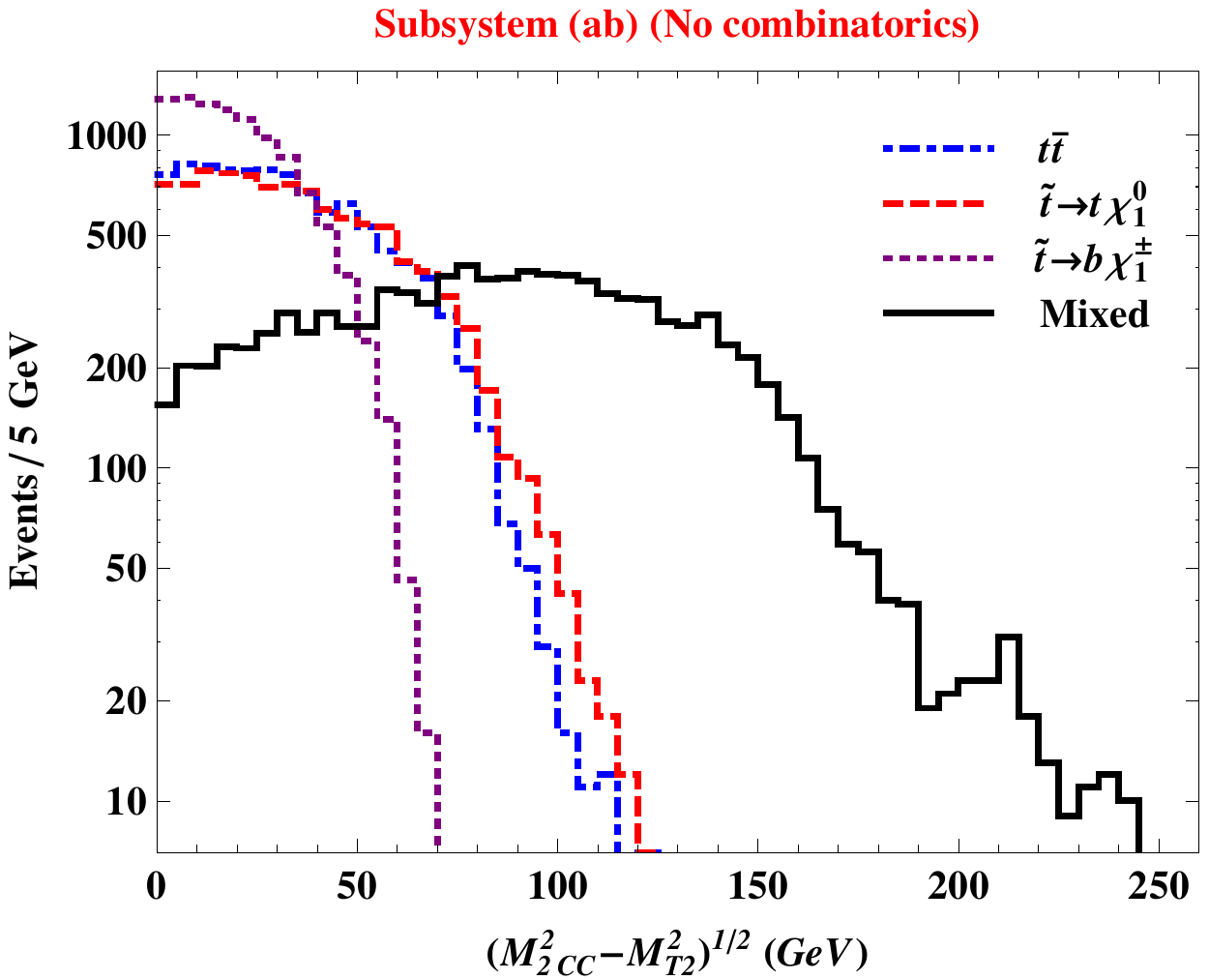}
\includegraphics[width=7.0cm]{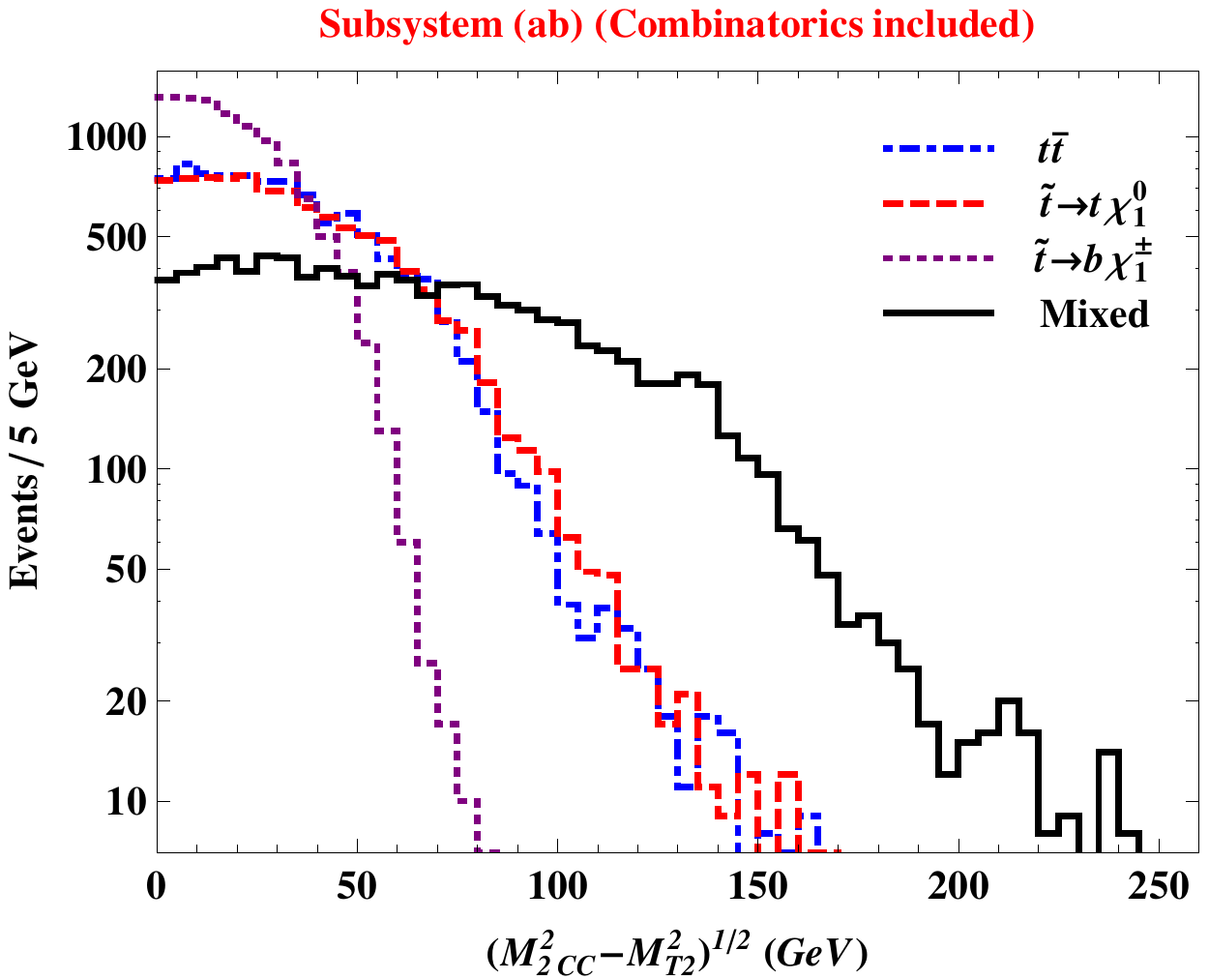} \\
\includegraphics[width=7.0cm]{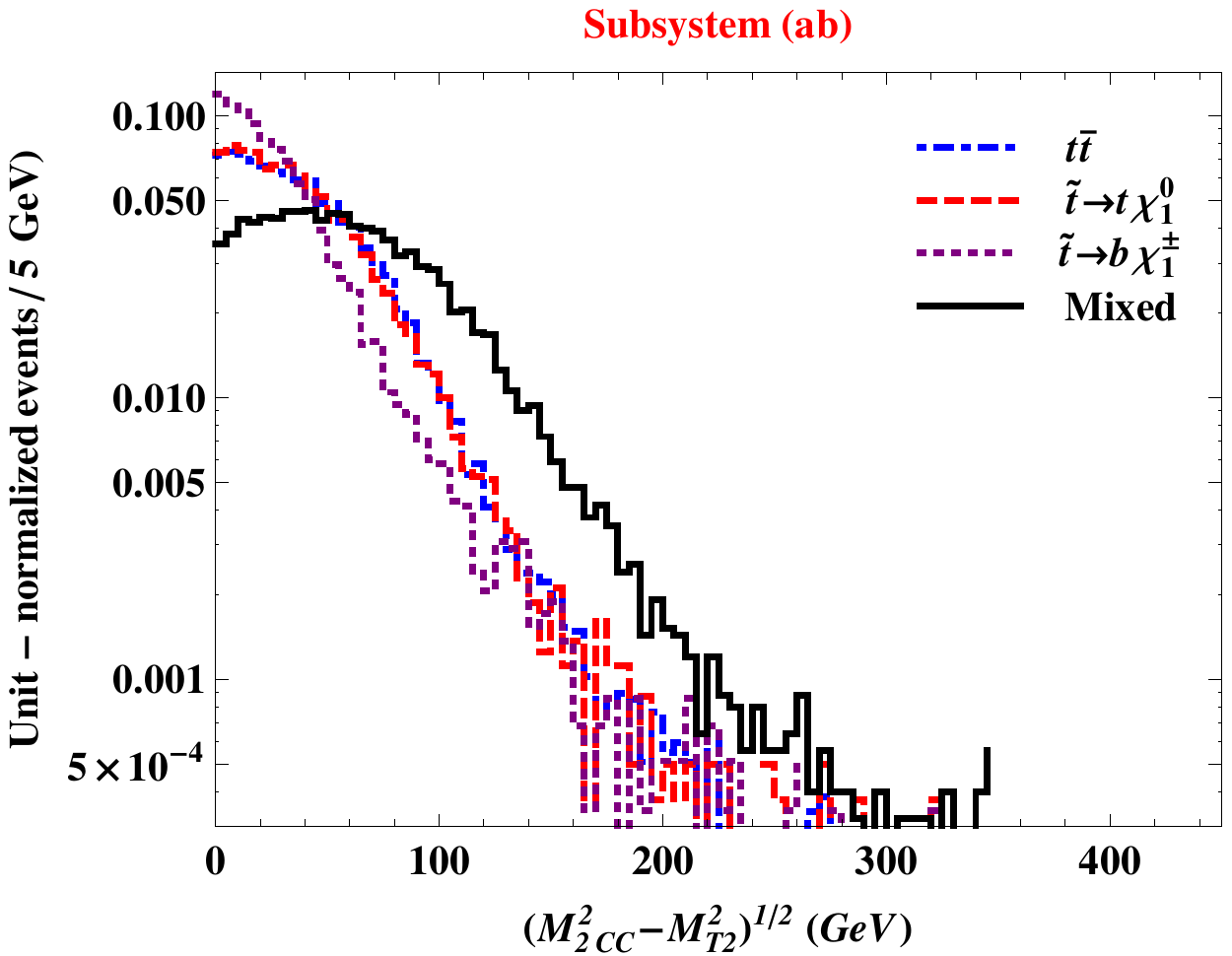}
\includegraphics[width=7.0cm]{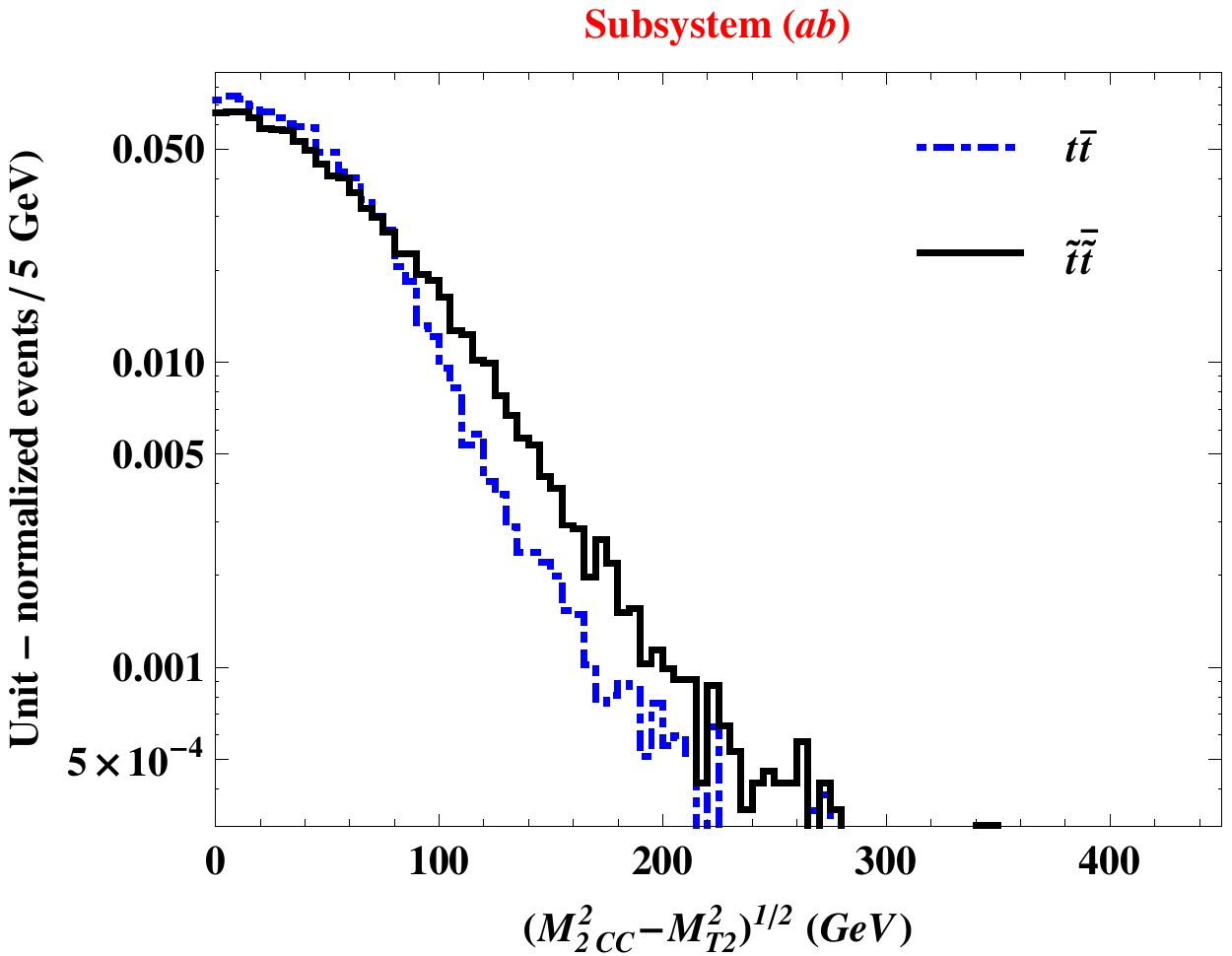}
\caption{\label{fig:Rvariable} Unit-normalized distributions of the 
$\Delta M(ab)$ variable for signal and background events, with the same
color coding scheme used in earlier figures. 
The top row shows parton level results before any selection cuts, 
while the bottom row includes detector simulation and selection cuts.
The upper right (upper left) panel does (does not) account for the two-fold combinatorial ambiguity. }
\end{figure}

For completeness, we also present results for the $\Delta M$ variable alone.
Fig.~\ref{fig:Rvariable} shows unit-normalized distributions for $\Delta M(ab)$
at the parton level (upper row) and after detector simulation and
selection cuts (lower row).
The upper left panel is done with perfect assignment for the
lepton--$b$-jet pairing,
while the upper right panel accounts for the two-fold combinatorial
ambiguity as before.
The lower right panel shows the observable total signal distribution, 
which is made up of the individual components identified on the lower
left panel.
Clearly, the variable $\Delta M(ab)$ performs quite well for signal events with 
a mixed event topology, and to some extent for signal events with Topology 1.
The effect is diluted, but still visible after detector simulation
(panels in the lower row).

\subsection{An alternative variable: the ``relative" mass difference}
\label{sec:relative}

In the previous section, we proposed the variable, $\Delta M$,
as a measure of the effect of the constraints 
(\ref{eq:parents-equal}) and (\ref{eq:relatives-equal}). 
The idea was to look at the change in the value of $M_2$ as a result
of enforcing these constraints.
Let us now look at a different way of capturing the same effect.

Recall that as a result of the minimization involved in calculating 
the {\em unconstrained} $M_{2XX}$ variable, one obtains values 
for the invisible momenta that minimize the maximal parent particle
invariant mass in the specified subsystem.
While these are not necessarily the true momenta of the 
invisible particles in the event, they do provide an useful ansatz 
and can be used to calculate various $3+1$-dimensional
kinematic quantities of interest~\cite{Cho:2014naa}.  (See also the 
MAOS method~\cite{Cho:2008tj,Park:2011uz}.)
In particular, we can compute the masses of the parent particles and the 
relative particles in the event and test whether the constraints 
(\ref{eq:parents-equal}) and (\ref{eq:relatives-equal}) are satisfied or not. 
However, there is one technical complication: the function
which is being minimized in order to compute $M_{2XX}$,
sometimes has a flat direction and does not lead to a unique ansatz for 
the invisible momenta~\cite{Cho:2014naa}. In order to avoid this problem, 
here we prefer to use the variable, $M_{CX}$, where the parent constraint
(\ref{eq:parents-equal}) is already applied.
Thus, we will be comparing the masses of the relative particles instead.
In analogy to (\ref{eq:deltaM}), we therefore define
\beq
\Delta M_R(S;\tilde m) \equiv \sqrt{|M^2_{R_1}(S;\tilde m) -
  M^2_{R_2}(S; \tilde m)|}.
\label{eq:deltaMR}
\eeq 
\begin{figure}[t]
\centering
\includegraphics[width=6.0cm]{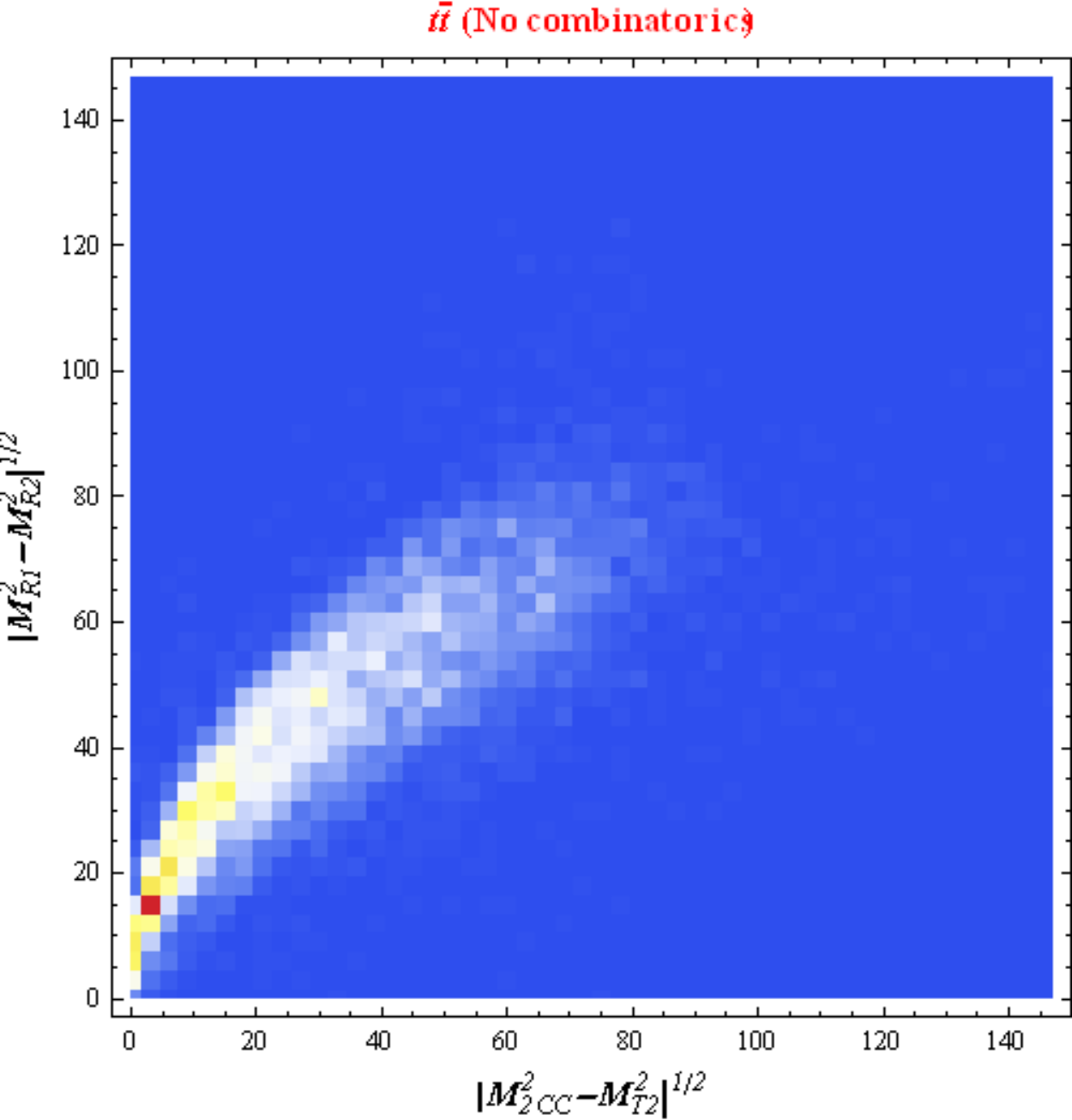}
\includegraphics[width=6.0cm]{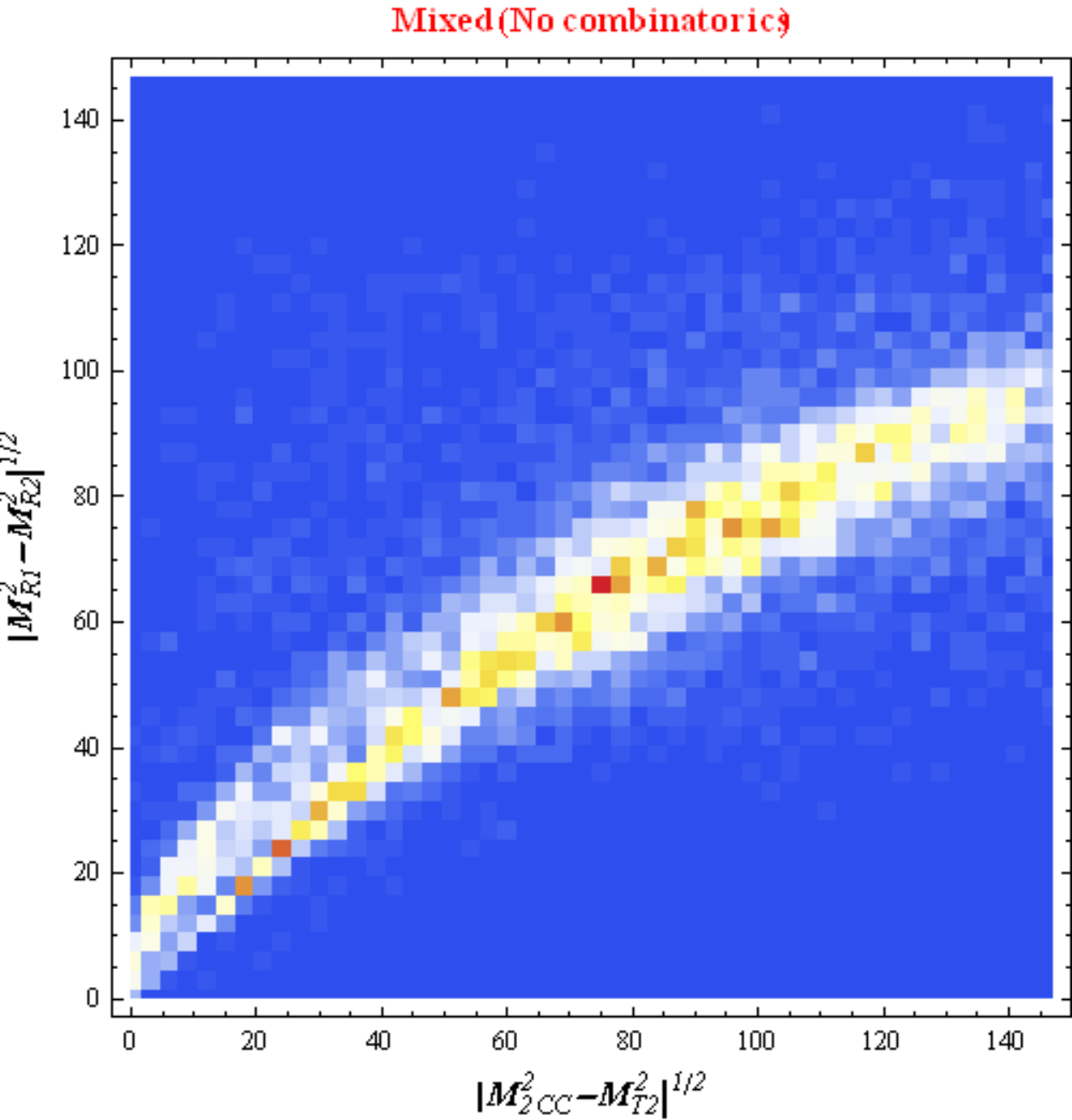}
\caption{\label{fig:MRdiff} 
The correlation between the parent mass difference $\Delta M_R$ defined in (\ref{eq:deltaM})
and the relative mass difference $\Delta M$ defined in (\ref{eq:deltaMR}) for background events (left) 
and signal events with the mixed event topology of Fig.~\ref{fig:DecayTopologies}(d) (right). } 
\end{figure}
Since they both measure the same effect, namely, the impact of the relative 
constraint (\ref{eq:relatives-equal}), we expect the two variables $\Delta M$
and $\Delta M_R$ to be correlated. This is illustrated in Fig.~\ref{fig:MRdiff},
where we compare $\Delta M(ab)$ and $\Delta M_R(ab)$
for background events (left panel) and signal events with the 
mixed event topology of Fig.~\ref{fig:DecayTopologies}(d) (right panel).
The correlation is very evident and suggests that $\Delta M_R$ can be used 
in place of $\Delta M$. The advantage of using $\Delta M_R$ is convenience:
in order to compute it, one needs to perform a single minimization
(that of the variable $M_{2CX}$), 
while to construct $\Delta M$, one needs to minimize twice: once for
$M_{T2}$ (or, equivalently,
$M_{2XX}$) and then once for $M_{2CC}$. We have also noticed that our
numerical minimization code finds the global minimum of
$M_{2CX}$ more reliably than it finds the minimum of the doubly constrained
variable $M_{2CC}$.

\section{Conclusions and outlook}
\label{sec:conclusions}

The search for ``top partners," like top squarks in SUSY, will be a
key component of the LHC research program in the next run of the LHC. 
This is due to several reasons.
First, particles which behave like top partners are theoretically
well-motivated since they are ubiquitous in models that try to
address the hierarchy problem. 
Second, the experimental limits on third generation partners are
generally weaker, leaving room for improvement in the next run. 
Third, the signatures of top partner production typically resemble
those of SM top production, a process which will continue to be under
close scrutiny because of the intrinsic interest in the top in its own
right.

The main goal of this paper was to tackle certain {\em difficult}
cases for stop discovery and propose new ideas for improving the
experimental sensitivity in the next LHC run.  We considered stop
signatures which led to an {\em identical} final state as the main
irreducible top background. Of special interest to us were corners of
parameter space which would evade easy detection by normal means,
either due to small mass splittings, which lead to soft jets and leptons, or 
because the new physics signature involves real SM top quarks. 
Thus we considered the two decay topologies of Fig.~\ref{fig:process},
which led to the three types of signal events depicted in
Fig.~\ref{fig:DecayTopologies}(b-d).

Given that the signal and background are so similar, discrimination is
only possible if we take full advantage of subtle kinematic
differences. This is why we focused on the recently proposed class of
on-shell constrained variables ($M_{2XX}$, $M_{2CX}$, $M_{2XC}$, and
$M_{2CC}$)~\cite{Mahbubani:2012kx,Cho:2014naa},
which can be suitably defined with the background event topology of
Fig.~\ref{fig:DecaySubsystem} in mind (see
Appendix~\ref{sec:appendix}). These variables have several useful
properties which can be used for isolating the signal over the
background:
\begin{itemize}
\item {\em Existence of upper kinematic endpoints.} While the
  background events obey the bounds (\ref{mtab}-\ref{mwb}), signal events may violate those bounds, depending on the new physics mass spectrum.
  Thus, by employing suitable high pass cuts on those variables, one can
  remove the majority of the background, leaving some fraction of
  the signal. In Sec.~\ref{subsec:regions-and-study-points}, we
  analyzed the relevant mass parameter space and classified the
  regions where given kinematic endpoints for signal
  events exceed those for the background. The ``easy" regions, where
  the signal endpoints are significantly above the background
  endpoints, should be the first targets in the next LHC runs.
\item {\em Endpoint violation in the case of the ``wrong" event
    topology.} The on-shell kinematic variables were defined with a
  specific background event topology in mind. If the signal events
  have a different event topology, either because they are asymmetric
  (e.g., the mixed event topology of
  Fig.~\ref{fig:DecayTopologies}(d)), or because they contain more
  invisible particles (as in the case of pure Topology 2 in
  Fig.~\ref{fig:DecayTopologies}(c)), they may again violate the
  background endpoints, see Figs.~\ref{fig:MT2cuts5} and
  \ref{fig:M2CCcuts5}.
  For concreteness, in this paper we only considered the two specific
  event topologies from Fig.~\ref{fig:process}, but in realistic
  models, there exist other well-motivated event topologies which
  would lead to the same final state.  (A couple of such examples are
  shown in Fig.~\ref{fig:process2}.) The multitude of possible stop
  decay modes increases the likelihood that the signal will include
  asymmetric events, which may manifest themselves through violations
  of the expected background endpoints. 
\begin{figure}[t]
\centering
\includegraphics[scale=0.9]{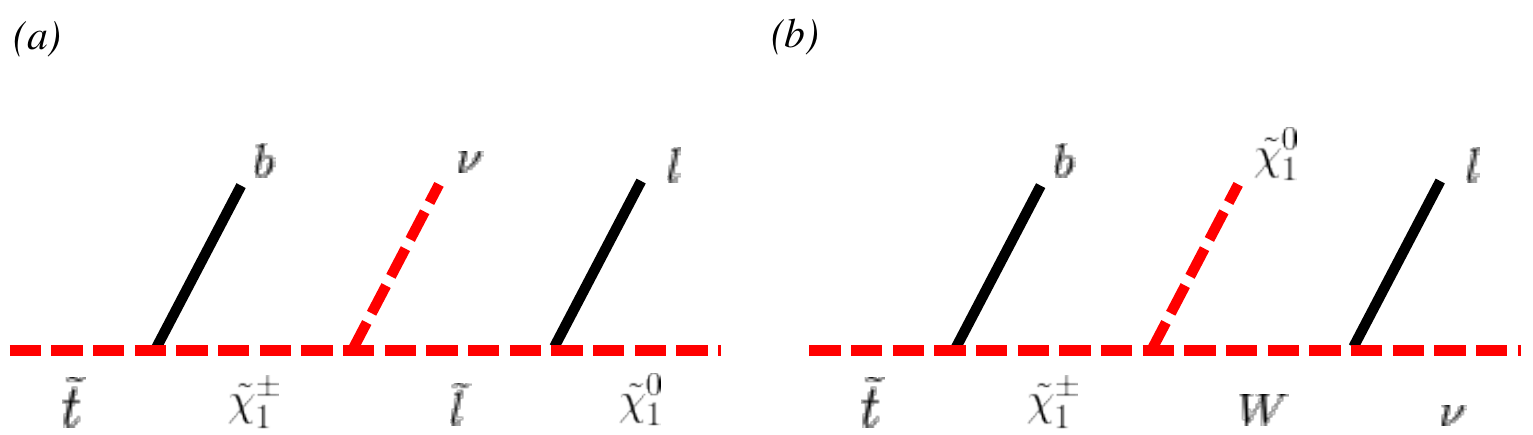}
\caption{\label{fig:process2} Two other possible decay topologies leading to the same final state as in Fig.~\ref{fig:process}.}
\end{figure}
\item {\em The existence of the hierarchy (\ref{eq:inequality})
    between the various $M_{2\sqcup\sqcup}$ variables.}
We showed that the hierarchy is relatively mild in the case of
symmetric events like the $t\bar{t}$ background and gets stronger as
the events become more asymmetric (as in the mixed topology of
Fig.~\ref{fig:DecayTopologies}(d)).  This observation allows us to
also target signal events which are {\em below} the background
kinematic endpoints. We believe that the related variables $\Delta M$
and $\Delta M_R$ defined in (\ref{eq:deltaM}) and (\ref{eq:deltaMR})
respectively, will be a useful addition to the experimenter's arsenal
of tools for new physics searches in missing energy events.
\end{itemize}

In spite of the advances proposed here, these searches remain extremely challenging.
Detector effects, jet combinatorics issues due to ISR and FSR,
and $b$-jet misidentification contribute to the degradation of the 
parton level significance\footnote{For a more complete discussion, see 
\cite{Kim:2014ana}. }. Nevertheless, we believe that the techniques presented here 
will prove useful in searches for top partners at the LHC. 

In this paper, we have employed a simplified model approach as shown in 
Fig.~\ref{fig:DecayTopologies} in order to best make contact with experimental efforts.
Of course, when interpreting such simplified model experimental limits or discoveries in 
terms of some complete theory, one must compute both signal 
\cite{Beenakker:1997ut,Beenakker:2010nq,Beenakker:2011fu}
and background \cite{Cacciari:2011hy,Czakon:2011xx,Czakon:2012zr,Czakon:2012pz}
production cross-sections and the relevant branching ratios
\cite{Djouadi:2006bz} to a high degree of precision. 

The on-shell constrained variables are suitable generalizations of the
stransverse mass variable, $M_{T2}$, which is being used extensively in
experimental searches, and for which several public codes exist.
In contrast, there is no public code which allows the computation of
the $M_2$ variables.  We are developing such a code for public
release in the near future to facilitate the wider use of $M_2$ variables
\cite{Cho:2015laa}.
 
\appendix

\section{The complete set of $M_2$ variables for the $t\bar{t}$ event topology}
\label{sec:appendix}

In this appendix, we collect the specific definitions of the fifteen
$M_2$ variables used in this paper. We consider the three subsystems
in Fig.~\ref{fig:DecaySubsystem} as applied to the $t\bar{t}$
background events of Fig.~\ref{fig:DecayTopologies}(a). We write the
equations in terms of the measured 4-momenta of the $b$ and $\bar{b}$
quarks, $p_b$ and $p_{\bar{b}}$, and the measured 4-momenta of the
lepton and antilepton, $p_{\ell^-}$ and $p_{\ell^+}$.  The invisible
neutrino 4-momenta will be denoted by $q_1$ and $q_2$, where we take
``1'' to refer to the decay chain initiated by a top quark and ``2''
to refer to the decay chain initiated by an anti-top. In each
subsystem, the test mass $\tilde m$ is taken to be the corresponding
true daughter mass.

In the $(ab)$ subsystem the daughter particles are the two neutrinos,
with mass $\tilde m=0$.
The definitions (\ref{M2CCdef}-\ref{MT2def}) imply
\bea
M_{2CC}^2(ab;\tilde m = 0) &\equiv& 
\min_{\substack{\vec{q}_{1},\vec{q}_{2}\\
q_1^2=q_2^2=\tilde m^2=0\\
\vec{q}_{1T}+\vec{q}_{2T} = \mpt  \\
(p_{b}+p_{\ell^+}+q_1)^2 =(p_{\bar b}+p_{\ell^-}+q_2)^2 \\
(p_{\ell^+}+q_1)^2 =(p_{\ell^-}+q_2)^2 }}
\left\{(p_{b}+p_{\ell^+}+q_1)^2 \right\},
\label{eq:m2CCab}
\\ [2mm]
M_{2CX}^2(ab;\tilde m = 0) &\equiv& 
\min_{\substack{\vec{q}_{1},\vec{q}_{2}\\
q_1^2=q_2^2=\tilde m^2=0\\
\vec{q}_{1T}+\vec{q}_{2T} = \mpt  \\
(p_{b}+p_{\ell^+}+q_1)^2 =(p_{\bar b}+p_{\ell^-}+q_2)^2 }}
\left\{(p_{b}+p_{\ell^+}+q_1)^2 \right\},
\label{eq:m2CXab}
\\ [2mm]
M_{2XC}^2(ab;\tilde m = 0) &\equiv& 
\!\!\!\!\!\!\!\!\!
\min_{\substack{\vec{q}_{1},\vec{q}_{2}\\
q_1^2=q_2^2=\tilde m^2=0\\
\vec{q}_{1T}+\vec{q}_{2T} = \mpt  \\
(p_{\ell^+}+q_1)^2 =(p_{\ell^-}+q_2)^2 }}
\!\!\!\!\!\!\!\!\!
\left\{
\max\left[(p_{b}+p_{\ell^+}+q_1)^2\;, (p_{\bar b}+p_{\ell^-}+q_2)^2
\right]\right\},~~~~
\label{eq:m2XCab}
\\ [2mm]
M_{2XX}^2(ab;\tilde m = 0) &\equiv& 
\!\!\!\!\!
\min_{\substack{\vec{q}_{1},\vec{q}_{2}\\
q_1^2=q_2^2=\tilde m^2=0\\
\vec{q}_{1T}+\vec{q}_{2T} = \mpt }}
\!\!\!\!\!
\left\{
\max\left[(p_{b}+p_{\ell^+}+q_1)^2\;, (p_{\bar b}+p_{\ell^-}+q_2)^2
\right]\right\},
\label{eq:m2XXab}
\\ [2mm]
M_{T2}^2(ab;\tilde m = 0) &\equiv&
\!\!\!\!\!\!\!
\min_{\substack{\vec{q}_{1T},\vec{q}_{2T}\\
\vec{q}_{1T}+\vec{q}_{2T} =\mpt }}
\!\!\!\!\!\!\!
\left\{\max\left[M^2_{Tt}(\vec{q}_{1T},\tilde m=0),\;
              M^2_{T\bar{t}}(\vec{q}_{2T},\tilde m=0)\right] \right\}.
\label{eq:mT2ab}
\eea 
In the last equation $M_{Tt}$ ($M_{T\bar{t}}$) is the transverse mass of the 
hypothesized top quark (anti-top quark):
\bea
M^2_{Tt}(\vec{q}_{1T},\tilde m) &\equiv&
 (E_{Tb}+E_{T\ell^+}+E_{T1})^2 -
 (\vec{p}_{Tb}+\vec{p}_{T\ell^+}+\vec{q}_{1T})^2,
\label{MTt} \\ [2mm]
M^2_{T\bar{t}}(\vec{q}_{2T},\tilde m) &\equiv&
 (E_{T\bar{b}}+E_{T\ell^-}+E_{T2})^2 -
 (\vec{p}_{T\bar{b}}+\vec{p}_{T\ell^-}+\vec{q}_{2T})^2,
 \label{MTt-bar} 
\eea
with the transverse energies defined as usual,
e.g.~$E_{Ti}=\sqrt{\tilde m^2 + \vec{q}_{Ti}^{\, 2}}$.

In the $(b)$ subsystem the daughter particles are again the massless
neutrinos, but this time we minimize the masses of the hypothesized
$W^\pm$ particles:
\bea
M_{2CC}^2(b;\tilde m = 0) &\equiv& 
\min_{\substack{\vec{q}_{1},\vec{q}_{2}\\
q_1^2=q_2^2=\tilde m^2=0\\
\vec{q}_{1T}+\vec{q}_{2T} = \mpt  \\
(p_{b}+p_{\ell^+}+q_1)^2 =(p_{\bar b}+p_{\ell^-}+q_2)^2 \\
(p_{\ell^+}+q_1)^2 =(p_{\ell^-}+q_2)^2 }}
\left\{(p_{\ell^+}+q_1)^2 \right\},
\label{eq:m2CCb}
\\ [2mm]
M_{2CX}^2(b;\tilde m = 0) &\equiv& 
\min_{\substack{\vec{q}_{1},\vec{q}_{2}\\
q_1^2=q_2^2=\tilde m^2=0\\
\vec{q}_{1T}+\vec{q}_{2T} = \mpt  \\
(p_{\ell^+}+q_1)^2 =(p_{\ell^-}+q_2)^2 }}
\left\{(p_{\ell^+}+q_1)^2 \right\},
\label{eq:m2CXb}
\\ [2mm]
M_{2XC}^2(b;\tilde m = 0) &\equiv& 
\!\!\!\!\!\!\!\!\!
\min_{\substack{\vec{q}_{1},\vec{q}_{2}\\
q_1^2=q_2^2=\tilde m^2=0\\
\vec{q}_{1T}+\vec{q}_{2T} = \mpt  \\
(p_{b}+p_{\ell^+}+q_1)^2 =(p_{\bar b}+p_{\ell^-}+q_2)^2 }}
\!\!\!\!\!\!\!\!\!
\left\{
\max\left[(p_{\ell^+}+q_1)^2\;, (p_{\ell^-}+q_2)^2 \right]\right\},~~~~
\label{eq:m2XCb}
\\ [2mm]
M_{2XX}^2(b;\tilde m = 0) &\equiv& 
\min_{\substack{\vec{q}_{1},\vec{q}_{2}\\
q_1^2=q_2^2=\tilde m^2=0\\
\vec{q}_{1T}+\vec{q}_{2T} = \mpt }}
\left\{
\max\left[(p_{\ell^+}+q_1)^2\;, (p_{\ell^-}+q_2)^2 \right]\right\},
\label{eq:m2XXb}
\\ [2mm]
M_{T2}^2(b;\tilde m = 0) &\equiv&
\!\!\!\!\!\!\!\!\!
\min_{\substack{\vec{q}_{1T},\vec{q}_{2T}\\
\vec{q}_{1T}+\vec{q}_{2T} =\mpt }}
\!\!\!\!\!\!\!\!\!
\left\{\max\left[M^2_{TW^+}(\vec{q}_{1T},\tilde m=0),\;
              M^2_{TW^-}(\vec{q}_{2T},\tilde m=0)\right] \right\},~~~~
\label{eq:mT2b}
\eea 
with
\bea
M^2_{TW^+}(\vec{q}_{1T},\tilde m) &\equiv&
 (E_{T\ell^+}+E_{T1})^2 - (\vec{p}_{T\ell^+}+\vec{q}_{1T})^2,
 \\ [2mm]
M^2_{TW^-}(\vec{q}_{2T},\tilde m) &\equiv&
 (E_{T\ell^-}+E_{T2})^2 - (\vec{p}_{T\ell^-}+\vec{q}_{2T})^2.
\eea

Finally, consider the $(a)$ subsystem, in which the the parents are
the top quarks,
while the daughters are the $W^\pm$ bosons, with masses $\tilde m=m_W$. 
Denoting now the 4-momenta of the two hypothesized 
$W^\pm$ bosons with $q_{W^+}$ and $q_{W^-}$, we have 
\bea
M_{2CC}^2(a;\tilde m = m_W) &\equiv& 
\min_{\substack{\vec{q}_{W^+},\vec{q}_{W^-}\\
q_{W^+}^2=q_{W^-}^2=\tilde m^2=m_W^2\\
\vec{q}_{TW^+}+\vec{q}_{TW^-} = \mpt +\vec{p}_{T\ell^+} +\vec{p}_{T\ell^-}  \\
(p_{b}+q_{W^+})^2 =(p_{\bar b}+q_{W^-})^2 \\
(q_{W^+}-p_{\ell^+})^2 =(q_{W^-}-p_{\ell^-})^2 }}
\left\{(p_{b}+q_{W^+})^2 \right\},
\label{eq:m2CCa}
\\ [2mm]
M_{2CX}^2(a;\tilde m = m_W) &\equiv& 
\min_{\substack{\vec{q}_{W^+},\vec{q}_{W^-}\\
q_{W^+}^2=q_{W^-}^2=\tilde m^2=m_W^2\\
\vec{q}_{TW^+}+\vec{q}_{TW^-} = \mpt +\vec{p}_{T\ell^+}
+\vec{p}_{T\ell^-}  \\
(p_{b}+q_{W^+})^2 =(p_{\bar b}+q_{W^-})^2 }}
\left\{(p_{b}+q_{W^+})^2 \right\},
\label{eq:m2CXa}
\\ [2mm]
M_{2XC}^2(a;\tilde m = m_W) &\equiv& 
\!\!\!\!\!\!\!\!\!\!\!\!
\min_{\substack{\vec{q}_{W^+},\vec{q}_{W^-}\\
q_{W^+}^2=q_{W^-}^2=\tilde m^2=m_W^2\\
\vec{q}_{TW^+}+\vec{q}_{TW^-} = \mpt +\vec{p}_{T\ell^+}
+\vec{p}_{T\ell^-}  \\
(q_{W^+}-p_{\ell^+})^2 =(q_{W^-}-p_{\ell^-})^2 }}
\!\!\!\!\!\!\!\!\!
\left\{
\max\left[(p_{b}+q_{W^+})^2\;, (p_{\bar b}+q_{W^-})^2 \right]\right\},~~~~
\label{eq:m2XCa}
\\ [2mm]
M_{2XX}^2(a;\tilde m = m_W) &\equiv& 
\!\!\!\!\!\!\!\!\!\!\!\!
\min_{\substack{\vec{q}_{W^+},\vec{q}_{W^-}\\
q_{W^+}^2=q_{W^-}^2=\tilde m^2=m_W^2\\
\vec{q}_{TW^+}+\vec{q}_{TW^-} = \mpt +\vec{p}_{T\ell^+} +\vec{p}_{T\ell^-} }}
\!\!\!\!\!\!\!\!\!
\left\{
\max\left[(p_{b}+q_{W^+})^2\;, (p_{\bar b}+q_{W^-})^2 \right]\right\},
\label{eq:m2XXa}
\\ [2mm]
M_{T2}^2(a;\tilde m = m_W) &\equiv&
\!\!\!\!\!\!\!\!\!\!\!\!\!\!\!\!\!\!\!\!\!\!
\min_{\substack{\vec{q}_{W^+},\vec{q}_{W^-}\\
\vec{q}_{TW^+}+\vec{q}_{TW^-} = \mpt +\vec{p}_{T\ell^+} +\vec{p}_{T\ell^-} }}
\!\!\!\!\!\!\!\!\!\!\!\!\!\!\!\!\!\!\!\!\!\!\!\!\!\!
\left\{\max\left[M^2_{Tt}(\vec{q}_{TW^+},m_W),\;
              M^2_{T\bar{t}}(\vec{q}_{TW^-},m_W)\right] \right\},
\label{eq:mT2a}
\eea 
where now the top quark transverse masses (\ref{MTt}-\ref{MTt-bar})
are rewritten in terms of the $W^\pm$ boson momenta, $q_{W^+}$ and
$q_{W^-}$, as
\bea
M^2_{Tt}(\vec{q}_{TW^+},\tilde m=m_W) &\equiv&
 (E_{Tb}+E_{TW^+})^2 - (\vec{p}_{Tb}+\vec{q}_{TW^+})^2, \\ [2mm]
M^2_{T\bar{t}}(\vec{q}_{TW^-},\tilde m=m_W) &\equiv&
 (E_{T\bar{b}}+E_{TW^-})^2 - (\vec{p}_{T\bar{b}}+\vec{q}_{TW^-})^2,
\eea
with $E_{TW^\pm}=\sqrt{ m_W^2 + \vec{q}_{TW^\pm}^{\, 2}}$.

\acknowledgments
We thank Kaustubh Agashe, Roberto Franceschini, and Kyoungchul Kong for
useful discussions.  
WC acknowledges support by Project Code (IBS-R018-D1);
WC and DK acknowledge support by LHC-TI
postdoctoral fellowships under grant NSF-PHY-0969510.
WC, JG, DK, and KM acknowledge support from U.S. Department of Energy
Grant ER41990.
JG, KM, FM, and LP thank their CMS colleagues for useful discussions.
MP was supported by World Premier International Research Center
Initiative (WPI), MEXT, Japan. MP also acknowledges the
Max-Planck-Gesellschaft, the Korea Ministry of Education, Science and
Technology, Gyeongsangbuk-Do and Pohang City for the support of the
Independent Junior Research Group at the APCTP.

\end{document}